\documentclass[twocolumn]{aastex63}

\pdfoutput=1 
\usepackage{appendix}
\usepackage{amsmath,amstext,amssymb}
\usepackage[T1]{fontenc}
\usepackage[figure,figure*]{hypcap}

\providecommand{\teff}{\ensuremath{T_{\rm eff}}}

\providecommand{\msun}{\ensuremath{\,M_{\odot}}}
\providecommand{\rsun}{\ensuremath{\,R_{\odot}}}

\providecommand{\mj}{\ensuremath{\,M_{\rm J}}}
\providecommand{\rj}{\ensuremath{\,R_{\rm J}}}

\providecommand{\minaus}{{\textsc{Minerva}}-Australis}

\newcommand{\mos}{\,m\,s$^{-1}$}
\newcommand{\kms}{\,km\,s$^{-1}$}

\submitjournal{The Astronomical Journal}
\received{2022 December 11}
\revised{2023 February 22}
\accepted{2023 March 8}
\published{2023 April 19}

\shorttitle{TOI-778~b}
\shortauthors{Clark et al.}

\graphicspath{{./}{figures/}}

\begin{document}

\title{Spinning up a Daze: \textit{TESS} Uncovers a Hot Jupiter orbiting the Rapid-Rotator TOI-778}

\correspondingauthor{Brett Addison}
\email{baddison2005@gmail.com}

\author[0000-0003-3964-4658]{Jake T. Clark}
\affiliation{University of Southern Queensland, Centre for Astrophysics, USQ Toowoomba, West Street, QLD 4350 Australia}

\author[0000-0003-3216-0626]{Brett C. Addison}
\affiliation{University of Southern Queensland, Centre for Astrophysics, USQ Toowoomba, West Street, QLD 4350 Australia}
\affiliation{Swinburne University of Technology, Centre for Astrophysics and Supercomputing, John Street, Hawthorn, VIC 3122, Australia}

\author[0000-0002-4876-8540]{Jack Okumura}
\affil{University of Southern Queensland, Centre for Astrophysics, USQ Toowoomba, West Street, QLD 4350 Australia}

\author[0000-0001-9158-9276]{Sydney Vach}
\affil{Center for Astrophysics \textbar \ Harvard \& Smithsonian, 60 Garden St, Cambridge, MA 02138, USA}
\affiliation{University of Southern Queensland, Centre for Astrophysics, USQ Toowoomba, West Street, QLD 4350 Australia}

\author[0000-0002-0839-4257]{Adriana Errico}
\affiliation{University of Southern Queensland, Centre for Astrophysics, USQ Toowoomba, West Street, QLD 4350 Australia}

\author[0000-0002-8091-7526]{Alexis Heitzmann}
\affiliation{University of Southern Queensland, Centre for Astrophysics, USQ Toowoomba, West Street, QLD 4350 Australia}
\affiliation{Astronomy Unit, Queen Mary University of London, Mile End Road, London E1 4NS, UK}

\author[0000-0001-8812-0565]{Joseph E. Rodriguez}
\affil{Department of Physics and Astronomy, Michigan State University, East Lansing, MI 48824, USA}

\author[0000-0001-7294-5386]{Duncan J. Wright}
\affiliation{University of Southern Queensland, Centre for Astrophysics, USQ Toowoomba, West Street, QLD 4350 Australia}

\author[0000-0001-8998-463X]{Mathieu~Clert{\'e}}
\affiliation{University of Southern Queensland, Centre for Astrophysics, USQ Toowoomba, West Street, QLD 4350 Australia}

\author[0000-0001-6649-4531]{Carolyn J. Brown}
\affil{University of Southern Queensland, Centre for Astrophysics, USQ Toowoomba, West Street, QLD 4350 Australia}

\author[0000-0002-3551-279X]{Tara Fetherolf}
\altaffiliation{UC Chancellor's Fellow}
\affiliation{Department of Earth and Planetary Sciences, University of California Riverside, 900 University Avenue, Riverside, CA 92521, USA}

\author[0000-0001-9957-9304]{Robert A. Wittenmyer}
\affiliation{University of Southern Queensland, Centre for Astrophysics, USQ Toowoomba, West Street, QLD 4350 Australia}

\author[0000-0002-8864-1667]{Peter Plavchan}
\affiliation{Department of Physics \& Astronomy, George Mason University, 4400 University Drive MS 3F3, Fairfax, VA 22030, USA}

\author[0000-0002-7084-0529]{Stephen R. Kane}
\affiliation{Department of Earth and Planetary Sciences, University of California, Riverside, CA 92521, USA}

\author[0000-0002-1160-7970]{Jonathan Horner}
\affiliation{University of Southern Queensland, Centre for Astrophysics, USQ Toowoomba, West Street, QLD 4350 Australia}

\author[0000-0003-0497-2651]{John F.\ Kielkopf}
\affiliation{Department of Physics and Astronomy, University of Louisville, Louisville, KY 40292, USA}

\author[0000-0002-1836-3120]{Avi Shporer}
\affil{Department of Physics and Kavli Institute for Astrophysics and Space Research, Massachusetts Institute of Technology, Cambridge, MA 02139, USA}

\author[0000-0002-7595-0970]{C.G. Tinney}
\affil{Exoplanetary Science at UNSW, School of Physics, UNSW Sydney, NSW 2052, Australia}

\author[0000-0001-5162-1753]{Liu Hui-Gen}
\affil{School of Astronomy and Space Science, Key Laboratory of Modern Astronomy and Astrophysics in Ministry of Education, Nanjing University, Nanjing 210046, Jiangsu, China}

\author[0000-0002-3247-5081]{Sarah Ballard}
\affiliation{Department of Astronomy, University of Florida, 211 Bryant Space Science Center, Gainesville, FL, 32611, USA}

\author[0000-0003-2649-2288]{Brendan P. Bowler}
\affil{Department of Astronomy, The University of Texas at Austin, TX 78712, USA}

\author[0000-0002-7830-6822]{Matthew W. Mengel}
\affil{University of Southern Queensland, Centre for Astrophysics, USQ Toowoomba, West Street, QLD 4350 Australia}

\author[0000-0002-4891-3517]{George Zhou}
\affiliation{University of Southern Queensland, Centre for Astrophysics, USQ Toowoomba, West Street, QLD 4350 Australia}

\author{Annette S. Lee}
\affil{The Native Skywatchers Initiative, MN USA}
\affil{Department of Physics and Astronomy, St. Cloud University, MN USA}
\affil{Department of Astronomy and Astrophysics, University of California, Santa Cruz, CA 95064, USA}
\affil{University of Southern Queensland, Centre for Astrophysics, USQ Toowoomba, West Street, QLD 4350 Australia}

\author[0000-0002-7224-5336]{Avelyn David}
\affil{Department of Physics and Astronomy, St. Cloud University, MN USA}
\affil{The Native Skywatchers Initiative, MN USA}

\author{Jessica Heim}
\affil{Department of Physics and Astronomy, St. Cloud University, MN USA}
\affil{The Native Skywatchers Initiative, MN USA}

\author{Michele E. Lee}
\affil{Jeremiah Horrocks Institute, University of Central Lancashire, Preston, PR1 2HE, UK}
\affil{The Native Skywatchers Initiative, MN USA}

\author{Ver{\'o}nica Sevilla}
\affil{Department of Physics and Astronomy, St. Cloud University, MN USA}
\affil{The Native Skywatchers Initiative, MN USA}

\author{Naqsh E. Zafar}
\affil{Department of Physics and Astronomy, St. Cloud University, MN USA}
\affil{The Native Skywatchers Initiative, MN USA}

\author[0000-0003-0595-5132]{Natalie R. Hinkel}
\affil{Space Science and Engineering Division, Southwest Research Institute, San Antonio, TX 78238, USA}


\author{Bridgette E. Allen}
\affil{NASA Exoplanet Science Institute - Caltech/IPAC, 1200 E. California Blvd, Pasadena, CA 91125 USA} \affil{University of Wisconsin Stout, 712 South Broadway Street, Menomonie, WI 54751 USA }

\author[0000-0001-6023-1335]{Daniel Bayliss}
\affil{Dept. of Physics, University of Warwick, Gibbet Hill Road, Coventry, CV4 7AL, UK}

\author{Arthur Berberyan}
\affil{NASA Exoplanet Science Institute - Caltech/IPAC, 1200 E. California Blvd, Pasadena, CA 91125 USA} \affil{College of the Canyons, 26455 Rockwell Canyon Rd., Santa Clarita, CA 91355, USA}

\author{Perry Berlind} 
\affil{Center for Astrophysics \textbar \ Harvard \& Smithsonian, 60 Garden St, Cambridge, MA 02138, USA}

\author[0000-0001-6637-5401]{Allyson Bieryla}
\affil{Center for Astrophysics \textbar \ Harvard \& Smithsonian, 60 Garden St, Cambridge, MA 02138, USA}

\author{François Bouchy}
\affil{Geneva Observatory, University of Geneva, Chemin Pegasi 51, 1290 Versoix, Switzerland}

\author[0000-0002-9158-7315]{Rafael Brahm}
\affiliation{Facultad de Ingeniería y Ciencias, Universidad Adolfo Ib\'a\~nez, Av.\ Diagonal las Torres 2640, Pe\~nalol\'en, Santiago, Chile}
\affiliation{Millennium Institute for Astrophysics, Chile}
\affiliation{Data Observatory Foundation, Chile}

\author[0000-0001-7904-4441]{Edward M. Bryant}
\affiliation{Dept.\ of Physics, University of Warwick, Gibbet Hill Road, Coventry CV4 7AL, UK}
\affiliation{Centre for Exoplanets and Habitability, University of Warwick, Gibbet Hill Road, Coventry CV4 7AL, UK}

\author[0000-0002-8035-4778]{Jessie L. Christiansen}
\affil{NASA Exoplanet Science Institute - Caltech/IPAC, 1200 E. California Blvd, Pasadena, CA 91125 USA}

\author[0000-0002-5741-3047]{David R. Ciardi}
\affil{NASA Exoplanet Science Institute - Caltech/IPAC, 1200 E. California Blvd, Pasadena, CA 91125 USA}

\author{Krys N. Ciardi}
\affil{NASA Exoplanet Science Institute - Caltech/IPAC, 1200 E. California Blvd, Pasadena, CA 91125 USA} \affil{Rhode Island College, 600 Mount Pleasant Avenue Providence, RI 02908 USA}

\author[0000-0001-6588-9574]{Karen A.\ Collins}
\affiliation{Center for Astrophysics \textbar \ Harvard \& Smithsonian, 60 Garden Street, Cambridge, MA 02138, USA}

\author[0000-0001-6763-4874]{Jules Dallant}
\affil{Geneva Observatory, University of Geneva, Chemin Pegasi 51, 1290 Versoix, Switzerland}

\author[0000-0002-5070-8395]{Allen B. Davis}
\affiliation{Department of Astronomy, Yale University, 52 Hillhouse Avenue, New Haven, CT 06511, USA}
\affiliation{International School of Boston, Cambridge, MA 02140, USA}

\author[0000-0002-2100-3257]{Mat\'ias R. D\'iaz} 
\affiliation{Las Campanas Observatory, Carnegie Institution of Washington, Colina el Pino, Casilla 601 La Serena, Chile}
\affiliation{Departamento de Astronom\'ia, Universidad de Chile, Camino El Observatorio 1515, Las Condes, Santiago, Chile}

\author[0000-0001-8189-0233]{Courtney D. Dressing}
\affil{Astronomy Department, University of California, Berkeley, CA 94720, USA}

\author[0000-0002-9789-5474]{Gilbert A. Esquerdo} 
\affil{Center for Astrophysics \textbar \ Harvard \& Smithsonian, 60 Garden St, Cambridge, MA 02138, USA}

\author[0000-0001-8935-2472]{Jan-Vincent Harre}
\affiliation{Institute of Planetary Research, German Aerospace Center (DLR), Rutherfordstraße 2, 12489 Berlin, Germany}

\author[0000-0002-2532-2853]{Steve B. Howell}
\affil{NASA Ames Research Center, Moffett Field, CA 94035, USA}

\author[0000-0002-4715-9460]{Jon~M.~Jenkins}
\affiliation{NASA Ames Research Center, Moffett Field, CA 94035, USA}

\author[0000-0002-4625-7333]{Eric L. N. Jensen}
\affiliation{Dept.\ of Physics \& Astronomy, Swarthmore College, Swarthmore PA 19081, USA}

\author{Mat{\'i}as I. Jones}
\affil{European Southern Observatory, Alonso de C{\'o}rdova 3107, Vitacura, Casilla 19001, Santiago, Chile}

\author[0000-0002-5389-3944]{Andr\'es Jord\'an} 
\affiliation{Facultad de Ingeniería y Ciencias, Universidad Adolfo Ib\'a\~nez, Av.\ Diagonal las Torres 2640, Pe\~nalol\'en, Santiago, Chile}
\affiliation{Millennium Institute for Astrophysics, Chile}
\affiliation{Data Observatory Foundation, Chile}

\author[0000-0001-9911-7388]{David W. Latham} 
\affil{Center for Astrophysics \textbar \ Harvard \& Smithsonian, 60 Garden St, Cambridge, MA 02138, USA}

\author[0000-0003-2527-1598]{Michael B. Lund}
\affil{NASA Exoplanet Science Institute - Caltech/IPAC, 1200 E. California Blvd, Pasadena, CA 91125 USA}

\author[0000-0003-1631-4170]{James McCormac}
\affiliation{Dept. Physics, University of Warwick, Gibbet Hill Road, CV4 7AL, UK}

\author[0000-0002-5254-2499]{Louise D. Nielsen}
\affil{Geneva Observatory, University of Geneva, Chemin Pegasi 51, 1290 Versoix, Switzerland}

\author{Jon Otegi}
\affil{Geneva Observatory, University of Geneva, Chemin Pegasi 51, 1290 Versoix, Switzerland}

\author[0000-0002-8964-8377]{Samuel N. Quinn} 
\affil{Center for Astrophysics \textbar \ Harvard \& Smithsonian, 60 Garden St, Cambridge, MA 02138, USA}

\author[0000-0002-3940-2360]{Don J. Radford}
\affiliation{Brierfield Observatory, New South Wales, Australia}

\author[0000-0003-2058-6662]{George~R.~Ricker}
\affiliation{Department of Physics and Kavli Institute for Astrophysics and Space Research, Massachusetts Institute of Technology, Cambridge, MA 02139, USA}

\author[0000-0001-8227-1020]{Richard P. Schwarz}
\affiliation{Center for Astrophysics \textbar \ Harvard \& Smithsonian, 60 Garden St, Cambridge, MA 02138, USA}

\author[0000-0002-6892-6948]{Sara~Seager}
\affiliation{Department of Physics and Kavli Institute for Astrophysics and Space Research, Massachusetts Institute of Technology, Cambridge, MA 02139, USA}
\affiliation{Department of Earth, Atmospheric and Planetary Sciences, Massachusetts Institute of Technology, Cambridge, MA 02139, USA}
\affiliation{Department of Aeronautics and Astronautics, MIT, 77 Massachusetts Avenue, Cambridge, MA 02139, USA}

\author[0000-0002-2386-4341]{Alexis M. S. Smith}
\affiliation{Institute of Planetary Research, German Aerospace Center (DLR), Rutherfordstraße 2, 12489 Berlin, Germany}

\author[0000-0003-2163-1437]{Chris Stockdale}
\affiliation{Hazelwood Observatory, Australia}

\author[0000-0001-5603-6895]{Thiam-Guan Tan}
\affiliation{Perth Exoplanet Survey Telescope, Perth, Western Australia}
\affiliation{Curtin Institute of Radio Astronomy, Curtin University, Bentley, Western Australia 6102}

\author[0000-0001-7576-6236]{Stéphane Udry}
\affil{Geneva Observatory, University of Geneva, Chemin Pegasi 51, 1290 Versoix, Switzerland}

\author[0000-0001-6763-6562]{Roland~Vanderspek}
\affiliation{Department of Physics and Kavli Institute for Astrophysics and Space Research, Massachusetts Institute of Technology, Cambridge, MA 02139, USA}

\author[0000-0002-3164-9086]{Maximilian N.\ G{\"u}nther}
\affiliation{European Space Agency (ESA), European Space Research and Technology Centre (ESTEC), Keplerlaan 1, 2201 AZ Noordwijk, The Netherlands}

\author[0000-0002-7846-6981]{Songhu Wang}
\affil{Department of Astronomy, Indiana University, Bloomington, IN 47405, USA}

\author{Geof Wingham}
\affiliation{Mt. Stuart Observatory, New Zealand}

\author[0000-0002-4265-047X]{Joshua N.\ Winn}
\affiliation{Department of Astrophysical Sciences, Princeton University, Princeton, NJ 08544, USA}

\begin{abstract}

NASA's \textit{Transiting Exoplanet Survey Satellite} (\textit{TESS}) mission, has been uncovering a growing number of exoplanets orbiting nearby, bright stars. Most exoplanets that have been discovered by \textit{TESS} orbit narrow-line, slow-rotating stars, facilitating the confirmation and mass determination of these worlds. We present the discovery of a hot Jupiter orbiting a rapidly rotating ($v\sin{(i)}= 35.1\pm1.0$\,\kms) early F3V-dwarf, HD\,115447 (TOI-778). The transit signal taken from Sectors 10 and 37 of \textit{TESS}'s initial detection of the exoplanet is combined with follow-up ground-based photometry and velocity measurements taken from \minaus, TRES, CORALIE and CHIRON to confirm and characterise TOI-778~b. A joint analysis of the light curves and the radial velocity measurements yield a mass, radius, and orbital period for TOI-778~b of $2.76^{+0.24}_{-0.23}$\,\mj, $1.370\pm0.043$\,\rj, and $\sim4.63$ days, respectively. The planet orbits a bright ($V = 9.1$\,mag) F3-dwarf with $M=1.40\pm0.05$\,\msun, $R=1.70\pm0.05$\,\rsun, and $\log g=4.05\pm0.17$. We observed a spectroscopic transit of TOI-778~b, which allowed us to derive a sky-projected spin-orbit angle of $18^{\circ}\pm11^{\circ}$, consistent with an aligned planetary system. This discovery demonstrates the capability of smaller aperture telescopes such as \minaus~to detect the radial velocity signals produced by planets orbiting broad-line, rapidly rotating stars.

\end{abstract}


\section{Introduction}\label{sec:intro}

In the late 1980s, the first exoplanetary candidates around main sequence stars were discovered orbiting Gamma Cephei \citep{GammaCeph} and HD\,114672 \citep{LathamsWorld}\footnote{This companion is likely a low-mass star in a face-on orbit \citep{LathamStar}.}. Soon after, \citet{51Peg} announced the discovery of 51\,Peg\,b, the first planet found orbiting a Sun-like star -- marking the start of the Exoplanet Era.

In the decade that followed that seminal discovery, the radial velocity technique dominated the search for exoplanets, revealing a plethora of ``hot Jupiters'' -- giant planets orbiting their host stars with periods of just a few days \citep[e.g.][]{HJ1,HJ2,HJ3}. Based solely on knowledge of the Solar System, it was broadly expected that planetary systems would feature giant planets on long period orbits, and small, rocky worlds on short period orbits\footnote{For a detailed overview of our knowledge of the Solar System, and its impact on our understanding of exoplanetary science, we direct the interested reader to \citet{SSRev}, and references therein.}. Instead, it became obvious that a significant number of stars \citep[$\sim$1\%, e.g.][]{HJOccur1,HJOccur2,HJOccur3} host scorching hot giant planets -- marking their planetary systems as being truly exotic when compared to our own. 

Such planets (commonly known as ``hot Jupiters'') are by far the easiest exoplanets to detect -- a fact made clear by the great success of the \textit{Kepler} mission. \textit{Kepler} launched in 2009 \citep[see, e.g.,][]{2010Sci...327..977B}, and spent slightly over four years staring continuously at a single patch of the night sky -- in the northern constellation of Cygnus -- monitoring the brightness of more than 150,000 stars. By recording minuscule dips in brightness exhibited by some of those stars, \textit{Kepler}'s primary mission led to the discovery of 3251 planets. However, only a fraction (373, $\sim11\%$) of those planets have both mass and radius measurements as the majority of the stars orbited by those planets are too faint for follow-up radial velocity mass measurements\footnote{As of 8th February, 2023; statistics taken from the NASA Exoplanet Archive counts page, at \url{https://exoplanetarchive.ipac.caltech.edu/docs/counts_detail.html.}}.

The successor to \textit{Kepler} is the \textit{Transiting Exoplanet Survey Satellite}, \textit{TESS} \citep{TESSRick}. Launched in April 2018, \textit{TESS} is currently in the process of scouring the sky, observing hundreds of thousands of the nearest and brightest stars, in an attempt to find short-period planets around them. \textit{TESS} observes the majority of its targets for two consecutive 13.7 day periods, separated by a short window where the spacecraft pivots to broadcast data back to Earth. This means that it is particularly well adapted for the discovery of hot Jupiters as transits of such planets are frequent (multiple transits likely to occur during the $\sim27\,\mathrm{day}$ observing window) and deep. Indeed, the majority of the 291 planets\footnote{As of 8th February, 2023; data courtesy of the NASA Exoplanet Archive's counts page.} confirmed by \textit{TESS} are either hot Jupiters or their smaller siblings, the ``hot Neptunes'' \citep[e.g.][]{TESS1,TESS2,TOI677,AUMic,TOI257,2021AJ....162..292A} and unlike the planets discovered by \textit{Kepler}, a majority (271, $\sim76\%$) of \textit{TESS} planets have both mass and radius measurements.

The origin of hot Jupiters and Neptunes have been the source of much debate. It is widely accepted that such planets cannot have formed on their current orbits, so close to their host stars. Instead, it is thought that they originate at much greater distances, beyond the ``ice-line'' -- the location in the protoplanetary disk around their host at which temperatures were sufficiently low for water ice to be present \citep[e.g.][]{Ice1,Ice2,cooljupiters}. 

Several different mechanisms have been proposed to explain this migration -- all of which likely occur in some, but not all, planetary systems. The current proposals include a smooth process of migration through the protoplanetary disks of their host stars, as the young planet interacts with the material from which it is feeding \citep[e.g.][]{Disk1,Disk3}; planet-planet scattering shifting the planet onto an extremely eccentric orbit, followed by a process of tidal circularisation \citep[e.g.][]{scatter1,scatter2,scatter3}; and secular perturbations imposed by a highly inclined unseen massive companion \citep[the Kozai-Lidov mechanism;][]{Kozai1,Kozai2,Kozai3,Kozai4}.



For individual hot Jupiters, tell-tale signs of their formation pathways may still remain. The orbits of planets that migrate purely as a result of interaction with their host star's protoplanetary disk are expected to remain co-planar with the star's equator, as long as the initial disk isn't tilted \citep[as is the case for < 100 Myr close-orbital giants AU\,Mic\,b, DS Tuc\,Ab, HIP\,67522\,b, V1298\,Tau\,b and c and TOI\,942 c][]{AUMic,DSTucAb,HIP67522b,V1298Tau}, whilst a process of planet-planet scattering can act to moderately incline a planet's orbit relative to that plane. Planets whose migration is driven by the Kozai-Lidov mechanism can become dramatically misaligned with the equators of their host stars -- sometimes even being injected to polar or retrograde orbits \citep[e.g.][]{polar1,polar2,retro1,retro2}.

Studies of the inclination of the orbits of hot Jupiters have revealed a correlation between planetary inclination and host star mass/temperature. The more massive (and hotter) the host star, the more likely it is that any short-period planets discovered in orbit will be strongly misaligned to the star's equatorial plane \citep[e.g.][]{HotMiss1,HotMiss2,HotMiss3}. 

Given that more massive stars are more likely to exhibit binarity \citep{bin1,bin2}, it is possible that the increased numbers of misaligned planets orbiting such stars is a direct result of those stars having undetected massive companions. Equally, more massive stars tend to form more massive planets \citep[e.g.][]{Big1,BigJohnson,Big2,Big3}, and so mechanisms involving planet-planet scattering are also more likely to play a role in determining the obliquities of short period planets. 

To determine the degree to which these different mechanisms contribute to the overall population of short-period planets, it is important to discover and characterise as many such planets, orbiting as wide a variety of stars, as possible. In this work, we present the discovery of a new hot Jupiter orbiting HD\,115447 (also known as TOI-778), an F3-dwarf with a mass of 1.39$\pm$0.02\,\msun\ and a surface temperature of 6875$\pm$190\,K. The candidate planet was detected by \textit{TESS} during Sector 10, in the first year of operation as it surveyed the southern sky. Here, we used follow-up observations from a variety of ground-based facilities to confirm the existence of TOI-778\,b, and characterise both the planet and its orbit around TOI-778.

In Section~\ref{sec:observations}, we describe the observations of TOI-778 followed by the latest characterisation of the star in Section~\ref{sec:thestar} from the observations we obtained. We then present the results of our analysis in Section~\ref{sec:Results}, discuss those results in Section~\ref{sec:Discussion}, and then conclude in Section~\ref{sec:Conclusion}.


\section{Observations and Data Reduction}\label{sec:observations}

In this section, we describe the photometric, spectroscopic, and imaging data sets used to validate the planetary nature of TOI-778~b. 

\subsection{Photometric Observations}\label{sec:light} 

\subsubsection{\textit{TESS} Light Curve} 

\begin{figure*}
  \centering
  \includegraphics[width=\textwidth]{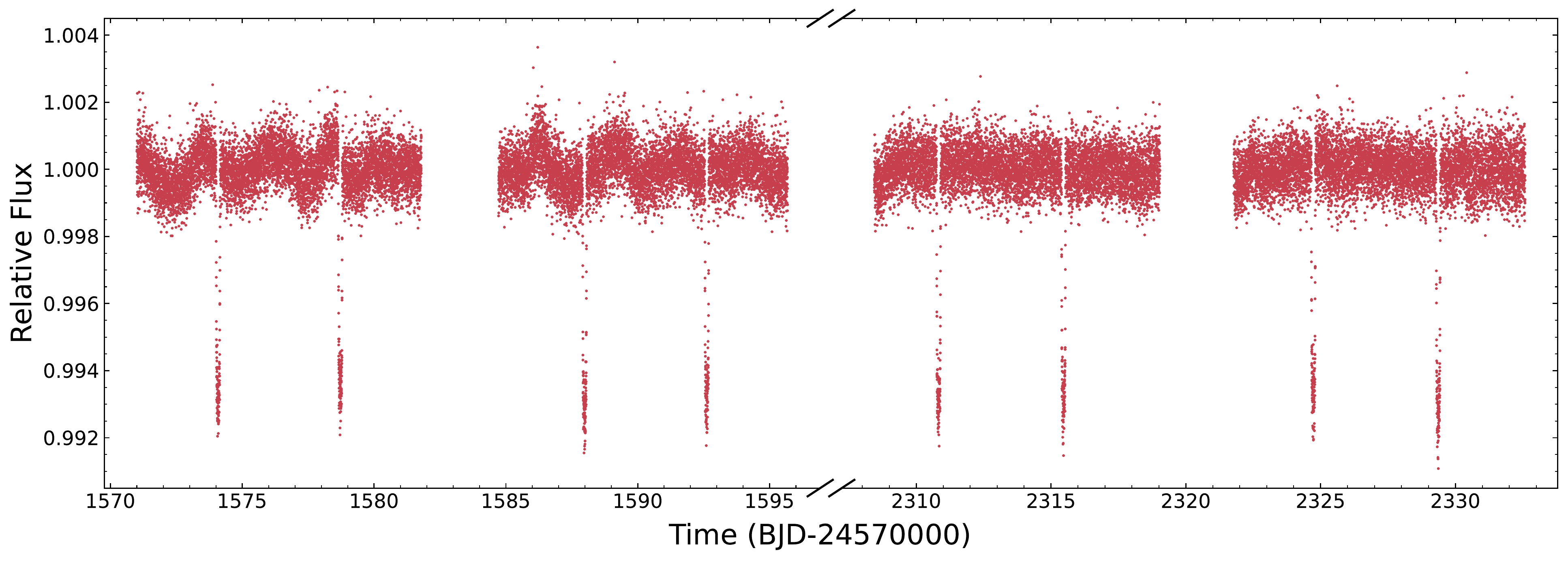}
  \caption{Full TESS PDCSAP median corrected light curves for TOI-778 from Sector 10 and Sector 37}\label{fig:tesslc}
\end{figure*}

TOI-778 (TIC\,335630746) was observed by \textit{TESS} during Sector 10 of its primary mission, from March 26 to April 21, 2019, and extended via Sector 37 observations, taken between April 2 and April 28, 2021. The target star was identified as a planet host candidate via the analyses of the Science Processing Observation Center \citep[SPOC,][]{2016SPIE.9913E..3EJ}, as described by \citet{2002ApJ...575..493J,2010SPIE.7740E..0DJ,Twicken:DVdiagnostics2018,Li:DVmodelFit2019,2020TPSkdph}, using light curves extracted from the two minute target pixel files. We used the Presearch Data Conditioning Simple Aperture Photometry flux values from the \textit{TESS} light curve, removing nonzero flagged data (such as momentum dumps) that could contaminate the light curve analysis. The light curves were then normalised by the median flux values and used in the analysis for TOI-778~b's confirmation. The resulting light curves are found in Figure \ref{fig:tesslc}.

\subsubsection{Next Generation Transit Survey}

A transit egress of TOI-778~b was observed using the Next Generation Transit Survey \citep[NGTS;][]{wheatley2018ngts} on UTC 2019 June 22. NGTS is an exoplanet hunting facility located at the ESO Paranal Observatory in Chile, which consists of twelve independently operated robotic telescopes. Each telescope has a 20\,cm diameter and an 8 square-degree field-of-view. The NGTS telescopes use the DONUTS auto-guiding algorithm \citep{mccormac13donuts} to achieve sub-pixel guiding. TOI-778 was observed simultaneously using two NGTS telescopes, and such multi-telescope observations have been shown to significantly improve the photometric precision of the observations \citep{smith2020multicam, bryant2020multicam}. A total of 1486 images were obtained using an exposure time of 10\,s and the custom NGTS filter (520 - 890\,nm). The airmass of the target was kept below 2 and the sky conditions were optimal throughout the observations.

The images were reduced using a custom aperture photometry pipeline \citep{bryant2020multicam} which uses the \texttt{SEP} library \citep{bertin96sextractor, Barbary2016} for source extraction and photometry. The pipeline automatically identifies comparison stars using \textit{Gaia} DR2 \citep{GAIA, GAIADR2}, ranking the stars in the field based on their similarity to TOI-778 in terms of brightness, colour and CCD position.


\subsubsection{Perth Exoplanet Survey Telescope}

We observed an egress of TOI-778 on UTC 2020 March 31 in V-band from the Perth Exoplanet Survey Telescope (PEST) near Perth, Australia. The 0.3 m telescope is equipped with a $1530\times1020$ SBIG ST-8XME camera with an image scale of 1$\farcs$2 pixel$^{-1}$ resulting in a $31\arcmin\times21\arcmin$ field of view. A custom pipeline based on {\tt C-Munipack}\footnote{http://c-munipack.sourceforge.net} was used to calibrate the images and extract the differential photometry, using an aperture with radius 7$\farcs$4. The images have typical stellar point spread functions (PSFs) with a full-width-half-maximum (FWHM) of $\sim4\arcsec$.

\subsubsection{LCO SAAO}

We observed a full transit of TOI-778 in Pan-STARSS $z$-short band on UTC 2020 May 30 from the LCOGT \citep{Brown:2013} 1.0\,m network node at South Africa Astronomical Observatory. We used the {\tt \textit{TESS} Transit Finder}, which is a customized version of the {\tt Tapir} software package \citep{Jensen:2013}, to schedule our transit observations. The $4096\times4096$ LCOGT SINISTRO cameras have an image scale of 0.389$\arcsec$ per pixel, resulting in a $26\arcmin\times26\arcmin$ field of view. The images were calibrated by the standard LCOGT {\tt BANZAI} pipeline \citep{McCully:2018}, and photometric data were extracted with {\tt AstroImageJ} \citep{Collins:2017}. The images were defocused to a FWHM of $\sim6\farcs7$, and circular apertures with radius $8\farcs2$ were used to extract the differential photometry.

\subsubsection{Mount Kent Observatory}

On 2020 June 5 at the Mount Kent Observatory, a photometric transit observation of TOI-778 was taken simultaneously with radial velocity observations from \minaus. The observation was performed with the Shared Skies Partnership's Planewave CDK700 telescope equipped with an Alta U16M Apogee camera. All data was taken using a Sloan i' filter with a 27.3'x27.3' field of view. All data reduction and analysis was completed using the {\tt AstroImageJ} software package.

\subsubsection{Mt. Stuart}

We observed TOI-778\,b on UTC 2020 April 28 in Sloan $r'$ band from the Mt. Stuart Observatory near Dunedin, New Zealand. The 0.32 m telescope is equipped with a $3072\times2048$ SBIG STXL6303E camera with an image scale of 0$\farcs$88 pixel$^{-1}$ resulting in a $44\arcmin\times30\arcmin$ field of view. The images were calibrated and photometric data were extracted with {\tt AstroImageJ} using a circular aperture with radius $3\farcs5$.

\subsection{Spectroscopic Observations} 

In order to obtain precise radial velocity follow-up data and stellar properties for TOI-778,  we carried out observations using four different facilities. 
Here we give details about the observations carried out by each instrument. 

\startlongtable
\begin{deluxetable}{ccccc}
\tabletypesize{\scriptsize}
\tablecaption{Radial Velocities for TOI-778}
\label{tab:allRV}
\tablehead{
\colhead{Time} & \colhead{Velocity} & \colhead{Uncertainty} & \colhead{SNRe} & \colhead{Instrument}\\
\colhead{[BJD]} & \colhead{[\mos]} & \colhead{[\mos]} & \colhead{ } & \colhead{ }}
\startdata
2458662.63164 & 227 & 129 & 46 & Chiron \\
2458662.63531 & -109 & 142 & 45 & Chiron \\
2458662.63899 & 32 & 130 & 48 & Chiron \\
2458664.59022 & -70 & 103 & 61 & Chiron \\
2458664.59389 & 166 & 110 & 59 & Chiron \\
2458652.69011 & -6229 & 133 & 7 & Coralie \\
2458653.67374 & -6090 & 131 & 19 & Coralie \\
2458654.68544 & -5971 & 106 & 5 & Coralie \\
2458670.48753 & -5466 & 56 & 53 & Coralie \\
2458676.51154 & -5935 & 66 & 48 & Coralie \\
2458647.92621 & -6528 & 149 & 24 & M-A T1 \\
2458647.94762 & -7042 & 228 & 35 & M-A T1 \\
2458654.02680 & -6794 & 189 & 29 & M-A T1 \\
2458665.04743 & -6480 & 249 & 22 & M-A T1 \\
2458673.95523 & -6876 & 96 & 30 & M-A T1 \\
2458647.92621 & -6303 & 179 & 36 & M-A T3 \\
2458647.94762 & -6453 & 200 & 19 & M-A T3 \\
2458654.02680 & -6898 & 177 & 26 & M-A T3 \\
2458662.02455 & -6579 & 229 & 27 & M-A T3 \\
2458662.04596 & -6457 & 318 & 30 & M-A T3 \\
2459004.87332 & -70 & 106 & 18 & M-A T4RM \\
2459004.88826 & 73 & 82 & 24 & M-A T4RM \\
2459004.90319 & 85 & 74 & 27 & M-A T4RM \\
2459004.91813 & 198 & 84 & 23 & M-A T4RM \\
2459004.93306 & 15 & 66 & 31 & M-A T4RM \\
2458649.66043 & 182 & 100 & 57 & TRES \\
2458651.66355 & 336 & 123 & 50 & TRES \\
2458653.65718 & -36 & 119 & 40 & TRES \\
2458653.66621 & -29 & 136 & 40 & TRES \\
2458656.66766 & 615 & 126 & 54 & TRES \\
\enddata
\tablecomments{\minaus\, is given as M-A. Table~\ref{tab:allRV} is published in its entirety in machine-readable format online. A portion is shown here for guidance regarding its form and content.}
\end{deluxetable}

\subsubsection{MINERVA-Australis}

We carried out the spectroscopic observations of TOI-778 using the \minaus~facility \citep{2018arXiv180609282W,addison2019,TOI257}. \minaus~consists of an array of four independently operated 0.7\,m CDK700 telescopes situated at the Mount Kent Observatory in Queensland, Australia \citep{addison2019}. Each telescope simultaneously feeds stellar light via fiber optic cables to a single KiwiSpec R4-100 high-resolution ($R=80,000$) spectrograph \citep{2012SPIE.8446E..88B} with wavelength coverage from 480 to 620\,nm. TOI-778 was observed on 71 epochs with three telescopes (labelled `T1, T3, T4') between 13 June 2019 and 4 June 2020. Each epoch consists of two 30-minute exposures and the resulting spectra had a signal-to-noise per resolution element (SNRe) range between 15 and 36. The resulting radial velocities and SNRe of each spectrum are given in Table~\ref{tab:allRV}. Radial velocities for the observations are derived for each telescope by cross-correlation, where the template being matched is the mean spectrum of each telescope. The instrumental variations are corrected by using simultaneous Thorium-Argon (ThAr) arc lamp observations. Radial velocities computed from different \minaus~telescopes are modeled in Section~\ref{sec:Results} as originating from independent instruments.

\subsubsection{TRES}

We obtained additional observations of TOI-778 via the Tillinghast Reflector Echelle Spectrograph \citep[TRES,][]{Furesz:2008} on the 1.5\,m reflector at the Fred Lawrence Whipple Observatory in Arizona, USA. TRES is a fiber-fed echelle with a resolving power of $R\sim 44,000$ over the wavelength range of $390-910$\,nm. Thirteen TRES radial velocities were collected between 15 June 2019 and 29 February 2020 using the standard observing procedure of obtaining a set of three science observations surrounded by ThAr calibration spectra. The science spectra are then combined to remove cosmic rays and wavelength calibrated using the ThAr spectra. The extraction technique follows procedures outlined in \citet{Buchhave:2010}. The spectra had a SNRe range between 41-79. The observation on the night of July 29, 2020 was discarded due to a telescope issue combined with partly cloudy skies. We derived multi-order radial velocities using the remaining 12 spectra by cross correlating each spectrum, order-by-order, against the strongest SNRe spectrum.


\begin{figure*}
  \centering
  \includegraphics[width=2\columnwidth]{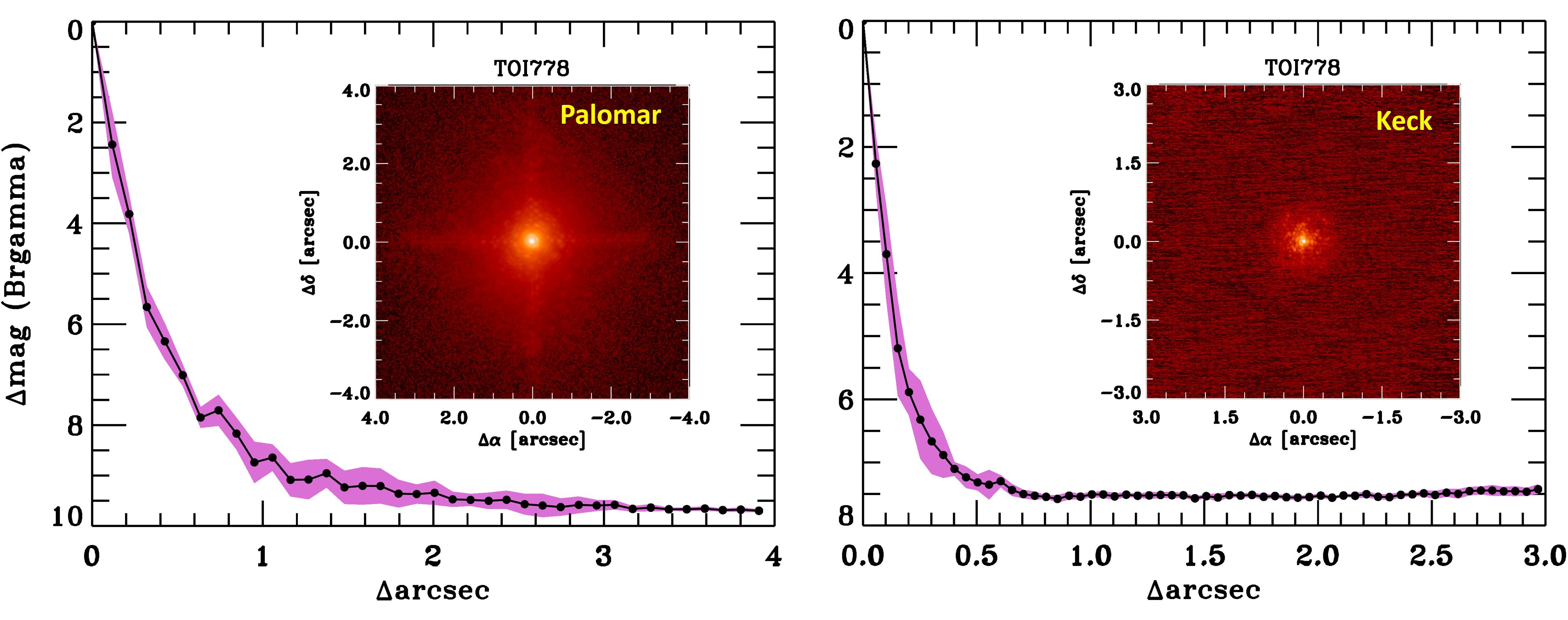}
  \caption{Companion sensitivity for the near-infrared adaptive optics imaging. The black points represent the 5$\sigma$ limits and are separated in steps of 1 FWHM; the purple represents the azimuthal dispersion (1$\sigma$) of the contrast determinations (see Section\,\ref{AO}). The inset image is of the primary target showing no additional close-in companions.\\}\label{fig:ao_contrast} 
\end{figure*}

\subsubsection{CORALIE}
The high resolution spectrograph CORALIE \citep{CORALIE} is mounted on the Swiss 1.2m Euler telescope at La Silla, Chile. The instrument is fed with a 2\arcsec\ on-sky science fibre and a simultaneous Fabry-P\'erot wavelength calibration fibre, with a resolution of $R=60,000$. 
A total of 28 spectra of TOI-778 were obtained between 17 June 2019 and 1 September 2019 by the Swiss CORALIE team and the WINE-collaboration. One epoch was discarded due to low S/N, leaving 27 remaining spectra with SNRe 10-50 at a wavelength 550\,nm. All spectra were extracted using the standard CORALIE data reduction pipeline. 

Radial velocities were extracted through the cross-correlation technique \citep{baranne1996}. We used a weighted binary mask corresponding to an A0-star dominated by H and Fe lines. This mask highly favours the strongest ~1000 absorption lines seen in hot stars \citep{2020A&A...638A..87W}. TOI-778 is a rapidly rotating star, resulting in non-Gaussian absorption lines. We therefore fit a rotational profile to the cross-correlation functions, as done for WASP-189 in \cite{WASP189}. Through this method we achieve a typical radial velocity precision of 80~$\mathrm{m\ s^{-1}}$.

\subsubsection{CHIRON}

We obtained 27 spectra of TOI-778 using the CHIRON spectrograph \citep{Tokovinin2013} on the Small and Moderate Aperture Research Telescope System (SMARTS) 1.5 m telescope at Cerro Tololo, Chile. The CHIRON spectra were obtained using the $R=80,000$ slicer mode, and each spectrum is bracketed by a pair of ThAr lamp exposures for wavelength calibration. 

This combination allows for higher throughput at the cost of some instrumental radial velocity precision. For this early-type star, however, the radial velocity uncertainties are dominated by the broad and sparse spectral lines, rather than wavelength calibration errors or line-spread function drifts (both of which are better addressed with the iodine cell, rather than ThAr). CHIRON's fiber has an on-sky radius of 1\farcs35 with individual exposure times set to 5 minutes in length. Three back-to-back exposures were taken per night that we observed TOI-778.

The radial velocities were derived following the procedure described in \citet{Jones2017,Wang2019,Davis2020}. To summarise, each set of observations of TOI-778 during a night was followed by a ThAr exposure, and the CHIRON pipeline \citep{Paredes2021} uses this lamp to compute a new wavelegth solution; this method yields a demonstrated long-term stability of $\sim$ 15~m~s$^{-1}$, on bright targets. Individual CHIRON spectra are shifted to a common rest frame and then stacked to form a template. We compute the  Cross-Correlation Function (CCF) between each observed spectrum and this template. We then fit a Gaussian function plus linear trend to the CCF, and take the maximum of the fit to be the radial velocity for that observation. This method is repeated for the 33 echelle orders between ${\sim}$470--650~nm where we have good wavelength calibration. The final radial velocity at each epoch is obtained from the median of the individual order velocities, after applying a 3$\sigma$ rejection method. radial velocity uncertainties are computed from the error in the mean of the non-rejected velocities (as in \citealt{Jones2017}). For this star, the typical radial velocity error found was about 150~$\mathrm{m\ s^{-1}}$.

\subsection{High Angular Resolution Imaging}\label{AO} 

As part of our standard process for validating transiting exoplanets to assess the possible contamination of bound or unbound companions on the derived planetary radii \citep{ciardi2015}, we observed TOI-778 with high-resolution near-infrared adaptive optics (AO) imaging at Palomar and Keck Observatories and with optical speckle interferometric imaging at Gemini-South. The infrared observations provide the deepest sensitivities to faint companions while the optical speckle observations provide the highest resolution imaging making the two techniques complementary. Additionally, since the speckle interferometry overlaps with the \textit{TESS} band pass, no assumptions were needed to be made about the SED of possible contaminants.

\subsubsection{Near-Infrared AO}
The Palomar Observatory observations were made with the Palomar High Angular Resolution Observer (PHARO) instrument \citep{hayward2001} behind the natural guide star AO system P3K \citep{dekany2013} on 2020~Jun~12 UT in a standard 5-point quincunx dither pattern with steps of 5\arcsec\ in the narrow-band $Br-\gamma$ filter $(\lambda_o = 2.1686; \Delta\lambda = 0.0326~\mu$m). Each dither position was observed three times, offset in position from each other by 0.5\arcsec\ for a total of 15 frames; with an integration time of 10 seconds per frame, the total on-source time was 150 seconds. PHARO has a pixel scale of $0.025\arcsec$ per pixel for a total field of view of $\sim25\arcsec$. 

The Keck Observatory observations were made with the NIRC2 instrument on Keck-II behind the natural guide star AO system \citep{wizinowich200} on June 25, 2019 in the standard 3-point dither pattern that is used with NIRC2 to avoid the left lower quadrant of the detector which is typically noisier than the other three quadrants. The dither pattern step size was $3\arcsec$ and was repeated twice, with each dither offset from the previous dither by $0.5\arcsec$. NIRC2 was used in the narrow-angle mode with a full field of view of $\sim10\arcsec$ and a pixel scale of approximately 10 milliarcsec per pixel. The Keck observations were made in both the narrow-band $Br-\gamma$ $(\lambda_o = 2.1686; \Delta\lambda = 0.0326~\mu$m) and the $J-cont$ $(\lambda_o = 1.2132; \Delta\lambda = 0.0198~\mu$m) filters with an integration time for each filter of 1 second for a total of 9 seconds on target.

The AO data were processed and analyzed with a custom set of IDL tools. The science frames were flat-fielded and sky-subtracted. The flat fields were generated from a median average of dark subtracted flats taken on-sky. The flats were normalised such that the median value of the flats is unity. The sky frames were generated from the median average of the 15 dithered science frames; each science image was then sky-subtracted and flat-fielded. The reduced science frames were combined into a single combined image using a intra-pixel interpolation that conserves flux, shifts the individual dithered frames by the appropriate fractional pixels, and median-coadds the frames. The final resolution of the combined dithers was determined from the full-width half-maximum of the point spread function; 0.105\arcsec\ and 0.050\arcsec\ for the Palomar and Keck observations respectively.

To within the limits of the AO observations, no stellar companions were detected. The sensitivities of the final combined AO image were determined by injecting simulated sources azimuthally around the primary target every $20^\circ $ at separations of integer multiples of the central source's FWHM \citep{furlan2017, lund2020}. The brightness of each injected source was scaled until standard aperture photometry detected it with $5\sigma $ significance. The resulting brightness of the injected sources relative to TOI-778 set the contrast limits at that injection location. The final $5\sigma $ limit at each separation was determined from the average of all of the determined limits at that separation and the uncertainty on the limit was set by the rms dispersion of the azimuthal slices at a given radial distance (Figure~\ref{fig:ao_contrast}).

\subsection{Optical Speckle Interferometry}

\begin{figure}[h]
  \centering
  \includegraphics[width=0.46\textwidth]{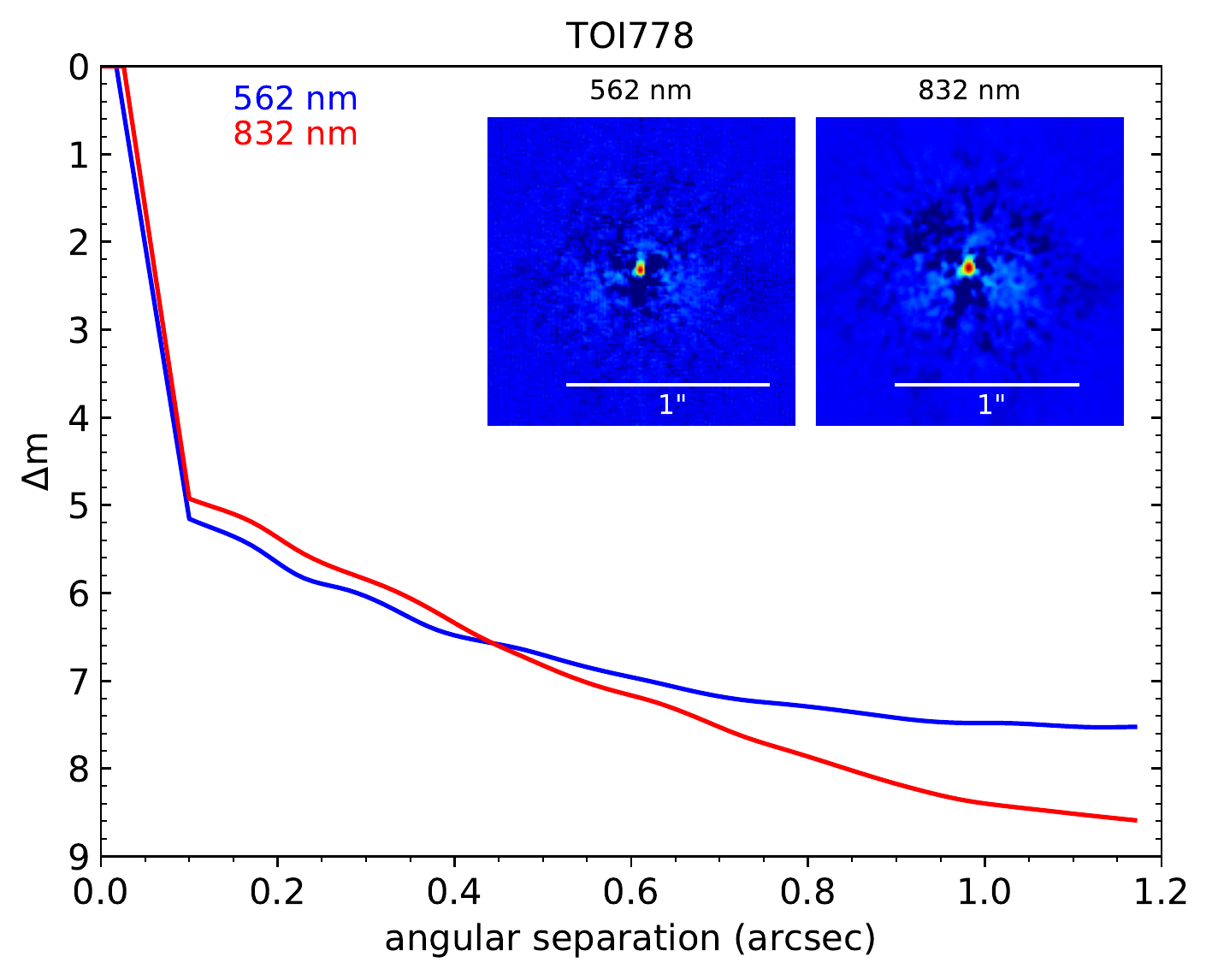}
  \caption{Speckle interferometric contrast curves and reconstructed images for the 562 nm and 832 nm observations. No companion was detected within the spatial limits of the diffraction limit and 1.2" equaling 3-4 to 196 au at the distance of TOI-778} \label{fig:speckle-final} 
	
\end{figure}

TOI-778 was observed on March 16, 2020, using the Zorro speckle interferometric instrument 
mounted on the 8 m Gemini South telescope on the summit of Cerro Pachon in Chile. Zorro simultaneously obtains diffraction-limited images at 562 and 832 nm. Our data set consisted of 3 minutes of total integration time on source taken as sets of 1000 x 0.06 s images plus a consecutive observation of a Point Spread Function (PSF) standard star. Following Howell et al. (2011), we combined all images and subjected them to Fourier analysis to produce speckle reconstructed imagery from which the 5$\sigma$ contrast curves are derived for each passband and nearby companion stars searched for (Figure~\ref{fig:speckle-final}). Our data reveal TOI-778 to be a single star to contrast limits of
5–8.5 magnitudes covering a spatial range of $\sim$3 to 196 au at the distance to TOI-778. 

\section{Stellar Properties of HD\,115447}\label{sec:thestar}

The planetary properties of TOI-778 b depend upon the stellar properties of its host star. We first combine the \minaus~spectra of TOI-778 to create a median spectrum to input into \texttt{iSpec} \citep{iSpec1,iSpec2}. \texttt{iSpec} uses a grid-modelling approach to calculate the effective temperature (\teff), surface gravity (${\rm \log}~g$) and overall metallicity ([M/H]) from the input spectra. These spectroscopic properties are then used along with other photometric and astrometric data as input for the Bayesian isochrone modelling program \texttt{isochrones}. We used the \textit{Gaia} DR2 \citep{GAIADR2} parallax and magnitudes ($G$, $G_R$ and $G_B$), 2MASS \citep{2MASS} magnitudes ($J$, $H$ and $K_s$) V-band magnitude ($V$) and colour excess (E(B-V)), with the \texttt{iSpec} values as input for the \texttt{isochrones} analysis. The resulting derived \texttt{isochrone} properties for TOI-778, including stellar mass, radius, luminosity, and age are given in Table~\ref{tab:star}. Our results are consistent with version 9 of the \textit{TESS} Input Catalog \citep{TIC19} and are the parameters used to further characterise the planetary nature of TOI-778~b. The above procedure of determining the stellar properties of TOI-778 is similar to that of \citet{TOI257}.

We also calculated the rotation period of TOI-778 through the light curve obtained by \textit{TESS} (discussed in more detail within Section \ref{sec:light}). Using \textsc{SciPy}'s Lomb-Scargle periodogram \citep{scipy} on the light curves collected by \textit{TESS} (plotted in Figure \ref{fig:tesslc}), we performed the analysis in two search ranges: 0.01--1.5\,days and 1--13\,days. The periodogram analysis was part of a larger Variability Catalog of TESS light curves \citep[see,][]{2022arXiv220811721F}, and TOI-778 did make it into the final Variability Catalog after several careful vetting steps to remove false positives that could be caused by spacecraft systematics (e.g., momentum dumps).

\begin{figure*}
  \centering
  \includegraphics[width=\textwidth]{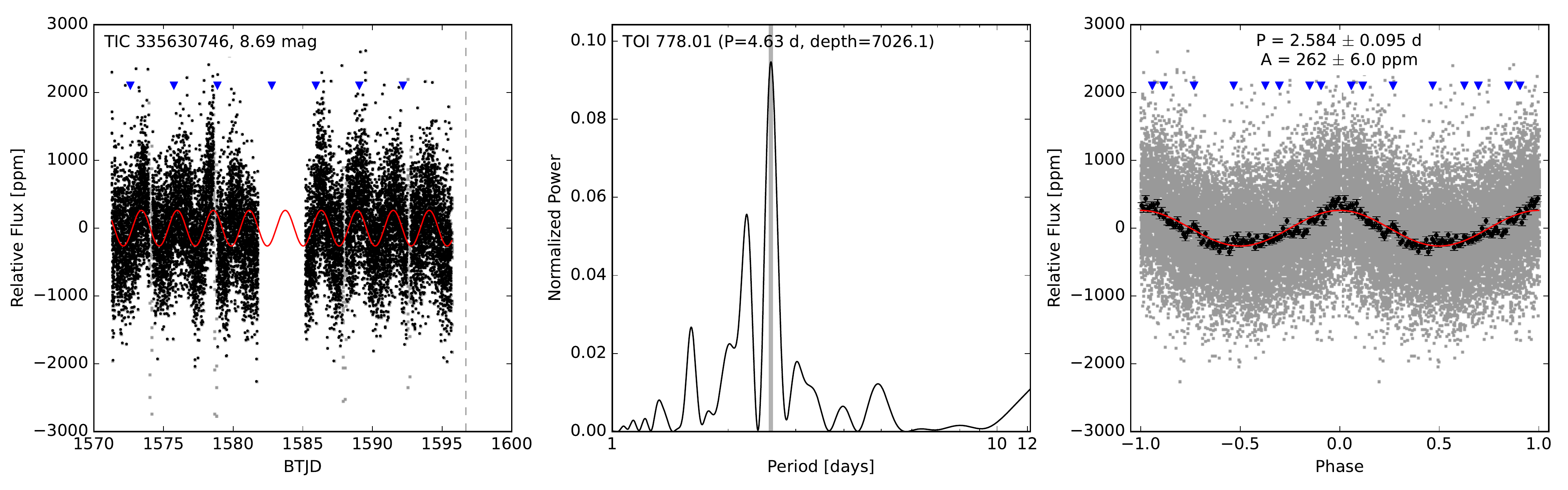}
  \caption{Left: Relative flux values from TOI-778 during Sector 10 (black dots). The transit events of TOI-778~b that have been removed for the periodogram analysis are shown as grey points with the periodic signal of 2.584\,days over-plotted in red. Momentum dumps from \textit{TESS} are shown by the blue triangles. Centre: The periodogram from the \textit{TESS} light curve. The transits have been removed from the light curve for the periodogram analysis such that the periodogram appears to be near zero at ~4.5 days. b. Right: A phase-folded version of Left figure, with the period and semi-amplitude of the periodic variations listed at the top of the figure.}\label{fig:period}
\end{figure*}

A significant signal was found in the 1--13\,day periodogram search corresponding to a stellar rotation period for TOI-778 to be $2.567\pm0.095$\,days. Figure \ref{fig:period} shows a phase-folded plot of TOI-778's light curve to the period of 2.567 days. There seems to be no correlation between the light curve modulation and the momentum dumps of TESS shown in Figure \ref{fig:period} and the normalized power of the periodogram is sufficiently high, such that we are confident that this modulation is astrophysical in nature. We include our derived stellar rotation value in Table \ref{tab:star}.

\begin{deluxetable}{lccc}[h]
\tablewidth{\columnwidth}
\tablecolumns{4}
\tablecaption{Stellar Parameters for HD\,115447. \textbf{Notes.}--$^{\dagger}$Derived using \minaus\, spectroscopic observations. $^{\ddagger}$Derived using spectroscopic observations from TRES. $^{\star}$Preferred solution and used as a prior in the \texttt{Allesfitter} analysis. $^{\wedge}$Lower limit placed on $v \sin i$ uncertainty of $\pm1.0$\,\kms\, due to contributions of other sources of absorption line-broadening. \label{tab:star}}
\tablehead{
\colhead{Stellar Parameters} & \colhead{Value} & \colhead{Source}}
\startdata
\hline
\textbf{Catalog Information} & & \\
Right Ascension (h:m:s)  & 13:17:20.189 & 1 \\
Declination (d:am:as)    & -15:16:24.944 & 1 \\
Parallax (mas)      & 6.15415~$\pm$~0.04231 & 1 \\
$\mu$\textsubscript{R.A} (mas yr\textsuperscript{-1}) &-60.600~$\pm$~0.083& 1\\
$\mu$\textsubscript{Dec.} (mas yr\textsuperscript{-1}) &-26.012~$\pm$~0.065& 1 \\
\textit{Gaia} DR2 ID & 3607877948613218304 & 1\\
\textit{2MASS} ID & J13172019-1516248 & 2\\
HD ID & 115447 & \\
TIC ID & 335630746 & 3 \\
TOI ID & 778 & \\
\textbf{Spectroscopic Properties} & & \\
Spectral type       & F2 & 4 \\
             & F3V & 5 \\
$T_{\rm eff}$ (K)     & $6715\pm128$ & 3 \\
             & $6875\pm190$\,$^{\dagger}$ & 7 \\
             & $6643\pm150$\,$^{\ddagger,\star}$ & 7 \\
$\log g$ (cgs)      & $4.144\pm0.085$ & 3 \\
             & $4.05\pm0.17$\,$^{\dagger}$ & 7 \\
             & $3.98\pm0.22$\,$^{\ddagger}$ & 7 \\
Metallicity, [m/H]    & $0.00\pm0.08$\,$^{\dagger}$ & 7 \\
Metallicity, [m/H]    & $0.03\pm0.08$\,$^{\ddagger}$ & 7 \\
$v \sin i$ (\kms)    & $28.9\pm3.7$\,$^{\dagger}$ & $7$ \\
   & $35.1\pm1.0$\,$^{\ddagger,\star,\wedge}$ & $7$ \\
\textbf{Photometric Properties} & & \\
$G$ (mag)         & 8.9944~$\pm$~0.0007 & 1 \\
$G_{BP}$ (mag)      & 9.226~$\pm$~0.002 & 1 \\
$G_{RP}$ (mag)      & 8.648~$\pm$~0.002 & 1 \\
$J$ (mag)         & 8.25~$\pm$~0.02 & 2 \\
$H$ (mag)         & 8.09~$\pm$~0.03 & 2 \\
$K_s$ (mag)        & 8.055~$\pm$~0.033 & 2 \\
$V$ (mag)         & 9.11~$\pm$~0.02 & 6 \\
$T$ (mag)         & 8.690~$\pm$~0.006 & 3 \\
\textbf{Derived Stellar Properties} & & \\
$M_{\star}$ ($M_{\odot}$) & 1.428~$\pm$~0.094 & 3 \\
             & $1.40\pm0.05$\,$^{\star}$ & 7 \\
$R_{\star}$ ($R_{\odot}$) & 1.677~$\pm$~0.068 & 3 \\
             & $1.71\pm0.05$\,$^{\star}$ & 7 \\
$\rho_{\star}$ (g cm$^{-3}$) & 0.40 ~$\pm$~ 0.03 & 7 \\
$L_{\star}$ ($L_{\odot}$) & 5.153~$\pm$~0.269 & 3 \\
             & 5.76~$\pm$~0.65 & 7 \\
Age (Gyr)         & $1.95^{+0.14}_{-0.13}$ & 7 \\
Distance (pc)       & $162\pm1$ & 1 \\
Rotation Period (days)  & 2.584$\pm$0.095 & 7 \\
\enddata
\raggedright
\tablerefs{1. \cite{GAIADR2}; 2. \cite{2MASS} ; 3. \cite{TIC19}; 4. \cite{CP93}; 5. \cite{1988Houk}, 
6. \cite{2000Hog}, 7. This work.}

\end{deluxetable}

\section{Joint Analysis and Results}\label{sec:Results}

\begin{figure*}
  \centering
  \includegraphics[width=1\columnwidth]{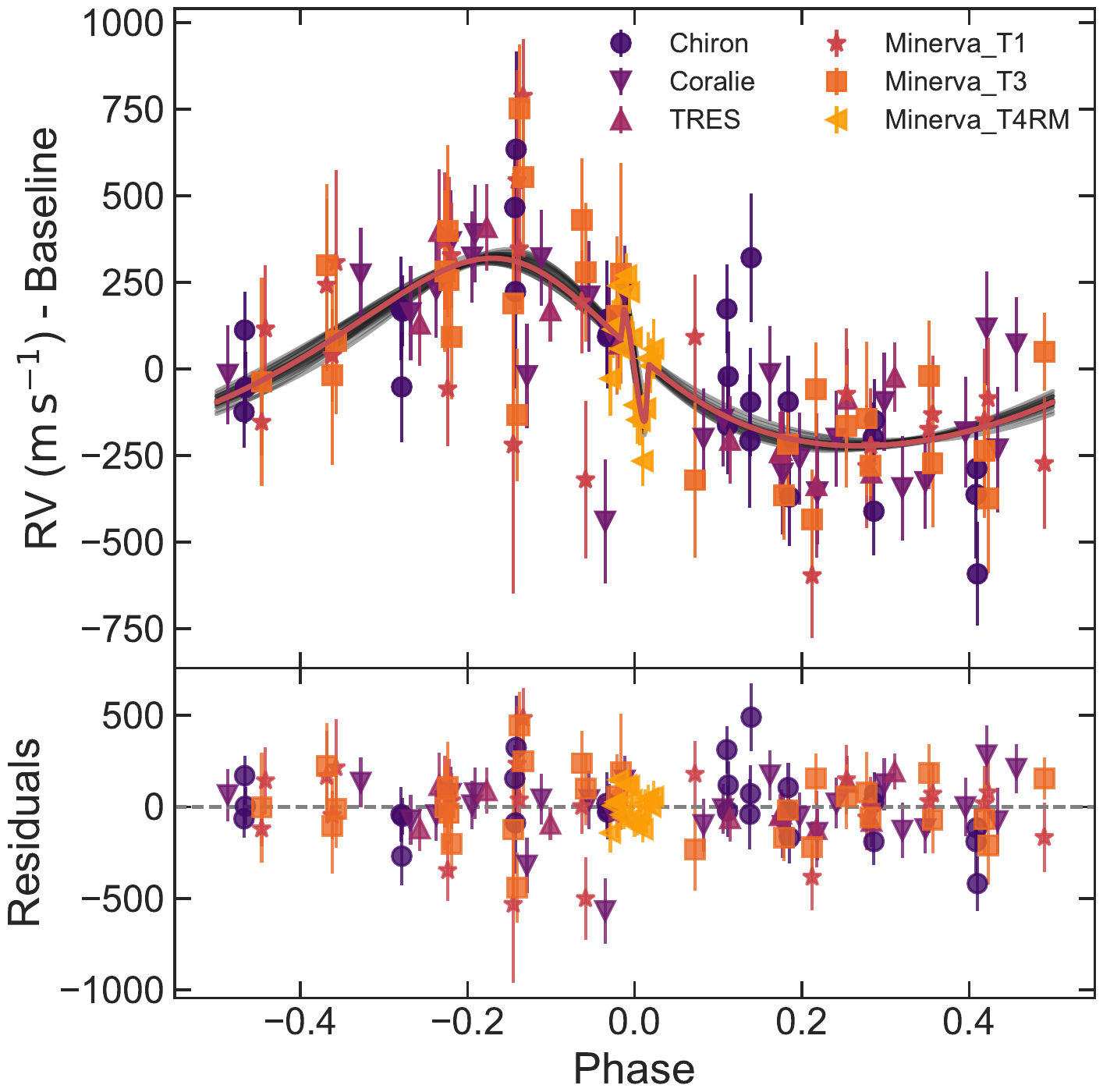}
  \includegraphics[width=1\columnwidth]{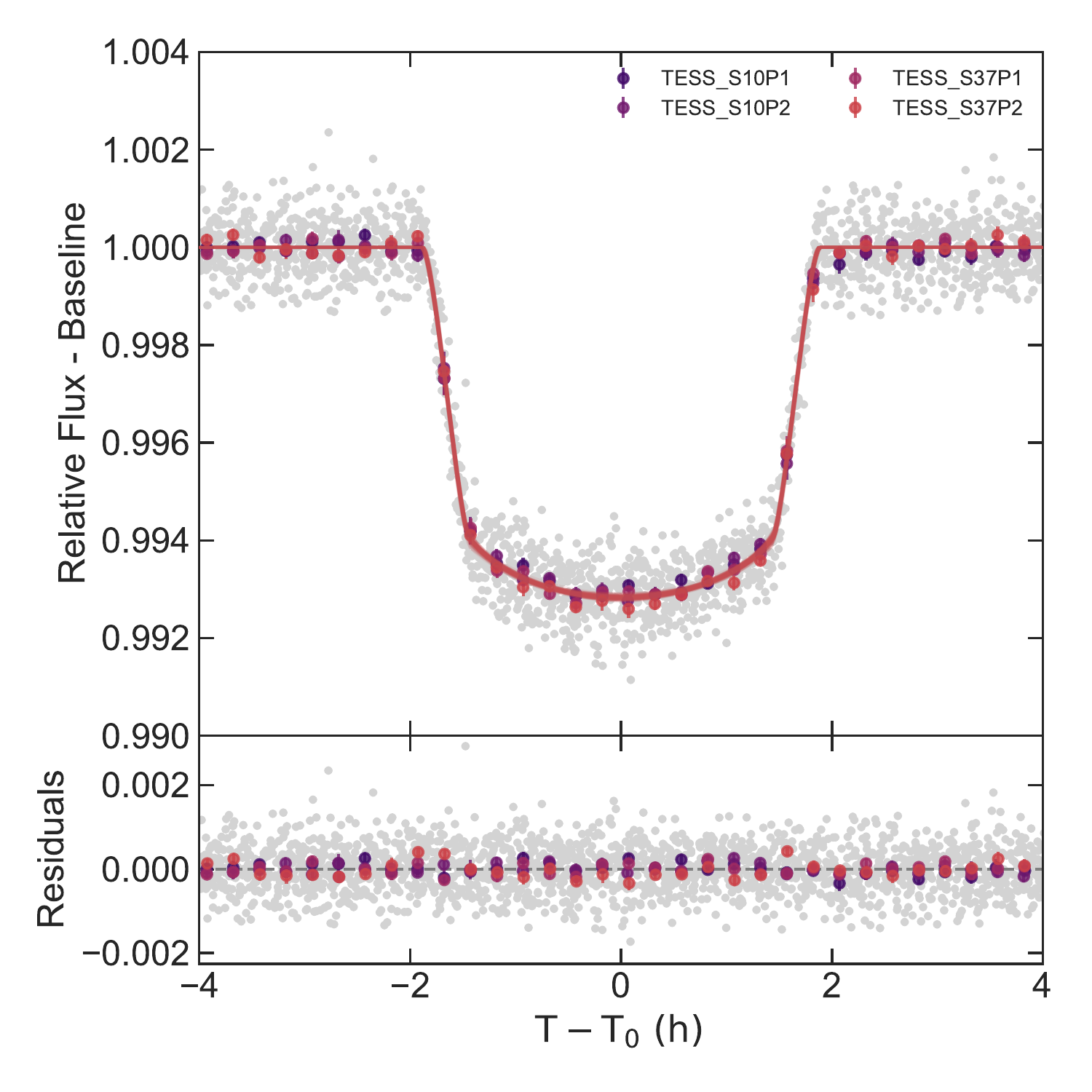}
  \caption{Left: A phase-folded radial velocity model of TOI-778\,b from our \texttt{AllesFitter} analysis with the Rossiter-Mclaughlin effect. Each radial velocity instrument's data is shown, with CHIRON in violet circles, CORALIE in purple down arrows, TRES in burgundy triangles, \minaus~Telescope 1 in pink stars, \minaus~Telescope 3 in orange squares and \minaus~Telescope 4 in yellow left arrows. 20 randomly drawn posterior radial velocity models are shown in grey lines while the red line is the best-fit model. Residual velocities to the best-fit model are shown below. Right: Phase-folded light curve model for TOI-778~b just from the \textit{TESS} data with the sectors (Sectors 10 and 37, respectively) and light curve segments (P1 and P2, respectively) indicated. Phase-folded light curve models for the ground-based follow-up is presented in Figure \ref{fig:Groundlc}.}\label{fig:RVandTESSlc} 
\end{figure*}

To determine the planetary nature of TOI-778~b, and its host star's obliquity, we used \texttt{Allesfitter} \citep{allesfitter-code,allesfitter-paper} to perform a joint analysis of the \textit{TESS} light curve segments, the photometric ground-based light curves, and the radial velocity measurements. We include both light curves from the \textit{TESS} Sectors 10 and 37, keeping them as separate observations with possible brightness offsets between the sectors. We also split the observations from each Sector into two parts, since there is a clear break in each sector during the data download. Thus, we have four independent \textit{TESS} light curves we use for the analysis. We use all ground-based photometric observations of TOI-778 that include at least 50\% of the transits of TOI-778~b. These facilities included the NGTS, PEST, LCO, Mt. Kent, and Mt. Stuart observations. We used all radial velocity measurements taken from \minaus, TRES, CORALIE, and CHIRON. These radial velocities include the data from \minaus~that were taken during the Rossiter-Mclaughlin observation on June 4, 2020. TOI-778's stellar radius, mass and effective temperature are used within the \texttt{Allesfitter} analysis, as these values are needed for deriving the planetary mass, radius, and equilibrium temperature.

The priors used for our analysis are given in Tables\,\ref{tab:star}, \ref{tab:planet}, and \ref{tab:fitted}, and described below. For each of our light curves, we calculated the quadratic limb darkening coefficients used by \citet{exofast}, an interpolation of the quadratic limb darkening tables derived by \citet{QLD}. These calculated values were then used as the starting values for the limb darkening coefficient parameters with uniform priors between 0 and 1 within the \texttt{Allesfitter} analysis. Since NGTS has a unique Band-pass filter, we set its initial quadratic limb darkening coefficients to 0.5 with a uniform prior between 0 and 1. We fixed the dilution parameter to 0 since the \textit{TESS} SPOC light curves used in the analysis already have any potential blending removed from the target star's flux from known sources. For the physical parameters used in the \texttt{Allesfitter} analysis, we applied uniform priors with starting values taken from or derived from NASA's Exoplanet Follow-up Observing Program database for the orbital period ($P_{p}$), transit mid-time ($T_{0;b}$), planet-to-star radius ratio ($R_{p}/R_{\star}$), the ratio of the sum of the planet and star radii to the semi-major axis ($(R_{p} +R_{\star})/a_{p}$), cosine of the inclination angle ($\cos{i_{p}}$), the radial velocity semi-amplitude ($K$), and eccentricity ($\sqrt{e_{p}} \cos{\omega_{p}}$ and $\sqrt{e_{p}} \sin{\omega_{p}}$). Uniform priors were used for the light curves flux error scaling ($\ln{\sigma_{F_{inst}}}$) as well as on the radial velocity baseline offsets ($\Delta RV_{inst}$) and jitter terms ($\ln{\sigma_{RV_{inst}}}$) for each instrument. Included in the joint fit was the Rossiter-McLaughlin effect and for this, we applied a weak ($5\sigma$) Gaussian prior on the stellar rotational velocity ($v\sin{(i)}$) as derived using spectroscopic observations from TRES (see Table\,\ref{tab:star}).

We utilise \texttt{Allesfitter}'s nested sampling approach to sample the model posteriors by implementing the \texttt{dynesty} package \citep{dynesty}. We used a dynamic nested sampling, with random walk sample, 500 live points and a tolerance of 0.01. We ran our analysis until a tolerance of 0.01 was achieved with the derived stellar, planetary and instrumental parameters shown in Table \ref{tab:planet} and \ref{tab:fitted}. Median values are shown in Table \ref{tab:planet} and \ref{tab:fitted} along with their associated 1-sigma errors.

\startlongtable
\begin{deluxetable}{lcc}
\tablewidth{0.95\columnwidth}
\tabletypesize{\scriptsize}
\tablecaption{Astrophysical parameters for TOI-778\,b as derived by \texttt{Allesfitter}. Priors are shown as uniform $\mathcal{U}$(a,b) or normal $\mathcal{N}$($\nu$,$\sigma$). Parameters used for the transit and radial velocity fits that are not located in this table can be found in Table~\ref{tab:fitted}.}
\label{tab:planet}
\tablehead{
\colhead{Parameter} & \colhead{Prior} & \colhead{Best-Fit}}
\startdata
\multicolumn{2}{l}{\textbf{Fitted Planetary Parameters}} & \\
$K_b$\,(\mos) & $\mathcal{U}$(10,1000) & $271\pm17$\\
$R_{p} / R_\star$ & $\mathcal{U}$(0.01,0.2) & $0.0825\pm0.0005$\\ 
$(R_\star + R_p) / a_p$ & $\mathcal{U}$(0.05,0.2) & $0.143\pm0.004$\\ 
$\cos{i}$ & $\mathcal{U}$(0,0.2) & $0.091^{+0.006}_{-0.005}$\\ 
$T_{0;b}$ - 2458000 ($\mathrm{BJD}$) & $\mathcal{U}$(577.7,579.7) & $578.7161\pm0.0001$\\ 
$P_b$ ($\mathrm{d}$) & $\mathcal{U}$(3.6,5.6) & $4.633611\pm0.000001$\\
$\sqrt{e_b} \cos{\omega_b}$ & $\mathcal{U}$(-0.9,0.9) & $0.40^{+0.06}_{-0.07}$\\ 
$\sqrt{e_b} \sin{\omega_b}$ & $\mathcal{U}$(-0.9,0.9) & $0.21\pm0.07$\\
\multicolumn{2}{l}{\textbf{Derived Planetary Parameters}} & \\
$a_\mathrm{p}/R_\star$ & & $7.6\pm0.2$\\ 
$R_\mathrm{p}$ ($\mathrm{R_{\oplus}}$) & & $15.4\pm0.5$\\
$M_\mathrm{p}$ ($\mathrm{M_{\oplus}}$) & & $878^{+77}_{-72}$\\
$i_\mathrm{p}$ (deg) & & $84.7^{+0.3}_{-0.4}$\\ 
$e_\mathrm{p}$ & & $0.21\pm0.04$\\ 
$w_\mathrm{p}$ (deg) & & $28_{-10}^{+12}$\\ 
$R_\mathrm{p}$ ($\mathrm{R_{jup}}$) & & $1.37\pm0.04$\\ 
$M_\mathrm{p}$ ($\mathrm{M_{jup}}$) & & $2.8\pm0.2$\\ 
$a_\mathrm{p}$ (AU) & & $0.060\pm0.003$\\ 
$b_\mathrm{tra;p}$ & & $0.61\pm0.02$\\ 
$T_\mathrm{tot;p}$ (hr) & & $3.76\pm0.01$\\ 
$T_\mathrm{full;p}$ (hr) & & $2.89\pm0.01$\\ 
$\rho_\mathrm{p}$ (cgs) & & $1.3\pm0.2$\\ 
$T_\mathrm{eq;p}$ (K) & & $1561^{+33}_{-32}$\\
\textbf{Stellar Parameters} & & \\
$v \sin{i}$ (\kms)& $\mathcal{N}$(35.1,5.0) & $39.9^{+4.5}_{-4.3}$\\ 
$\lambda$ ($\mathrm{deg}$) & $\mathcal{U}$(-180,180) & $18\pm11$\\ 
$\rho_\mathrm{\star}$ (cgs, derived) & & $0.38\pm0.03$\\
\enddata
\end{deluxetable}

\begin{figure}
  \centering
  \includegraphics[width=0.95\columnwidth]{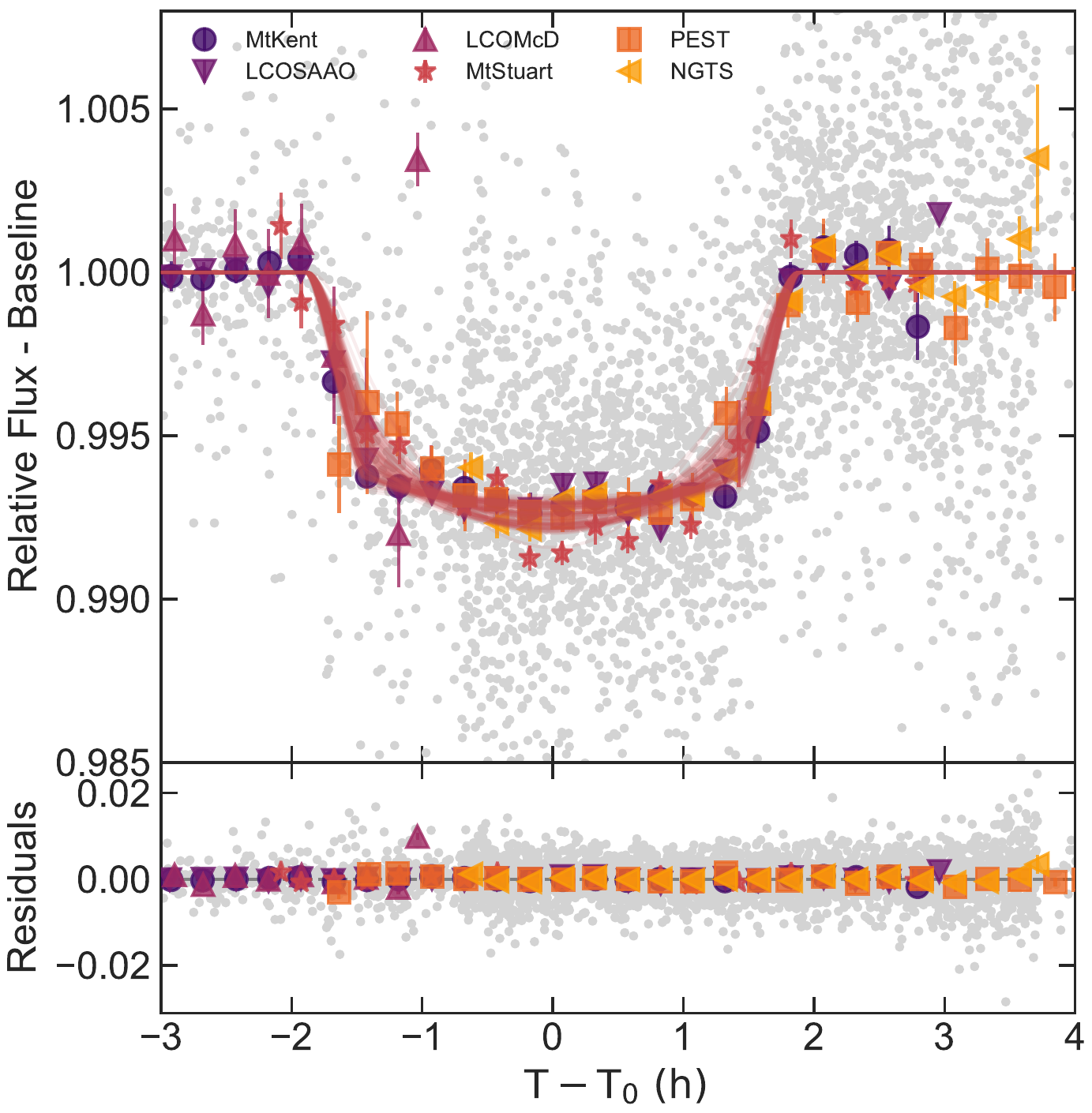}
  \caption{Phase-folded Light curve model for TOI-778\,b from the ground-based observations, the colored symbols are binned observations from Mt. Kent (violet circles), LCO McDonald (burgundy up triangles), PEST (orange squares), LCO SAAO (purple down triangles), Mt. Stuart (red stars), and NGTS (yellow left triangles. The grey points are the unbinned observations from all the ground-based facilities. 100 randomly drawn posteriors are shown in red, with the residuals of the fits shown in bottom plot.}\label{fig:Groundlc}
\end{figure}

The resulting transit light curve model for the TESS and ground-based photometry can be found in Figure~\ref{fig:RVandTESSlc} and Figure~\ref{fig:Groundlc}, respectively. Figure \ref{fig:RVandTESSlc} also includes the radial velocity model for TOI-778\,b with our Doppler spectroscopy data. From the global analysis, TOI-778\,b's orbital period is $4.633611\pm0.000001$\,days, in line with the $4.63361\pm0.00011$\,days found by \textit{TESS} in Sector 10. It has a relatively large radial velocity semi-amplitude, with a $16\,\sigma$ detection of $K=271^{+18}_{-17}$\,\mos. Given these parameter posteriors, and our estimates of the stellar mass, radius and their uncertainties given in Table \ref{tab:star}, we derive a planetary mass, radius, and bulk density for TOI-778\,b to be $2.8\pm0.2$\,\mj, $1.37\pm0.04$\,\rj, and $\rho_\mathrm{p}=1.3\pm0.2$\,cgs, respectively. With its radius, and orbital period, we confirm the planetary nature of TOI-778\,b as a hot Jupiter.

TOI-778~b was found to have a statistically significant non-zero orbital eccentricity, of $0.21\pm0.04$. With previous research showing that eccentric orbits can exist due to hidden planetary companions \citep[e.g.][]{witt13,trifonov17,boisvert18,witt19}, we further inspected the radial velocity residuals using a general Lomb-Scargle periodograms, but found no significant signals. We also performed an independent analysis of our radial velocity and photometric data through \texttt{EXOFASTv2} \citep{exofastV2,exofastsoftware}, and found consistent results at the 1-$\sigma$ level.

\begin{figure}
  \centering
  \includegraphics[width=\columnwidth]{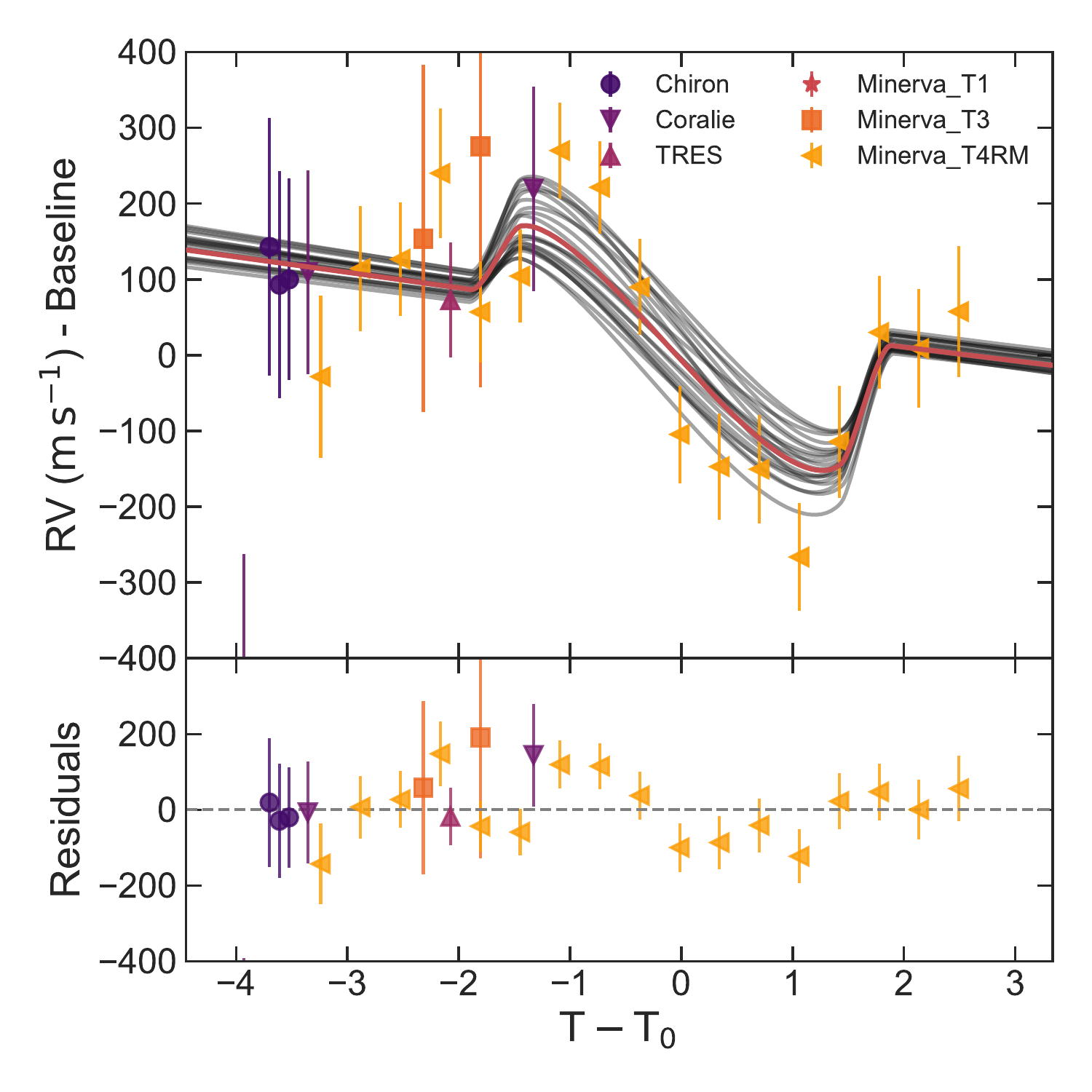}
  \caption{A phase-folded radial velocity model of the Rossiter-McLaughlin effect of TOI-778. Each radial velocity instrument's data is shown with CHIRON in violet circles, CORALIE in purple down arrows, TRES in burgundy triangles, \minaus~Telescope 1 in pink stars, \minaus~Telescope 3 in orange squares and \minaus~Telescope 4 in yellow left arrows, similar to Figure~\ref{fig:RVandTESSlc}. 20 randomly drawn posterior radial velocity models are shown as grey lines while the best-fit model is plotted as the red line. The residuals in the velocities from the best-fit model are shown in the bottom panel.}\label{fig:RMobs} 
\end{figure}

We also conducted a joint analysis of the Rossiter-Mclaughlin observation (Minerva T4RM) with the global fit to measure the sky-projected spin-orbit alignment of the system. The resulting effect can clearly be seen in Figures~\ref{fig:RVandTESSlc} and \ref{fig:RMobs}. The global fit analysis yields a sky-projected spin-orbit angle of $18\pm11^{\circ}$. In addition, following \citet{2020AJ....159...81M}, we find the stellar inclination to be well aligned to the line of sight, with $I_\star > 50^\circ$ at $3\sigma$ significance. This result is consistent with an aligned system and is discussed further in Section \ref{sec:Discussion}.

\section{TOI-778~b in Context}\label{sec:Discussion}


We have confirmed the planetary nature of TOI-778~b, detected by \textit{TESS} in Sector 10 of its primary mission. It is a hot Jupiter with radius 1.37$\pm$0.04\,\rj\ and mass 2.8$\pm$0.2\,\mj\ orbiting a rapidly-rotating early-F star. TOI-778~b appears to be somewhat inflated when compared to other hot Jupiters of similar masses.

\begin{figure}
  \centering
  \includegraphics[width=0.5\textwidth]{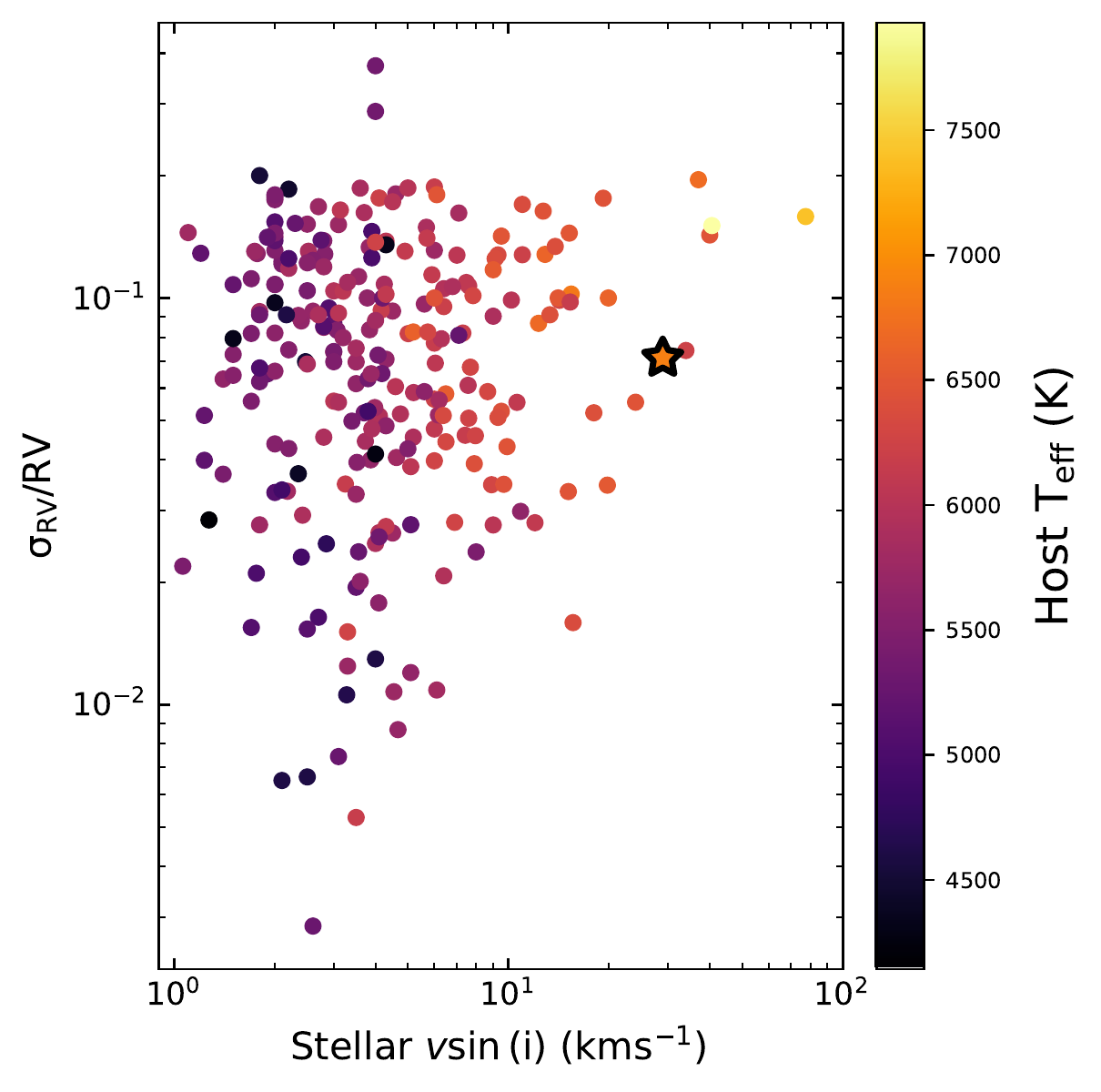}
  \caption{Host-star $v \sin{i}$ as a function of radial velocity amplitude precision, for 842 planets for which masses have been measured with a precision of better than 20\%.}
  \label{fig:speedy}
\end{figure}

The confirmation and mass measurement of TOI-778~b was challenging due to the relatively rapid $\sim$35\kms\ rotation of its early-F type host star. Figure \ref{fig:speedy} shows the stellar rotational velocity against the semi-amplitude radial velocity precision of all known exoplanets with a mass measurement precision better than 20\%. 

There are only five other planets that have been discovered with a mass precision of better than 20\%, orbiting around a more rapidly rotating stars. These include CoRoT-11 b \citep{spinny:CoRoT11}, HAT-P-69 b \citep{spinny:HAT-P-69}, HATS-70 b \citep{spinny:HATS-70b}, Kepler-1658 b \citep{spinny:Kepler-1658} and WASP-93 b \citep{spinny:WASP93b}. In the most extreme case, HAT-P-69 b was found around the rapidly rotating A star ($v \sin{i} = 77.44\pm0.56$\kms), and achieved a semi-amplitude radial velocity precision of K = 309$\pm$49\mos. Our results demonstrates how a facility like \minaus, with an effective telescope radius of 1.20m, can be utilised to follow up and confirm planetary candidates around such rapidly rotating stars.

Hot Jupiters appear to be less frequent around early type stars than solar-like stars \citep{2022MNRAS.516...75B,2022MNRAS.516..636S}, with \citet{zhou19} discovering an occurrence rate in \textit{TESS} data of only 0.43$\pm$0.15\% for hot Jupiters orbiting main-sequence F stars, and 0.26$\pm$0.11\% for A-type stars. These low occurrence rates for earlier-type stars are consistent with those found for their evolved kin \citep{grunblatt19}.

The rapid rotation, brightness ($V=9.1$ mag), and large radius of the planet all work in favour of measuring the spin-orbit angle for this system. From Rossiter-McLaughlin observations of a transit of TOI-778~b, we find the planet's orbit to be close to being aligned with the stellar equator, with a host star sky-projected obliquity of $19\pm10$\,degrees.

An obliquity measurement can aid in better determining the origin and formation history of exoplanets, especially large and relatively close-in orbiting ones like TOI-778~b. Since the star's effective temperature ($\teff\sim 6700$\,K) is beyond the Kraft break temperature of $\sim6200$\,K, it is unlikely to have realigned from a high obliquity orbit \citep{KraftBreak}. Thus we may be seeing the exoplanet's primordial obliquity, rather than the result of a secondary realignment. It therefore seems most likely that TOI-778~b sedately migrated through its host's disk, rather than reaching its current location through more chaotic means. Indeed, with a low obliquity angle and a stellar age of $\sim2$\,Gyr, migration mechanisms such as high-eccentricity, planet-planet scattering, Kozai–Lidov tidal and secular chaos migrations are disfavored \citep{HJF:Masset2003,HJF:2008Nagasawa,HJF:Dawson18}.

While the radial velocity data are limited and noisy, we can also exclude the presence of perturbing objects more massive than $\sim$6\,\mj\ within 0.3\,AU. Additionally, based on the estimated sensitivity of the high angular resolution imaging from Gemini, Keck, and Palomar (Section~\ref{AO}) and the 
\citet{2013ApJS..208....9P} table of stellar properties, we can place upper mass limits for potential stellar companions in this system as $0.73$\,\msun\ at a distance of 9.2\,AU (separation of $0.057$\,\arcsec, equivalent to 1\,FWHM from the star for Keck NIRC2), $0.16$\,\msun\ at 40.4\,AU ($0.25$\,\arcsec), and $0.10$\,\msun\ at 80.9\,AU ($0.50$\,\arcsec). Combined, the radial velocity and direct imaging data cannot fully rule out massive planetary, substellar, or low mass stellar objects with masses $<0.16$\,\msun\ within $\sim40$\,AU, therefore, planet-planet scattering and Kozai-Lidov tidal migration remain potentially viable migration mechanisms for TOI-778\,b.

\begin{figure}
  \centering
  \includegraphics[width=\columnwidth]{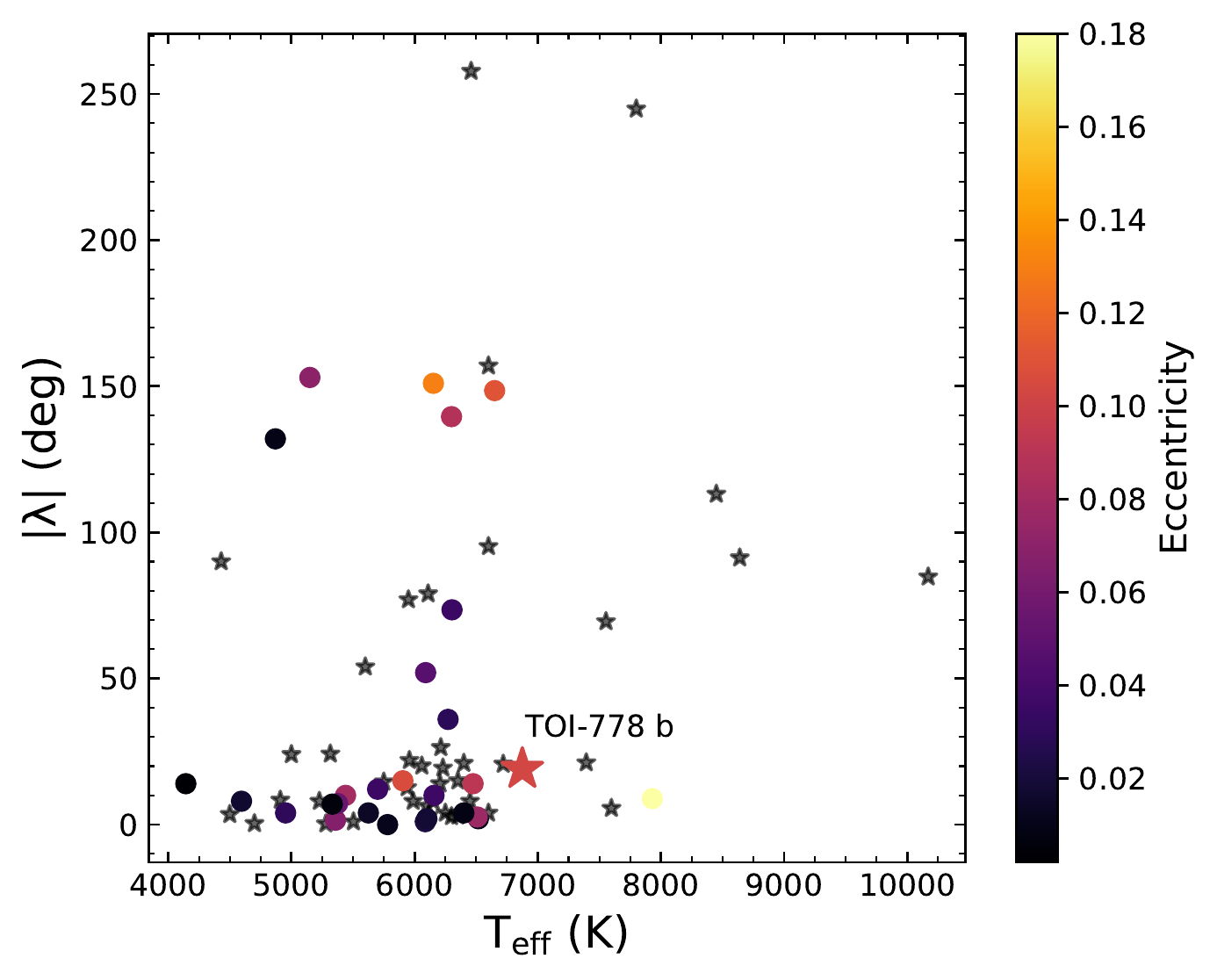}
  \caption{We show all known hot Jupiters (${\rm R_p > 0.7~R_J, P_P < 10~days}$) with obliquity measurements and plot those measurements against their host star's effective temperature. Points are coloured by the planet's orbital eccentricity, albeit planet's in circular orbits shown in grey stars. TOI-778~b is shown as the large red filled in star.}
  \label{fig:teffdeg}
\end{figure}

Figure~\ref{fig:teffdeg} shows the planetary obliquity as a function of host star temperature with TOI-778\,b and the sample of known hot Jupiters with obliquity measurements. From Figure \ref{fig:teffdeg}, TOI-778\,b joins a cohort of other hot Jupiters with well-aligned orbits, suggesting that disk migration is the likely case of their evolution to their current positions.

\section{Conclusion}\label{sec:Conclusion}

During Sector 10 of \textit{TESS}'s primary mission, an exoplanet candidate was discovered around the rapidly rotating star HD\,115447, also known as TOI-778. This 4.63\,day signal was subsequently followed up by the exoplanetary community through transit observations from PEST, LCO, NGTS, Mount Kent and Mt. Stuart. Coinciding with these efforts, radial velocity measurements from TOI-778 were collected from \minaus, TRES, CORALIE and CHIRON to then determine the exoplanetary nature of TOI-778\,b. From our \texttt{AllesFitter} global model, we confirm the presence of the hot Jupiter TOI~778~b ($1.37\pm0.04$\,\rj, $2.8\pm0.2$\,\mj). We conducted a Rossiter-Mclaughlin observation of TOI-778\,b and discovered its spin-orbit angle to its host is $18\pm11^{\circ}$, which is consistent with an aligned planetary system. These results also highlight how smaller telescope arrays such as \minaus can lead the charge of confirming and characterising exoplanets around rapidly rotating stars.

\acknowledgments
We respectfully acknowledge the traditional custodians of all lands throughout Australia, and recognise their continued cultural and spiritual connection to the land, waterways, cosmos, and community. We pay our deepest respects to all Elders, ancestors and descendants of the Giabal, Jarowair, and Kambuwal nations, upon whose lands the \minaus~facility at Mt Kent is situated.

\minaus~is supported by Australian Research Council LIEF Grant LE160100001, Discovery Grants DP180100972 and DP220100365, Mount Cuba Astronomical Foundation, and institutional partners University of Southern Queensland, UNSW Sydney, MIT, Nanjing University, George Mason University, University of Louisville, University of California Riverside, University of Florida, and The University of Texas at Austin.

This paper includes data collected at the Keck Telescopes. We recognise and acknowledge the cultural role and reverence that the summit of Maunakea has within the indigenous Hawaiian community. We are deeply grateful to have the opportunity to conduct observations from this mountain.

Funding for the \textit{TESS} mission is provided by NASA's Science Mission directorate. We acknowledge the use of public \textit{TESS} Alert data from pipelines at the \textit{TESS} Science Office and at the \textit{TESS} Science Processing Operations Center. This research has made use of the Exoplanet Follow-up Observation Program website, which is operated by the California Institute of Technology, under contract with the National Aeronautics and Space Administration under the Exoplanet Exploration Program. Resources supporting this work were provided by the NASA High-End Computing (HEC) Program through the NASA Advanced Supercomputing (NAS) Division at Ames Research Center for the production of the SPOC data products. This paper includes data collected by the \textit{TESS} mission, which are publicly available from the Mikulski Archive for Space Telescopes (MAST).

Some of the observations in the paper made use of the High-Resolution Imaging instrument Zorro. Zorro was funded by the NASA Exoplanet Exploration Program and built at the NASA Ames Research Center by Steve B. Howell, Nic Scott, Elliott P. Horch, and Emmett Quigley. Data were reduced using a software pipeline originally written by Elliott Horch and Mark Everett. Zorro was mounted on the Gemini South telescope of the international Gemini Observatory, a program of NSF’s OIR Lab, which is managed by the Association of Universities for Research in Astronomy (AURA) under a cooperative agreement with the National Science Foundation. on behalf of the Gemini partnership: the National Science Foundation (United States), National Research Council (Canada), Agencia Nacional de Investigación y Desarrollo (Chile), Ministerio de Ciencia, Tecnología e Innovación (Argentina), Ministério da Ciência, Tecnologia, Inovações e Comunicações (Brazil), and Korea Astronomy and Space Science Institute (Republic of Korea).

We thank the Swiss National Science Foundation (SNSF) and the Geneva University for their continuous support to our planet search programs. This work has been in particular carried out in the frame of the National Centre for Competence in Research {\it PlanetS} supported by the Swiss National Science Foundation (SNSF). \\ 
This publication makes use of The Data \& Analysis Center for Exoplanets (DACE), which is a facility based at the University of Geneva (CH) dedicated to extrasolar planets data visualisation, exchange and analysis. DACE is a platform of the Swiss National Centre of Competence in Research (NCCR) PlanetS, federating the Swiss expertise in Exoplanet research. The DACE platform is available at \url{https://dace.unige.ch}. 

This work makes use of observations from the LCOGT network.\\ 

J.T.C would like to thank BC and DN, and is supported by the Australian Government Research Training Program (RTP) Scholarship.
T.F. acknowledges support from the University of California President's Postdoctoral Fellowship Program.
A.J.\ and R.B.\ acknowledge support from ANID -- Millennium Science Initiative -- ICN12\_009. A.J.\ acknowledges additional support from FONDECYT project 1210718. R.B.\ acknowledges additional support from FONDECYT Project 1120075.

%

\vspace{5mm}
\facilities{\minaus, Euler1.2m (CORALIE), LCOGT, CHIRON. All of the \textit{TESS} data used in this paper can be found in MAST: \dataset[10.17909/t9-nmc8-f686]{http://dx.doi.org/10.17909/t9-nmc8-f686}.}


\software{
     \texttt{dynesty} \citep{dynesty},
     \texttt{ExoFAST} \citep{exofastsoftware},
     \texttt{isochrones} \citep{isochrones},
     \texttt{Allesfitter} \citep{allesfitter-code,allesfitter-paper}
     }



\appendix

\section{Extra Fitting Information}

\startlongtable
\begin{deluxetable*}{llcc}
\tabletypesize{\scriptsize}
\tablecaption{Median values and 68\% confidence interval of the fitted and derived parameters for TOI-778 from the joint Nested Sampling \texttt{Allesfitter} analysis of the photometry and radial velocity data.\label{tab:fitted}}
\tablehead{
\colhead{Parameter} & \colhead{Description} & \colhead{Prior} & \colhead{Best-Fit}
}
\startdata
\textbf{Radial Velocity Model Parameters} & & &\\
$\ln{\sigma_{\mathrm{jitter}}}$ ($\mathrm{RV_{Chiron}}$) & RV jitter ($\ln$ \mos) & $\mathcal{U}$(-3,9) & $1.68^{+1.58}_{-1.61}$\\ 
$\ln{\sigma_{\mathrm{jitter}}}$ ($\mathrm{RV_{Coralie}}$) & RV jitter ($\ln$ \mos) & $\mathcal{U}$(-3,9) & $4.72_{-3.14}^{+3.09}$\\ 
$\ln{\sigma_{\mathrm{jitter}}}$ ($\mathrm{RV_{TRES}}$) & RV jitter ($\ln$ \mos) & $\mathcal{U}$(-3,9) & $1.06_{-0.97}^{+0.93}$\\ 
$\ln{\sigma_{\mathrm{jitter}}}$ ($\mathrm{RV_{Minerva;T1}}$) & RV jitter ($\ln$ \mos) & $\mathcal{U}$(-3,9) & $1.72_{-1.68}^{+1.66}$\\ 
$\ln{\sigma_{\mathrm{jitter}}}$ ($\mathrm{RV_{Minerva;T3}}$) & RV jitter ($\ln$ \mos) & $\mathcal{U}$(-3,9) & $1.64_{-1.57}^{+1.53}$\\ 
$\ln{\sigma_{\mathrm{jitter}}}$ ($\mathrm{RV_{Minerva;T4RM}}$) & RV jitter ($\ln$ \mos) & $\mathcal{U}$(-3,9) & $1.56_{-1.50}^{+1.44}$\\ 
$\Delta \mathrm{RV_{CHIRON}}$ & RV offset (\mos) & $\mathcal{U}$(-1500,1500) & $54\pm25$\\ 
$\Delta \mathrm{RV_{CORALIE}}$ & RV offset (\mos) & $\mathcal{U}$(-7300,-4300) & $-5788_{-27}^{+26}$\\ 
$\Delta \mathrm{RV_{TRES}}$ & RV offset (\mos) & $\mathcal{U}$(-1300,1600) & $206\pm34$\\ 
$\Delta \mathrm{RV_{Minerva_{T1}}}$ & RV offset (\mos) & $\mathcal{U}$(-8300,-5300) & $-6722_{-34}^{+33}$\\ 
$\Delta \mathrm{RV_{Minerva_{T3}}}$ & RV offset (\mos) & $\mathcal{U}$(-8300,-5300) & $-6732^{+31}_{-30}$\\ 
$\Delta \mathrm{RV_{Minerva_{T4RM}}}$ & RV offset (\mos) & $\mathcal{U}$(-1500,1500) & $-42\pm29$\\
\textbf{Photometric Model Parameters} & &\\
$q_{\mathrm{1; TESS}}$ & Transformed limb darkening & $\mathcal{U}(0,1)$ & $0.15_{-0.03}^{+0.04}$\\ 
$q_{\mathrm{2; TESS}}$ & Transformed limb darkening & $\mathcal{U}(0,1)$ & $0.46_{-0.18}^{+0.21}$\\ 
$q_{\mathrm{1; MtKent}}$ & Transformed limb darkening & $\mathcal{U}(0,1)$ & $0.06_{-0.04}^{+0.06}$\\
$q_{\mathrm{2; MtKent}}$ & Transformed limb darkening & $\mathcal{U}(0,1)$ & $0.43_{-0.28}^{+0.34}$\\
$q_{\mathrm{1; LCOSAAO}}$ & Transformed limb darkening & $\mathcal{U}(0,1)$ & $0.16_{-0.06}^{+0.08}$\\
$q_{\mathrm{2; LCOSAAO}}$ & Transformed limb darkening & $\mathcal{U}(0,1)$ & $0.30_{-0.21}^{+0.31}$\\
$q_{\mathrm{1; LCOMcD}}$ & Transformed limb darkening & $\mathcal{U}(0,1)$ & $0.61_{-0.33}^{+0.25}$\\
$q_{\mathrm{2; LCOMcD}}$ & Transformed limb darkening & $\mathcal{U}(0,1)$ & $0.44_{-0.28}^{+0.33}$\\
$q_{\mathrm{1; MtStuart}}$ & Transformed limb darkening & $\mathcal{U}(0,1)$ & $0.84_{-0.14}^{+0.11}$\\
$q_{\mathrm{2; MtStuart}}$ & Transformed limb darkening & $\mathcal{U}(0,1)$ & $0.39_{-0.17}^{+0.18}$\\
$q_{\mathrm{1; PEST}}$ & Transformed limb darkening & $\mathcal{U}(0,1)$ & $0.35_{-0.16}^{+0.19}$\\
$q_{\mathrm{2; PEST}}$ & Transformed limb darkening & $\mathcal{U}(0,1)$ & $0.66_{-0.31}^{+0.24}$\\
$q_{\mathrm{1; NGTS}}$ & Transformed limb darkening & $\mathcal{U}(0,1)$ & $0.31_{-0.12}^{+0.17}$\\ 
$q_{\mathrm{2; NGTS}}$ & Transformed limb darkening & $\mathcal{U}(0,1)$ & $0.42_{-0.28}^{+0.33}$\\
$q_{\mathrm{1; Minerva_{T4RM}}}$ & Transformed limb darkening & $\mathcal{U}(0,1)$ & $0.62_{-0.30}^{+0.25}$\\ 
$q_{\mathrm{2; Minerva_{T4RM}}}$ & Transformed limb darkening & $\mathcal{U}(0,1)$ & $0.39_{-0.26}^{+0.33}$\\
$\ln{\sigma_\mathrm{TESS}}$ & Flux error scaling ($\ln{ \mathrm{rel. flux.} }$) & $\mathcal{U}(-10,-3)$ & $-7.49\pm0.01$\\ 
$\ln{\sigma_\mathrm{Mount Kent}}$ & Flux error scaling ($\ln{ \mathrm{rel. flux.} }$) & $\mathcal{U}(-10,-3)$ & $-6.39\pm0.04$\\ 
$\ln{\sigma_\mathrm{LCO AAO}}$ & Flux error scaling ($\ln{ \mathrm{rel. flux.} }$) & $\mathcal{U}(-10,-3)$ & $-6.65\pm0.03$\\ 
$\ln{\sigma_\mathrm{LCO McD}}$ & Flux error scaling ($\ln{ \mathrm{rel. flux.} }$) & $\mathcal{U}(-10,-3)$ & $-5.10\pm0.03$\\ 
$\ln{\sigma_\mathrm{Mount Stuart}}$ & Flux error scaling ($\ln{ \mathrm{rel. flux.} }$) & $\mathcal{U}(-10,-3)$ & $-6.06\pm0.03$\\ 
$\ln{\sigma_\mathrm{PEST}}$ & Flux error scaling ($\ln{ \mathrm{rel. flux.} }$) & $\mathcal{U}(-10,-3)$ & $-5.58\pm0.03$\\ 
$\ln{\sigma_\mathrm{NGTS}}$ & Flux error scaling ($\ln{ \mathrm{rel. flux.} }$) & $\mathcal{U}(-10,-3)$ & $-5.34\pm0.01$\\ 
$\delta_\mathrm{tr; TESS;S10P1}$ & Transit depth (ppt) & \textit{Derived parameter} & $7.17\pm0.04$\\ 
$\delta_\mathrm{tr; TESS;S10P2}$ & Transit depth (ppt) & \textit{Derived parameter} & $7.17\pm0.04$\\ 
$\delta_\mathrm{tr; TESS;S37P1}$ & Transit depth (ppt) & \textit{Derived parameter} & $7.17_{-0.04}^{+0.03}$\\ 
$\delta_\mathrm{tr; TESS;S37P2}$ & Transit depth (ppt) & \textit{Derived parameter} & $7.17\pm0.04$\\
$\delta_\mathrm{tr; MtKent}$ & Transit depth (ppt) & \textit{Derived parameter} & $7.01\pm0.10$\\ 
$\delta_\mathrm{tr; LCOSAAO}$ & Transit depth (ppt) & \textit{Derived parameter} & $7.16\pm0.09$\\ 
$\delta_\mathrm{tr; LCOMcD}$ & Transit depth (ppt) & \textit{Derived parameter} & $7.66_{-0.33}^{+0.29}$\\ 
$\delta_\mathrm{tr; MtStuart}$ & Transit depth (ppt) & \textit{Derived parameter} & $7.85_{-0.11}^{+0.12}$\\ 
$\delta_\mathrm{tr; PEST}$ & Transit depth (ppt) & \textit{Derived parameter} & $7.46_{-0.22}^{+0.20}$\\ 
$\delta_\mathrm{tr; NGTS}$ & Transit depth (ppt) & \textit{Derived parameter} & $7.36_{-0.15}^{+0.16}$
\enddata
\end{deluxetable*}



\bibliography{bibby}{}

\begin{thebibliography}{}
\expandafter\ifx\csname natexlab\endcsname\relax\def\natexlab#1{#1}\fi
\providecommand{\url}[1]{\href{#1}{#1}}
\providecommand{\dodoi}[1]{doi:~\href{http://doi.org/#1}{\nolinkurl{#1}}}
\providecommand{\doeprint}[1]{\href{http://ascl.net/#1}{\nolinkurl{http://ascl.net/#1}}}
\providecommand{\doarXiv}[1]{\href{https://arxiv.org/abs/#1}{\nolinkurl{https://arxiv.org/abs/#1}}}

\bibitem[{{Addison} {et~al.}(2019){Addison}, {Wright}, {Wittenmyer}, {Horner},
  {Mengel}, {Johns}, {Marti}, {Nicholson}, {Soutter}, {Bowler}, {Crossfield},
  {Kane}, {Kielkopf}, {Plavchan}, {Tinney}, {Zhang}, {Clark}, {Clerte},
  {Eastman}, {Swift}, {Bottom}, {Muirhead}, {McCrady}, {Herzig}, {Hogstrom},
  {Wilson}, {Sliski}, {Johnson}, {Wright}, {Johnson}, {Blake}, {Riddle}, {Lin},
  {Cornachione}, {Bedding}, {Stello}, {Huber}, {Marsden}, \&
  {Carter}}]{addison2019}
{Addison}, B., {Wright}, D.~J., {Wittenmyer}, R.~A., {et~al.} 2019, \pasp, 131,
  115003, \dodoi{10.1088/1538-3873/ab03aa}

\bibitem[{{Addison} {et~al.}(2013){Addison}, {Tinney}, {Wright}, {Bayliss},
  {Zhou}, {Hartman}, {Bakos}, \& {Schmidt}}]{HotMiss2}
{Addison}, B.~C., {Tinney}, C.~G., {Wright}, D.~J., {et~al.} 2013, \apjl, 774,
  L9, \dodoi{10.1088/2041-8205/774/1/L9}

\bibitem[{{Addison} {et~al.}(2018){Addison}, {Wang}, {Johnson}, {Tinney},
  {Wright}, \& {Bayliss}}]{polar2}
{Addison}, B.~C., {Wang}, S., {Johnson}, M.~C., {et~al.} 2018, \aj, 156, 197,
  \dodoi{10.3847/1538-3881/aade91}

\bibitem[{{Addison} {et~al.}(2021{\natexlab{a}}){Addison}, {Wright},
  {Nicholson}, {Cale}, {Mocnik}, {Huber}, {Plavchan}, {Wittenmyer},
  {Vanderburg}, {Chaplin}, {Chontos}, {Clark}, {Eastman}, {Ziegler}, {Brahm},
  {Carter}, {Clerte}, {Espinoza}, {Horner}, {Bentley}, {Jord{\'a}n}, {Kane},
  {Kielkopf}, {Laychock}, {Mengel}, {Okumura}, {Stassun}, {Bedding}, {Bowler},
  {Burnelis}, {Blanco-Cuaresma}, {Collins}, {Crossfield}, {Davis},
  {Evensberget}, {Heitzmann}, {Howell}, {Law}, {Mann}, {Marsden}, {Matson},
  {O'Connor}, {Shporer}, {Stevens}, {Tinney}, {Tylor}, {Wang}, {Zhang},
  {Henning}, {Kossakowski}, {Ricker}, {Sarkis}, {Schlecker}, {Torres},
  {Vanderspek}, {Latham}, {Seager}, {Winn}, {Jenkins}, {Mireles}, {Rowden},
  {Pepper}, {Daylan}, {Schlieder}, {Collins}, {Collins}, {Tan}, {Ball}, {Basu},
  {Buzasi}, {Campante}, {Corsaro}, {Gonz{\'a}lez-Cuesta}, {Davies}, {de
  Almeida}, {do Nascimento}, {Garc{\'\i}a}, {Guo}, {Handberg}, {Hekker}, {Hey},
  {Kallinger}, {Kawaler}, {Kayhan}, {Kuszlewicz}, {Lund}, {Lyttle}, {Mathur},
  {Miglio}, {Mosser}, {Nielsen}, {Serenelli}, {Aguirre}, \&
  {Theme{\ss}l}}]{TOI257}
{Addison}, B.~C., {Wright}, D.~J., {Nicholson}, B.~A., {et~al.}
  2021{\natexlab{a}}, \mnras, 502, 3704, \dodoi{10.1093/mnras/staa3960}

\bibitem[{{Addison} {et~al.}(2021{\natexlab{b}}){Addison}, {Knudstrup}, {Wong},
  {H{\'e}brard}, {Dorval}, {Snellen}, {Albrecht}, {Bello-Arufe}, {Almenara},
  {Boisse}, {Bonfils}, {Dalal}, {Demangeon}, {Hoyer}, {Kiefer}, {Santos},
  {Nowak}, {Luque}, {Stangret}, {Palle}, {Tronsgaard}, {Antoci}, {Buchhave},
  {G{\"u}nther}, {Daylan}, {Murgas}, {Parviainen}, {Esparza-Borges}, {Crouzet},
  {Narita}, {Fukui}, {Kawauchi}, {Watanabe}, {Rabus}, {Johnson}, {Otten}, {Jan
  Talens}, {Cabot}, {Fischer}, {Grundahl}, {Fredslund Andersen},
  {Jessen-Hansen}, {Pall{\'e}}, {Shporer}, {Ciardi}, {Clark}, {Wittenmyer},
  {Wright}, {Horner}, {Collins}, {Jensen}, {Kielkopf}, {Schwarz}, {Srdoc},
  {Yilmaz}, {Senavci}, {Diamond}, {Harbeck}, {Komacek}, {Smith}, {Wang},
  {Eastman}, {Stassun}, {Latham}, {Vanderspek}, {Seager}, {Winn}, {Jenkins},
  {Louie}, {Bouma}, {Twicken}, {Levine}, \& {McLean}}]{2021AJ....162..292A}
{Addison}, B.~C., {Knudstrup}, E., {Wong}, I., {et~al.} 2021{\natexlab{b}},
  \aj, 162, 292, \dodoi{10.3847/1538-3881/ac224e}

\bibitem[{{Albrecht} {et~al.}(2012){Albrecht}, {Winn}, {Johnson}, {Howard},
  {Marcy}, {Butler}, {Arriagada}, {Crane}, {Shectman}, {Thompson}, {Hirano},
  {Bakos}, \& {Hartman}}]{polar1}
{Albrecht}, S., {Winn}, J.~N., {Johnson}, J.~A., {et~al.} 2012, \apj, 757, 18,
  \dodoi{10.1088/0004-637X/757/1/18}

\bibitem[{{Anderson} {et~al.}(2018){Anderson}, {Temple}, {Nielsen}, {Burdanov},
  {Hellier}, {Bouchy}, {Brown}, {Collier Cameron}, {Gillon}, {Jehin}, {Maxted},
  {Pepe}, {Pollacco}, {Pozuelos}, {Queloz}, {S{\'e}gransan}, {Smalley},
  {Triaud}, {Turner}, {Udry}, \& {West}}]{WASP189}
{Anderson}, D.~R., {Temple}, L.~Y., {Nielsen}, L.~D., {et~al.} 2018, arXiv
  e-prints, arXiv:1809.04897.
\newblock \doarXiv{1809.04897}

\bibitem[{{Baranne} {et~al.}(1996){Baranne}, {Queloz}, {Mayor}, {Adrianzyk},
  {Knispel}, {Kohler}, {Lacroix}, {Meunier}, {Rimbaud}, \& {Vin}}]{baranne1996}
{Baranne}, A., {Queloz}, D., {Mayor}, M., {et~al.} 1996, \aaps, 119, 373

\bibitem[{{Barbary}(2016)}]{Barbary2016}
{Barbary}, K. 2016, The Journal of Open Source Software, 1, 58,
  \dodoi{10.21105/joss.00058}

\bibitem[{{Barnes} {et~al.}(2012){Barnes}, {Gibson}, {Nield}, \&
  {Cochrane}}]{2012SPIE.8446E..88B}
{Barnes}, S.~I., {Gibson}, S., {Nield}, K., \& {Cochrane}, D. 2012, in Society
  of Photo-Optical Instrumentation Engineers (SPIE) Conference Series, Vol.
  8446, Ground-based and Airborne Instrumentation for Astronomy IV, ed. I.~S.
  {McLean}, S.~K. {Ramsay}, \& H.~{Takami}, 844688, \dodoi{10.1117/12.926527}

\bibitem[{{Beaug{\'e}} \& {Nesvorn{\'y}}(2012)}]{scatter2}
{Beaug{\'e}}, C., \& {Nesvorn{\'y}}, D. 2012, \apj, 751, 119,
  \dodoi{10.1088/0004-637X/751/2/119}

\bibitem[{{Beleznay} \& {Kunimoto}(2022)}]{2022MNRAS.516...75B}
{Beleznay}, M., \& {Kunimoto}, M. 2022, \mnras, 516, 75,
  \dodoi{10.1093/mnras/stac2179}

\bibitem[{{Bertin} \& {Arnouts}(1996)}]{bertin96sextractor}
{Bertin}, E., \& {Arnouts}, S. 1996, \aaps, 117, 393,
  \dodoi{10.1051/aas:1996164}

\bibitem[{{Blanco-Cuaresma}(2019)}]{iSpec2}
{Blanco-Cuaresma}, S. 2019, \mnras, 486, 2075, \dodoi{10.1093/mnras/stz549}

\bibitem[{{Blanco-Cuaresma} {et~al.}(2014){Blanco-Cuaresma}, {Soubiran},
  {Heiter}, \& {Jofr{\'e}}}]{iSpec1}
{Blanco-Cuaresma}, S., {Soubiran}, C., {Heiter}, U., \& {Jofr{\'e}}, P. 2014,
  \aap, 569, A111, \dodoi{10.1051/0004-6361/201423945}

\bibitem[{{B{\"o}hm-Vitense}(2007)}]{bin2}
{B{\"o}hm-Vitense}, E. 2007, \aj, 133, 1903, \dodoi{10.1086/512124}

\bibitem[{{Boisvert} {et~al.}(2018){Boisvert}, {Nelson}, \&
  {Steffen}}]{boisvert18}
{Boisvert}, J.~H., {Nelson}, B.~E., \& {Steffen}, J.~H. 2018, \mnras, 480,
  2846, \dodoi{10.1093/mnras/sty2023}

\bibitem[{{Borucki} {et~al.}(2010){Borucki}, {Koch}, {Basri}, {Batalha},
  {Brown}, {Caldwell}, {Caldwell}, {Christensen-Dalsgaard}, {Cochran},
  {DeVore}, {Dunham}, {Dupree}, {Gautier}, {Geary}, {Gilliland}, {Gould},
  {Howell}, {Jenkins}, {Kondo}, {Latham}, {Marcy}, {Meibom}, {Kjeldsen},
  {Lissauer}, {Monet}, {Morrison}, {Sasselov}, {Tarter}, {Boss}, {Brownlee},
  {Owen}, {Buzasi}, {Charbonneau}, {Doyle}, {Fortney}, {Ford}, {Holman},
  {Seager}, {Steffen}, {Welsh}, {Rowe}, {Anderson}, {Buchhave}, {Ciardi},
  {Walkowicz}, {Sherry}, {Horch}, {Isaacson}, {Everett}, {Fischer}, {Torres},
  {Johnson}, {Endl}, {MacQueen}, {Bryson}, {Dotson}, {Haas}, {Kolodziejczak},
  {Van Cleve}, {Chandrasekaran}, {Twicken}, {Quintana}, {Clarke}, {Allen},
  {Li}, {Wu}, {Tenenbaum}, {Verner}, {Bruhweiler}, {Barnes}, \&
  {Prsa}}]{2010Sci...327..977B}
{Borucki}, W.~J., {Koch}, D., {Basri}, G., {et~al.} 2010, Science, 327, 977,
  \dodoi{10.1126/science.1185402}

\bibitem[{{Bowler} {et~al.}(2010){Bowler}, {Johnson}, {Marcy}, {Henry}, {Peek},
  {Fischer}, {Clubb}, {Liu}, {Reffert}, {Schwab}, \& {Lowe}}]{Big2}
{Bowler}, B.~P., {Johnson}, J.~A., {Marcy}, G.~W., {et~al.} 2010, \apj, 709,
  396, \dodoi{10.1088/0004-637X/709/1/396}

\bibitem[{{Brown} {et~al.}(2013){Brown}, {Baliber}, {Bianco}, {Bowman},
  {Burleson}, {Conway}, {Crellin}, {Depagne}, {De Vera}, {Dilday}, {Dragomir},
  {Dubberley}, {Eastman}, {Elphick}, {Falarski}, {Foale}, {Ford}, {Fulton},
  {Garza}, {Gomez}, {Graham}, {Greene}, {Haldeman}, {Hawkins}, {Haworth},
  {Haynes}, {Hidas}, {Hjelstrom}, {Howell}, {Hygelund}, {Lister}, {Lobdill},
  {Martinez}, {Mullins}, {Norbury}, {Parrent}, {Paulson}, {Petry}, {Pickles},
  {Posner}, {Rosing}, {Ross}, {Sand}, {Saunders}, {Shobbrook}, {Shporer},
  {Street}, {Thomas}, {Tsapras}, {Tufts}, {Valenti}, {Vander Horst}, {Walker},
  {White}, \& {Willis}}]{Brown:2013}
{Brown}, T.~M., {Baliber}, N., {Bianco}, F.~B., {et~al.} 2013, \pasp, 125,
  1031, \dodoi{10.1086/673168}

\bibitem[{{Bryant} {et~al.}(2020){Bryant}, {Bayliss}, {McCormac}, {Wheatley},
  {Acton}, {Anderson}, {Armstrong}, {Bouchy}, {Belardi}, {Burleigh},
  {Tilbrook}, {Casewell}, {Cooke}, {Gill}, {Goad}, {Jenkins}, {Lendl},
  {Pollacco}, {Queloz}, {Raynard}, {Smith}, {Vines}, {West}, \&
  {Udry}}]{bryant2020multicam}
{Bryant}, E.~M., {Bayliss}, D., {McCormac}, J., {et~al.} 2020, \mnras, 494,
  5872, \dodoi{10.1093/mnras/staa1075}

\bibitem[{{Buchhave} {et~al.}(2010){Buchhave}, {Bakos}, {Hartman}, {Torres},
  {Kov{\'a}cs}, {Latham}, {Noyes}, {Esquerdo}, {Everett}, {Howard}, {Marcy},
  {Fischer}, {Johnson}, {Andersen}, {F{\H{u}}r{\'e}sz}, {Perumpilly},
  {Sasselov}, {Stefanik}, {B{\'e}ky}, {L{\'a}z{\'a}r}, {Papp}, \&
  {S{\'a}ri}}]{Buchhave:2010}
{Buchhave}, L.~A., {Bakos}, G.~{\'A}., {Hartman}, J.~D., {et~al.} 2010, \apj,
  720, 1118, \dodoi{10.1088/0004-637X/720/2/1118}

\bibitem[{{Butler} {et~al.}(1997){Butler}, {Marcy}, {Williams}, {Hauser}, \&
  {Shirts}}]{HJ1}
{Butler}, R.~P., {Marcy}, G.~W., {Williams}, E., {Hauser}, H., \& {Shirts}, P.
  1997, \apjl, 474, L115, \dodoi{10.1086/310444}

\bibitem[{{Campbell} {et~al.}(1988){Campbell}, {Walker}, \& {Yang}}]{GammaCeph}
{Campbell}, B., {Walker}, G.~A.~H., \& {Yang}, S. 1988, \apj, 331, 902,
  \dodoi{10.1086/166608}

\bibitem[{{Cannon} \& {Pickering}(1993)}]{CP93}
{Cannon}, A.~J., \& {Pickering}, E.~C. 1993, VizieR Online Data Catalog,
  III/135A

\bibitem[{{Chatterjee} {et~al.}(2008){Chatterjee}, {Ford}, {Matsumura}, \&
  {Rasio}}]{scatter1}
{Chatterjee}, S., {Ford}, E.~B., {Matsumura}, S., \& {Rasio}, F.~A. 2008, \apj,
  686, 580, \dodoi{10.1086/590227}

\bibitem[{{Chontos} {et~al.}(2019){Chontos}, {Huber}, {Latham}, {Bieryla}, {Van
  Eylen}, {Bedding}, {Berger}, {Buchhave}, {Campante}, {Chaplin}, {Colman},
  {Coughlin}, {Davies}, {Hirano}, {Howard}, \& {Isaacson}}]{spinny:Kepler-1658}
{Chontos}, A., {Huber}, D., {Latham}, D.~W., {et~al.} 2019, \aj, 157, 192,
  \dodoi{10.3847/1538-3881/ab0e8e}

\bibitem[{{Ciardi} {et~al.}(2015){Ciardi}, {Beichman}, {Horch}, \&
  {Howell}}]{ciardi2015}
{Ciardi}, D.~R., {Beichman}, C.~A., {Horch}, E.~P., \& {Howell}, S.~B. 2015,
  \apj, 805, 16, \dodoi{10.1088/0004-637X/805/1/16}

\bibitem[{{Claret} \& {Bloemen}(2011)}]{QLD}
{Claret}, A., \& {Bloemen}, S. 2011, \aap, 529, A75,
  \dodoi{10.1051/0004-6361/201116451}

\bibitem[{{Collier Cameron} {et~al.}(2010){Collier Cameron}, {Guenther},
  {Smalley}, {McDonald}, {Hebb}, {Andersen}, {Augusteijn}, {Barros}, {Brown},
  {Cochran}, {Endl}, {Fossey}, {Hartmann}, {Maxted}, {Pollacco}, {Skillen},
  {Telting}, {Waldmann}, \& {West}}]{HotMiss1}
{Collier Cameron}, A., {Guenther}, E., {Smalley}, B., {et~al.} 2010, \mnras,
  407, 507, \dodoi{10.1111/j.1365-2966.2010.16922.x}

\bibitem[{{Collins} {et~al.}(2017){Collins}, {Kielkopf}, {Stassun}, \&
  {Hessman}}]{Collins:2017}
{Collins}, K.~A., {Kielkopf}, J.~F., {Stassun}, K.~G., \& {Hessman}, F.~V.
  2017, \aj, 153, 77, \dodoi{10.3847/1538-3881/153/2/77}

\bibitem[{{Cutri} {et~al.}(2003){Cutri}, {Skrutskie}, {van Dyk}, {Beichman},
  {Carpenter}, {Chester}, {Cambresy}, {Evans}, {Fowler}, {Gizis}, {Howard},
  {Huchra}, {Jarrett}, {Kopan}, {Kirkpatrick}, {Light}, {Marsh}, {McCallon},
  {Schneider}, {Stiening}, {Sykes}, {Weinberg}, {Wheaton}, {Wheelock}, \&
  {Zacarias}}]{2MASS}
{Cutri}, R.~M., {Skrutskie}, M.~F., {van Dyk}, S., {et~al.} 2003, VizieR Online
  Data Catalog, II/246

\bibitem[{{Davis} {et~al.}(2020){Davis}, {Wang}, {Jones}, {Eastman},
  {G{\"u}nther}, {Stassun}, {Addison}, {Collins}, {Quinn}, {Latham},
  {Trifonov}, {Shahaf}, {Mazeh}, {Kane}, {Narita}, {Wang}, {Tan}, {Ciardi},
  {Tokovinin}, {Ziegler}, {Tronsgaard}, {Millholland}, {Cruz}, {Berlind},
  {Calkins}, {Esquerdo}, {Collins}, {Conti}, {Murgas}, {Evans}, {Lewin},
  {Radford}, {Paredes}, {Henry}, {Hodari-Sadiki}, {Lund}, {Christiansen},
  {Law}, {Mann}, {Brice{\~n}o}, {Parviainen}, {Palle}, {Watanabe}, {Ricker},
  {Vanderspek}, {Seager}, {Winn}, {Jenkins}, {Krishnamurthy}, {Batalha},
  {Burt}, {Col{\'o}n}, {Dynes}, {Caldwell}, {Morris}, {Henze}, \&
  {Fischer}}]{Davis2020}
{Davis}, A.~B., {Wang}, S., {Jones}, M., {et~al.} 2020, \aj, 160, 229,
  \dodoi{10.3847/1538-3881/aba49d}

\bibitem[{{Dawson} \& {Johnson}(2018)}]{HJF:Dawson18}
{Dawson}, R.~I., \& {Johnson}, J.~A. 2018, \araa, 56, 175,
  \dodoi{10.1146/annurev-astro-081817-051853}

\bibitem[{{Dekany} {et~al.}(2013){Dekany}, {Roberts}, {Burruss}, {Bouchez},
  {Truong}, {Baranec}, {Guiwits}, {Hale}, {Angione}, {Trinh}, {Zolkower},
  {Shelton}, {Palmer}, {Henning}, {Croner}, {Troy}, {McKenna}, {Tesch},
  {Hildebrandt}, \& {Milburn}}]{dekany2013}
{Dekany}, R., {Roberts}, J., {Burruss}, R., {et~al.} 2013, \apj, 776, 130,
  \dodoi{10.1088/0004-637X/776/2/130}

\bibitem[{{Eastman}(2017)}]{exofastsoftware}
{Eastman}, J. 2017, {EXOFASTv2: Generalized publication-quality exoplanet
  modeling code}, Astrophysics Source Code Library.
\newblock \doeprint{1710.003}

\bibitem[{{Eastman} {et~al.}(2013){Eastman}, {Gaudi}, \& {Agol}}]{exofast}
{Eastman}, J., {Gaudi}, B.~S., \& {Agol}, E. 2013, \pasp, 125, 83,
  \dodoi{10.1086/669497}

\bibitem[{{Eastman} {et~al.}(2019){Eastman}, {Rodriguez}, {Agol}, {Stassun},
  {Beatty}, {Vanderburg}, {Gaudi}, {Collins}, \& {Luger}}]{exofastV2}
{Eastman}, J.~D., {Rodriguez}, J.~E., {Agol}, E., {et~al.} 2019, arXiv
  e-prints, arXiv:1907.09480.
\newblock \doarXiv{1907.09480}

\bibitem[{{Fetherolf} {et~al.}(2022){Fetherolf}, {Pepper}, {Simpson}, {Kane},
  {Mocnik}, {Antoci}, {Huber}, {Jenkins}, {Stassun}, {Twicken}, {Vanderspek},
  \& {Winn}}]{2022arXiv220811721F}
{Fetherolf}, T., {Pepper}, J., {Simpson}, E., {et~al.} 2022, arXiv e-prints,
  arXiv:2208.11721.
\newblock \doarXiv{2208.11721}

\bibitem[{{{F{\H u}r{\'e}sz}}(2008)}]{Furesz:2008}
{{F{\H u}r{\'e}sz}}, G. 2008, PhD thesis, Univ. of Szeged, Hungary

\bibitem[{{Furlan} {et~al.}(2017){Furlan}, {Ciardi}, {Everett}, {Saylors},
  {Teske}, {Horch}, {Howell}, {van Belle}, {Hirsch}, {Gautier}, {Adams},
  {Barrado}, {Cartier}, {Dressing}, {Dupree}, {Gilliland}, {Lillo-Box},
  {Lucas}, \& {Wang}}]{furlan2017}
{Furlan}, E., {Ciardi}, D.~R., {Everett}, M.~E., {et~al.} 2017, \aj, 153, 71,
  \dodoi{10.3847/1538-3881/153/2/71}

\bibitem[{{Gaia Collaboration} {et~al.}(2016){Gaia Collaboration}, {Prusti},
  {de Bruijne}, {Brown}, {Vallenari}, {Babusiaux}, {Bailer-Jones}, {Bastian},
  {Biermann}, {Evans}, {Eyer}, {Jansen}, {Jordi}, {Klioner}, {Lammers},
  {Lindegren}, {Luri}, {Mignard}, {Milligan}, {Panem}, {Poinsignon},
  {Pourbaix}, {Randich}, {Sarri}, {Sartoretti}, {Siddiqui}, {Soubiran},
  {Valette}, {van Leeuwen}, {Walton}, {Aerts}, {Arenou}, {Cropper}, {Drimmel},
  {H{\o}g}, {Katz}, {Lattanzi}, {O'Mullane}, {Grebel}, {Holland}, {Huc},
  {Passot}, {Bramante}, {Cacciari}, {Casta{\~n}eda}, {Chaoul}, {Cheek}, {De
  Angeli}, {Fabricius}, {Guerra}, {Hern{\'a}ndez}, {Jean-Antoine-Piccolo},
  {Masana}, {Messineo}, {Mowlavi}, {Nienartowicz}, {Ord{\'o}{\~n}ez-Blanco},
  {Panuzzo}, {Portell}, {Richards}, {Riello}, {Seabroke}, {Tanga},
  {Th{\'e}venin}, {Torra}, {Els}, {Gracia-Abril}, {Comoretto},
  {Garcia-Reinaldos}, {Lock}, {Mercier}, {Altmann}, {Andrae}, {Astraatmadja},
  {Bellas-Velidis}, {Benson}, {Berthier}, {Blomme}, {Busso}, {Carry},
  {Cellino}, {Clementini}, {Cowell}, {Creevey}, {Cuypers}, {Davidson}, {De
  Ridder}, {de Torres}, {Delchambre}, {Dell'Oro}, {Ducourant}, {Fr{\'e}mat},
  {Garc{\'\i}a-Torres}, {Gosset}, {Halbwachs}, {Hambly}, {Harrison}, {Hauser},
  {Hestroffer}, {Hodgkin}, {Huckle}, {Hutton}, {Jasniewicz}, {Jordan},
  {Kontizas}, {Korn}, {Lanzafame}, {Manteiga}, {Moitinho}, {Muinonen},
  {Osinde}, {Pancino}, {Pauwels}, {Petit}, {Recio-Blanco}, {Robin}, {Sarro},
  {Siopis}, {Smith}, {Smith}, {Sozzetti}, {Thuillot}, {van Reeven}, {Viala},
  {Abbas}, {Abreu Aramburu}, {Accart}, {Aguado}, {Allan}, {Allasia},
  {Altavilla}, {{\'A}lvarez}, {Alves}, {Anderson}, {Andrei}, {Anglada Varela},
  {Antiche}, {Antoja}, {Ant{\'o}n}, {Arcay}, {Atzei}, {Ayache}, {Bach},
  {Baker}, {Balaguer-N{\'u}{\~n}ez}, {Barache}, {Barata}, {Barbier}, {Barblan},
  {Baroni}, {Barrado y Navascu{\'e}s}, {Barros}, {Barstow}, {Becciani},
  {Bellazzini}, {Bellei}, {Bello Garc{\'\i}a}, {Belokurov}, {Bendjoya},
  {Berihuete}, {Bianchi}, {Bienaym{\'e}}, {Billebaud}, {Blagorodnova},
  {Blanco-Cuaresma}, {Boch}, {Bombrun}, {Borrachero}, {Bouquillon}, {Bourda},
  {Bouy}, {Bragaglia}, {Breddels}, {Brouillet}, {Br{\"u}semeister},
  {Bucciarelli}, {Budnik}, {Burgess}, {Burgon}, {Burlacu}, {Busonero}, {Buzzi},
  {Caffau}, {Cambras}, {Campbell}, {Cancelliere}, {Cantat-Gaudin}, {Carlucci},
  {Carrasco}, {Castellani}, {Charlot}, {Charnas}, {Charvet}, {Chassat},
  {Chiavassa}, {Clotet}, {Cocozza}, {Collins}, {Collins}, {Costigan}, {Crifo},
  {Cross}, {Crosta}, {Crowley}, {Dafonte}, {Damerdji}, {Dapergolas}, {David},
  {David}, {De Cat}, {de Felice}, {de Laverny}, {De Luise}, {De March}, {de
  Martino}, {de Souza}, {Debosscher}, {del Pozo}, {Delbo}, {Delgado},
  {Delgado}, {di Marco}, {Di Matteo}, {Diakite}, {Distefano}, {Dolding}, {Dos
  Anjos}, {Drazinos}, {Dur{\'a}n}, {Dzigan}, {Ecale}, {Edvardsson}, {Enke},
  {Erdmann}, {Escolar}, {Espina}, {Evans}, {Eynard Bontemps}, {Fabre},
  {Fabrizio}, {Faigler}, {Falc{\~a}o}, {Farr{\`a}s Casas}, {Faye}, {Federici},
  {Fedorets}, {Fern{\'a}ndez-Hern{\'a}ndez}, {Fernique}, {Fienga}, {Figueras},
  {Filippi}, {Findeisen}, {Fonti}, {Fouesneau}, {Fraile}, {Fraser}, {Fuchs},
  {Furnell}, {Gai}, {Galleti}, {Galluccio}, {Garabato}, {Garc{\'\i}a-Sedano},
  {Gar{\'e}}, {Garofalo}, {Garralda}, {Gavras}, {Gerssen}, {Geyer}, {Gilmore},
  {Girona}, {Giuffrida}, {Gomes}, {Gonz{\'a}lez-Marcos},
  {Gonz{\'a}lez-N{\'u}{\~n}ez}, {Gonz{\'a}lez-Vidal}, {Granvik}, {Guerrier},
  {Guillout}, {Guiraud}, {G{\'u}rpide}, {Guti{\'e}rrez-S{\'a}nchez}, {Guy},
  {Haigron}, {Hatzidimitriou}, {Haywood}, {Heiter}, {Helmi}, {Hobbs},
  {Hofmann}, {Holl}, {Holland}, {Hunt}, {Hypki}, {Icardi}, {Irwin}, {Jevardat
  de Fombelle}, {Jofr{\'e}}, {Jonker}, {Jorissen}, {Julbe}, {Karampelas},
  {Kochoska}, {Kohley}, {Kolenberg}, {Kontizas}, {Koposov}, {Kordopatis},
  {Koubsky}, {Kowalczyk}, {Krone-Martins}, {Kudryashova}, {Kull}, {Bachchan},
  {Lacoste-Seris}, {Lanza}, {Lavigne}, {Le Poncin-Lafitte}, {Lebreton},
  {Lebzelter}, {Leccia}, {Leclerc}, {Lecoeur-Taibi}, {Lemaitre}, {Lenhardt},
  {Leroux}, {Liao}, {Licata}, {Lindstr{\o}m}, {Lister}, {Livanou}, {Lobel},
  {L{\"o}ffler}, {L{\'o}pez}, {Lopez-Lozano}, {Lorenz}, {Loureiro},
  {MacDonald}, {Magalh{\~a}es Fernandes}, {Managau}, {Mann}, {Mantelet},
  {Marchal}, {Marchant}, {Marconi}, {Marie}, {Marinoni}, {Marrese},
  {Marschalk{\'o}}, {Marshall}, {Mart{\'\i}n-Fleitas}, {Martino}, {Mary},
  {Matijevi{\v{c}}}, {Mazeh}, {McMillan}, {Messina}, {Mestre}, {Michalik},
  {Millar}, {Miranda}, {Molina}, {Molinaro}, {Molinaro}, {Moln{\'a}r},
  {Moniez}, {Montegriffo}, {Monteiro}, {Mor}, {Mora}, {Morbidelli}, {Morel},
  {Morgenthaler}, {Morley}, {Morris}, {Mulone}, {Muraveva}, {Musella},
  {Narbonne}, {Nelemans}, {Nicastro}, {Noval}, {Ord{\'e}novic},
  {Ordieres-Mer{\'e}}, {Osborne}, {Pagani}, {Pagano}, {Pailler}, {Palacin},
  {Palaversa}, {Parsons}, {Paulsen}, {Pecoraro}, {Pedrosa}, {Pentik{\"a}inen},
  {Pereira}, {Pichon}, {Piersimoni}, {Pineau}, {Plachy}, {Plum}, {Poujoulet},
  {Pr{\v{s}}a}, {Pulone}, {Ragaini}, {Rago}, {Rambaux}, {Ramos-Lerate},
  {Ranalli}, {Rauw}, {Read}, {Regibo}, {Renk}, {Reyl{\'e}}, {Ribeiro},
  {Rimoldini}, {Ripepi}, {Riva}, {Rixon}, {Roelens}, {Romero-G{\'o}mez},
  {Rowell}, {Royer}, {Rudolph}, {Ruiz-Dern}, {Sadowski}, {Sagrist{\`a}
  Sell{\'e}s}, {Sahlmann}, {Salgado}, {Salguero}, {Sarasso}, {Savietto},
  {Schnorhk}, {Schultheis}, {Sciacca}, {Segol}, {Segovia}, {Segransan},
  {Serpell}, {Shih}, {Smareglia}, {Smart}, {Smith}, {Solano}, {Solitro},
  {Sordo}, {Soria Nieto}, {Souchay}, {Spagna}, {Spoto}, {Stampa}, {Steele},
  {Steidelm{\"u}ller}, {Stephenson}, {Stoev}, {Suess}, {S{\"u}veges}, {Surdej},
  {Szabados}, {Szegedi-Elek}, {Tapiador}, {Taris}, {Tauran}, {Taylor},
  {Teixeira}, {Terrett}, {Tingley}, {Trager}, {Turon}, {Ulla}, {Utrilla},
  {Valentini}, {van Elteren}, {Van Hemelryck}, {van Leeuwen}, {Varadi},
  {Vecchiato}, {Veljanoski}, {Via}, {Vicente}, {Vogt}, {Voss}, {Votruba},
  {Voutsinas}, {Walmsley}, {Weiler}, {Weingrill}, {Werner}, {Wevers},
  {Whitehead}, {Wyrzykowski}, {Yoldas}, {{\v{Z}}erjal}, {Zucker}, {Zurbach},
  {Zwitter}, {Alecu}, {Allen}, {Allende Prieto}, {Amorim},
  {Anglada-Escud{\'e}}, {Arsenijevic}, {Azaz}, {Balm}, {Beck}, {Bernstein},
  {Bigot}, {Bijaoui}, {Blasco}, {Bonfigli}, {Bono}, {Boudreault}, {Bressan},
  {Brown}, {Brunet}, {Bunclark}, {Buonanno}, {Butkevich}, {Carret}, {Carrion},
  {Chemin}, {Ch{\'e}reau}, {Corcione}, {Darmigny}, {de Boer}, {de Teodoro}, {de
  Zeeuw}, {Delle Luche}, {Domingues}, {Dubath}, {Fodor}, {Fr{\'e}zouls},
  {Fries}, {Fustes}, {Fyfe}, {Gallardo}, {Gallegos}, {Gardiol}, {Gebran},
  {Gomboc}, {G{\'o}mez}, {Grux}, {Gueguen}, {Heyrovsky}, {Hoar}, {Iannicola},
  {Isasi Parache}, {Janotto}, {Joliet}, {Jonckheere}, {Keil}, {Kim},
  {Klagyivik}, {Klar}, {Knude}, {Kochukhov}, {Kolka}, {Kos}, {Kutka}, {Lainey},
  {LeBouquin}, {Liu}, {Loreggia}, {Makarov}, {Marseille}, {Martayan},
  {Martinez-Rubi}, {Massart}, {Meynadier}, {Mignot}, {Munari}, {Nguyen},
  {Nordlander}, {Ocvirk}, {O'Flaherty}, {Olias Sanz}, {Ortiz}, {Osorio},
  {Oszkiewicz}, {Ouzounis}, {Palmer}, {Park}, {Pasquato}, {Peltzer}, {Peralta},
  {P{\'e}turaud}, {Pieniluoma}, {Pigozzi}, {Poels}, {Prat}, {Prod'homme},
  {Raison}, {Rebordao}, {Risquez}, {Rocca-Volmerange}, {Rosen}, {Ruiz-Fuertes},
  {Russo}, {Sembay}, {Serraller Vizcaino}, {Short}, {Siebert}, {Silva},
  {Sinachopoulos}, {Slezak}, {Soffel}, {Sosnowska}, {Strai{\v{z}}ys}, {ter
  Linden}, {Terrell}, {Theil}, {Tiede}, {Troisi}, {Tsalmantza}, {Tur},
  {Vaccari}, {Vachier}, {Valles}, {Van Hamme}, {Veltz}, {Virtanen}, {Wallut},
  {Wichmann}, {Wilkinson}, {Ziaeepour}, \& {Zschocke}}]{GAIA}
{Gaia Collaboration}, {Prusti}, T., {de Bruijne}, J.~H.~J., {et~al.} 2016,
  \aap, 595, A1, \dodoi{10.1051/0004-6361/201629272}

\bibitem[{{Gaia Collaboration} {et~al.}(2018){Gaia Collaboration}, {Brown},
  {Vallenari}, {Prusti}, {de Bruijne}, {Babusiaux}, {Bailer-Jones}, {Biermann},
  {Evans}, {Eyer}, {Jansen}, {Jordi}, {Klioner}, {Lammers}, {Lindegren},
  {Luri}, {Mignard}, {Panem}, {Pourbaix}, {Randich}, {Sartoretti}, {Siddiqui},
  {Soubiran}, {van Leeuwen}, {Walton}, {Arenou}, {Bastian}, {Cropper},
  {Drimmel}, {Katz}, {Lattanzi}, {Bakker}, {Cacciari}, {Casta{\~n}eda},
  {Chaoul}, {Cheek}, {De Angeli}, {Fabricius}, {Guerra}, {Holl}, {Masana},
  {Messineo}, {Mowlavi}, {Nienartowicz}, {Panuzzo}, {Portell}, {Riello},
  {Seabroke}, {Tanga}, {Th{\'e}venin}, {Gracia-Abril}, {Comoretto},
  {Garcia-Reinaldos}, {Teyssier}, {Altmann}, {Andrae}, {Audard},
  {Bellas-Velidis}, {Benson}, {Berthier}, {Blomme}, {Burgess}, {Busso},
  {Carry}, {Cellino}, {Clementini}, {Clotet}, {Creevey}, {Davidson}, {De
  Ridder}, {Delchambre}, {Dell'Oro}, {Ducourant},
  {Fern{\'a}ndez-Hern{\'a}ndez}, {Fouesneau}, {Fr{\'e}mat}, {Galluccio},
  {Garc{\'\i}a-Torres}, {Gonz{\'a}lez-N{\'u}{\~n}ez}, {Gonz{\'a}lez-Vidal},
  {Gosset}, {Guy}, {Halbwachs}, {Hambly}, {Harrison}, {Hern{\'a}ndez},
  {Hestroffer}, {Hodgkin}, {Hutton}, {Jasniewicz}, {Jean-Antoine-Piccolo},
  {Jordan}, {Korn}, {Krone-Martins}, {Lanzafame}, {Lebzelter}, {L{\"o}ffler},
  {Manteiga}, {Marrese}, {Mart{\'\i}n-Fleitas}, {Moitinho}, {Mora}, {Muinonen},
  {Osinde}, {Pancino}, {Pauwels}, {Petit}, {Recio-Blanco}, {Richards},
  {Rimoldini}, {Robin}, {Sarro}, {Siopis}, {Smith}, {Sozzetti}, {S{\"u}veges},
  {Torra}, {van Reeven}, {Abbas}, {Abreu Aramburu}, {Accart}, {Aerts},
  {Altavilla}, {{\'A}lvarez}, {Alvarez}, {Alves}, {Anderson}, {Andrei},
  {Anglada Varela}, {Antiche}, {Antoja}, {Arcay}, {Astraatmadja}, {Bach},
  {Baker}, {Balaguer-N{\'u}{\~n}ez}, {Balm}, {Barache}, {Barata}, {Barbato},
  {Barblan}, {Barklem}, {Barrado}, {Barros}, {Barstow}, {Bartholom{\'e}
  Mu{\~n}oz}, {Bassilana}, {Becciani}, {Bellazzini}, {Berihuete}, {Bertone},
  {Bianchi}, {Bienaym{\'e}}, {Blanco-Cuaresma}, {Boch}, {Boeche}, {Bombrun},
  {Borrachero}, {Bossini}, {Bouquillon}, {Bourda}, {Bragaglia}, {Bramante},
  {Breddels}, {Bressan}, {Brouillet}, {Br{\"u}semeister}, {Brugaletta},
  {Bucciarelli}, {Burlacu}, {Busonero}, {Butkevich}, {Buzzi}, {Caffau},
  {Cancelliere}, {Cannizzaro}, {Cantat-Gaudin}, {Carballo}, {Carlucci},
  {Carrasco}, {Casamiquela}, {Castellani}, {Castro-Ginard}, {Charlot},
  {Chemin}, {Chiavassa}, {Cocozza}, {Costigan}, {Cowell}, {Crifo}, {Crosta},
  {Crowley}, {Cuypers}, {Dafonte}, {Damerdji}, {Dapergolas}, {David}, {David},
  {de Laverny}, {De Luise}, {De March}, {de Martino}, {de Souza}, {de Torres},
  {Debosscher}, {del Pozo}, {Delbo}, {Delgado}, {Delgado}, {Di Matteo},
  {Diakite}, {Diener}, {Distefano}, {Dolding}, {Drazinos}, {Dur{\'a}n},
  {Edvardsson}, {Enke}, {Eriksson}, {Esquej}, {Eynard Bontemps}, {Fabre},
  {Fabrizio}, {Faigler}, {Falc{\~a}o}, {Farr{\`a}s Casas}, {Federici},
  {Fedorets}, {Fernique}, {Figueras}, {Filippi}, {Findeisen}, {Fonti},
  {Fraile}, {Fraser}, {Fr{\'e}zouls}, {Gai}, {Galleti}, {Garabato},
  {Garc{\'\i}a-Sedano}, {Garofalo}, {Garralda}, {Gavel}, {Gavras}, {Gerssen},
  {Geyer}, {Giacobbe}, {Gilmore}, {Girona}, {Giuffrida}, {Glass}, {Gomes},
  {Granvik}, {Gueguen}, {Guerrier}, {Guiraud}, {Guti{\'e}rrez-S{\'a}nchez},
  {Haigron}, {Hatzidimitriou}, {Hauser}, {Haywood}, {Heiter}, {Helmi}, {Heu},
  {Hilger}, {Hobbs}, {Hofmann}, {Holland}, {Huckle}, {Hypki}, {Icardi},
  {Jan{\ss}en}, {Jevardat de Fombelle}, {Jonker}, {Juh{\'a}sz}, {Julbe},
  {Karampelas}, {Kewley}, {Klar}, {Kochoska}, {Kohley}, {Kolenberg},
  {Kontizas}, {Kontizas}, {Koposov}, {Kordopatis}, {Kostrzewa-Rutkowska},
  {Koubsky}, {Lambert}, {Lanza}, {Lasne}, {Lavigne}, {Le Fustec}, {Le
  Poncin-Lafitte}, {Lebreton}, {Leccia}, {Leclerc}, {Lecoeur-Taibi},
  {Lenhardt}, {Leroux}, {Liao}, {Licata}, {Lindstr{\o}m}, {Lister}, {Livanou},
  {Lobel}, {L{\'o}pez}, {Managau}, {Mann}, {Mantelet}, {Marchal}, {Marchant},
  {Marconi}, {Marinoni}, {Marschalk{\'o}}, {Marshall}, {Martino}, {Marton},
  {Mary}, {Massari}, {Matijevi{\v{c}}}, {Mazeh}, {McMillan}, {Messina},
  {Michalik}, {Millar}, {Molina}, {Molinaro}, {Moln{\'a}r}, {Montegriffo},
  {Mor}, {Morbidelli}, {Morel}, {Morris}, {Mulone}, {Muraveva}, {Musella},
  {Nelemans}, {Nicastro}, {Noval}, {O'Mullane}, {Ord{\'e}novic},
  {Ord{\'o}{\~n}ez-Blanco}, {Osborne}, {Pagani}, {Pagano}, {Pailler},
  {Palacin}, {Palaversa}, {Panahi}, {Pawlak}, {Piersimoni}, {Pineau}, {Plachy},
  {Plum}, {Poggio}, {Poujoulet}, {Pr{\v{s}}a}, {Pulone}, {Racero}, {Ragaini},
  {Rambaux}, {Ramos-Lerate}, {Regibo}, {Reyl{\'e}}, {Riclet}, {Ripepi}, {Riva},
  {Rivard}, {Rixon}, {Roegiers}, {Roelens}, {Romero-G{\'o}mez}, {Rowell},
  {Royer}, {Ruiz-Dern}, {Sadowski}, {Sagrist{\`a} Sell{\'e}s}, {Sahlmann},
  {Salgado}, {Salguero}, {Sanna}, {Santana-Ros}, {Sarasso}, {Savietto},
  {Schultheis}, {Sciacca}, {Segol}, {Segovia}, {S{\'e}gransan}, {Shih},
  {Siltala}, {Silva}, {Smart}, {Smith}, {Solano}, {Solitro}, {Sordo}, {Soria
  Nieto}, {Souchay}, {Spagna}, {Spoto}, {Stampa}, {Steele},
  {Steidelm{\"u}ller}, {Stephenson}, {Stoev}, {Suess}, {Surdej}, {Szabados},
  {Szegedi-Elek}, {Tapiador}, {Taris}, {Tauran}, {Taylor}, {Teixeira},
  {Terrett}, {Teyssandier}, {Thuillot}, {Titarenko}, {Torra Clotet}, {Turon},
  {Ulla}, {Utrilla}, {Uzzi}, {Vaillant}, {Valentini}, {Valette}, {van Elteren},
  {Van Hemelryck}, {van Leeuwen}, {Vaschetto}, {Vecchiato}, {Veljanoski},
  {Viala}, {Vicente}, {Vogt}, {von Essen}, {Voss}, {Votruba}, {Voutsinas},
  {Walmsley}, {Weiler}, {Wertz}, {Wevers}, {Wyrzykowski}, {Yoldas},
  {{\v{Z}}erjal}, {Ziaeepour}, {Zorec}, {Zschocke}, {Zucker}, {Zurbach}, \&
  {Zwitter}}]{GAIADR2}
{Gaia Collaboration}, {Brown}, A.~G.~A., {Vallenari}, A., {et~al.} 2018, \aap,
  616, A1, \dodoi{10.1051/0004-6361/201833051}

\bibitem[{{Gaidos} {et~al.}(2022){Gaidos}, {Hirano}, {Beichman}, {Livingston},
  {Harakawa}, {Hodapp}, {Ishizuka}, {Jacobson}, {Konishi}, {Kotani}, {Kudo},
  {Kurokawa}, {Kuzuhara}, {Nishikawa}, {Omiya}, {Serizawa}, {Tamura}, {Ueda},
  \& {Vievard}}]{V1298Tau}
{Gaidos}, E., {Hirano}, T., {Beichman}, C., {et~al.} 2022, \mnras, 509, 2969,
  \dodoi{10.1093/mnras/stab3107}

\bibitem[{{Gandolfi} {et~al.}(2010){Gandolfi}, {H{\'e}brard}, {Alonso},
  {Deleuil}, {Guenther}, {Fridlund}, {Endl}, {Eigm{\"u}ller}, {Csizmadia},
  {Havel}, {Aigrain}, {Auvergne}, {Baglin}, {Barge}, {Bonomo}, {Bord{\'e}},
  {Bouchy}, {Bruntt}, {Cabrera}, {Carpano}, {Carone}, {Cochran}, {Deeg},
  {Dvorak}, {Eisl{\"o}ffel}, {Erikson}, {Ferraz-Mello}, {Gazzano}, {Gibson},
  {Gillon}, {Gondoin}, {Guillot}, {Hartmann}, {Hatzes}, {Jorda}, {Kabath},
  {L{\'e}ger}, {Llebaria}, {Lammer}, {MacQueen}, {Mayor}, {Mazeh}, {Moutou},
  {Ollivier}, {P{\"a}tzold}, {Pepe}, {Queloz}, {Rauer}, {Rouan}, {Samuel},
  {Schneider}, {Stecklum}, {Tingley}, {Udry}, \& {Wuchterl}}]{spinny:CoRoT11}
{Gandolfi}, D., {H{\'e}brard}, G., {Alonso}, R., {et~al.} 2010, \aap, 524, A55,
  \dodoi{10.1051/0004-6361/201015132}

\bibitem[{{Grunblatt} {et~al.}(2019){Grunblatt}, {Huber}, {Gaidos}, {Hon},
  {Zinn}, \& {Stello}}]{grunblatt19}
{Grunblatt}, S.~K., {Huber}, D., {Gaidos}, E., {et~al.} 2019, \aj, 158, 227,
  \dodoi{10.3847/1538-3881/ab4c35}

\bibitem[{{G{\"u}nther} \& {Daylan}(2019)}]{allesfitter-code}
{G{\"u}nther}, M.~N., \& {Daylan}, T. 2019, {Allesfitter: Flexible Star and
  Exoplanet Inference From Photometry and Radial Velocity}, Astrophysics Source
  Code Library.
\newblock \doeprint{1903.003}

\bibitem[{{G{\"u}nther} \& {Daylan}(2021)}]{allesfitter-paper}
---. 2021, \apjs, 254, 13, \dodoi{10.3847/1538-4365/abe70e}

\bibitem[{{Hay} {et~al.}(2016){Hay}, {Collier-Cameron}, {Doyle}, {H{\'e}brard},
  {Skillen}, {Anderson}, {Barros}, {Brown}, {Bouchy}, {Busuttil}, {Delorme},
  {Delrez}, {Demangeon}, {D{\'\i}az}, {Gillon}, {G{\'o}mez Maqueo Chew},
  {Gonz{\`a}lez}, {Hellier}, {Holmes}, {Jarvis}, {Jehin}, {Joshi}, {Kolb},
  {Lendl}, {Maxted}, {McCormac}, {Miller}, {Mortier}, {Pall{\'e}}, {Pollacco},
  {Prieto-Arranz}, {Queloz}, {S{\'e}gransan}, {Simpson}, {Smalley},
  {Southworth}, {Triaud}, {Turner}, {Udry}, {Vanhuysse}, {West}, \&
  {Wilson}}]{spinny:WASP93b}
{Hay}, K.~L., {Collier-Cameron}, A., {Doyle}, A.~P., {et~al.} 2016, \mnras,
  463, 3276, \dodoi{10.1093/mnras/stw2090}

\bibitem[{{Hayward} {et~al.}(2001){Hayward}, {Brandl}, {Pirger}, {Blacken},
  {Gull}, {Schoenwald}, \& {Houck}}]{hayward2001}
{Hayward}, T.~L., {Brandl}, B., {Pirger}, B., {et~al.} 2001, \pasp, 113, 105,
  \dodoi{10.1086/317969}

\bibitem[{{Heitzmann} {et~al.}(2021){Heitzmann}, {Zhou}, {Quinn}, {Marsden},
  {Wright}, {Petit}, {Vanderburg}, {Bouma}, {Mann}, \& {Rizzuto}}]{HIP67522b}
{Heitzmann}, A., {Zhou}, G., {Quinn}, S.~N., {et~al.} 2021, \apjl, 922, L1,
  \dodoi{10.3847/2041-8213/ac3485}

\bibitem[{{Henry} {et~al.}(2000){Henry}, {Marcy}, {Butler}, \& {Vogt}}]{HJ2}
{Henry}, G.~W., {Marcy}, G.~W., {Butler}, R.~P., \& {Vogt}, S.~S. 2000, \apjl,
  529, L41, \dodoi{10.1086/312458}

\bibitem[{{H{\o}g} {et~al.}(2000){H{\o}g}, {Fabricius}, {Makarov}, {Urban},
  {Corbin}, {Wycoff}, {Bastian}, {Schwekendiek}, \& {Wicenec}}]{2000Hog}
{H{\o}g}, E., {Fabricius}, C., {Makarov}, V.~V., {et~al.} 2000, \aap, 355, L27

\bibitem[{{Horner} {et~al.}(2020){Horner}, {Kane}, {Marshall}, {Dalba}, {Holt},
  {Wood}, {Maynard-Casely}, {Wittenmyer}, {Lykawka}, {Hill}, {Salmeron},
  {Bailey}, {L{\"o}hne}, {Agnew}, {Carter}, \& {Tylor}}]{SSRev}
{Horner}, J., {Kane}, S.~R., {Marshall}, J.~P., {et~al.} 2020, \pasp, 132,
  102001, \dodoi{10.1088/1538-3873/ab8eb9}

\bibitem[{{Houk} \& {Smith-Moore}(1988)}]{1988Houk}
{Houk}, N., \& {Smith-Moore}, M. 1988, {Michigan Catalogue of Two-dimensional
  Spectral Types for the HD Stars. Volume 4, Declinations -26.0 to -12.0.},
  Vol.~4

\bibitem[{{Ida} \& {Lin}(2004)}]{Ice2}
{Ida}, S., \& {Lin}, D.~N.~C. 2004, \apj, 616, 567, \dodoi{10.1086/424830}

\bibitem[{{Ida} \& {Lin}(2005)}]{Big1}
---. 2005, \apj, 626, 1045, \dodoi{10.1086/429953}

\bibitem[{{Jenkins}(2002)}]{2002ApJ...575..493J}
{Jenkins}, J.~M. 2002, \apj, 575, 493, \dodoi{10.1086/341136}

\bibitem[{{Jenkins} {et~al.}(2020){Jenkins}, {Tenenbaum}, {Seader}, {Burke},
  {McCauliff}, {Smith}, {Twicken}, \& {Chandrasekaran}}]{2020TPSkdph}
{Jenkins}, J.~M., {Tenenbaum}, P., {Seader}, S., {et~al.} 2020, {Kepler Data
  Processing Handbook: Transiting Planet Search}, Kepler Science Document
  KSCI-19081-003

\bibitem[{{Jenkins} {et~al.}(2010){Jenkins}, {Chandrasekaran}, {McCauliff},
  {Caldwell}, {Tenenbaum}, {Li}, {Klaus}, {Cote}, \&
  {Middour}}]{2010SPIE.7740E..0DJ}
{Jenkins}, J.~M., {Chandrasekaran}, H., {McCauliff}, S.~D., {et~al.} 2010, in
  Society of Photo-Optical Instrumentation Engineers (SPIE) Conference Series,
  Vol. 7740, Software and Cyberinfrastructure for Astronomy, ed. N.~M.
  {Radziwill} \& A.~{Bridger}, 77400D, \dodoi{10.1117/12.856764}

\bibitem[{{Jenkins} {et~al.}(2016){Jenkins}, {Twicken}, {McCauliff},
  {Campbell}, {Sanderfer}, {Lung}, {Mansouri-Samani}, {Girouard}, {Tenenbaum},
  {Klaus}, {Smith}, {Caldwell}, {Chacon}, {Henze}, {Heiges}, {Latham},
  {Morgan}, {Swade}, {Rinehart}, \& {Vanderspek}}]{2016SPIE.9913E..3EJ}
{Jenkins}, J.~M., {Twicken}, J.~D., {McCauliff}, S., {et~al.} 2016, in
  \procspie, Vol. 9913, Software and Cyberinfrastructure for Astronomy IV,
  99133E, \dodoi{10.1117/12.2233418}

\bibitem[{{Jensen}(2013)}]{Jensen:2013}
{Jensen}, E. 2013, {Tapir: A web interface for transit/eclipse observability}.
\newblock \doeprint{1306.007}

\bibitem[{{Johnson} {et~al.}(2007){Johnson}, {Butler}, {Marcy}, {Fischer},
  {Vogt}, {Wright}, \& {Peek}}]{BigJohnson}
{Johnson}, J.~A., {Butler}, R.~P., {Marcy}, G.~W., {et~al.} 2007, \apj, 670,
  833, \dodoi{10.1086/521720}

\bibitem[{{Jones} {et~al.}(2017){Jones}, {Brahm}, {Wittenmyer}, {Drass},
  {Jenkins}, {Melo}, {Vos}, \& {Rojo}}]{Jones2017}
{Jones}, M.~I., {Brahm}, R., {Wittenmyer}, R.~A., {et~al.} 2017, \aap, 602,
  A58, \dodoi{10.1051/0004-6361/201630278}

\bibitem[{{Jones} {et~al.}(2016){Jones}, {Jenkins}, {Brahm}, {Wittenmyer},
  {Olivares E.}, {Melo}, {Rojo}, {Jord{\'a}n}, {Drass}, {Butler}, \&
  {Wang}}]{Big3}
{Jones}, M.~I., {Jenkins}, J.~S., {Brahm}, R., {et~al.} 2016, \aap, 590, A38,
  \dodoi{10.1051/0004-6361/201628067}

\bibitem[{{Jord{\'a}n} {et~al.}(2020){Jord{\'a}n}, {Brahm}, {Espinoza},
  {Henning}, {Jones}, {Kossakowski}, {Sarkis}, {Trifonov}, {Rojas}, {Torres},
  {Drass}, {Nandakumar}, {Barbieri}, {Davis}, {Wang}, {Bayliss}, {Bouma},
  {Dragomir}, {Eastman}, {Daylan}, {Guerrero}, {Barclay}, {Ting}, {Henze},
  {Ricker}, {Vanderspek}, {Latham}, {Seager}, {Winn}, {Jenkins}, {Wittenmyer},
  {Bowler}, {Crossfield}, {Horner}, {Kane}, {Kielkopf}, {Morton}, {Plavchan},
  {Tinney}, {Addison}, {Mengel}, {Okumura}, {Shahaf}, {Mazeh}, {Rabus},
  {Shporer}, {Ziegler}, {Mann}, \& {Hart}}]{TOI677}
{Jord{\'a}n}, A., {Brahm}, R., {Espinoza}, N., {et~al.} 2020, \aj, 159, 145,
  \dodoi{10.3847/1538-3881/ab6f67}

\bibitem[{{Kiefer}(2019)}]{LathamStar}
{Kiefer}, F. 2019, \aap, 632, L9, \dodoi{10.1051/0004-6361/201936942}

\bibitem[{{Kossakowski} {et~al.}(2019){Kossakowski}, {Espinoza}, {Brahm},
  {Jord{\'a}n}, {Henning}, {Rojas}, {K{\"u}rster}, {Sarkis}, {Schlecker},
  {Pozuelos}, {Barkaoui}, {Jehin}, {Gillon}, {Matthews}, {Horch}, {Ciardi},
  {Crossfield}, {Gonzales}, {Howell}, {Matson}, {Schlieder}, {Jenkins},
  {Ricker}, {Seager}, {Winn}, {Li}, {Rose}, {Smith}, {Dynes}, {Morgan},
  {Villasenor}, {Charbonneau}, {Jaffe}, {Yu}, {Bakos}, {Bhatti}, {Bouchy},
  {Collins}, {Collins}, {Csubry}, {Evans}, {Jensen}, {Lovis}, {Marmier},
  {Nielsen}, {Osip}, {Pepe}, {Relles}, {S{\'e}gransan}, {Shporer}, {Stockdale},
  {Suc}, {Turner}, \& {Udry}}]{TESS2}
{Kossakowski}, D., {Espinoza}, N., {Brahm}, R., {et~al.} 2019, \mnras, 490,
  1094, \dodoi{10.1093/mnras/stz2433}

\bibitem[{{Kozai}(1962)}]{Kozai1}
{Kozai}, Y. 1962, \aj, 67, 591, \dodoi{10.1086/108790}

\bibitem[{{Kraft}(1967)}]{KraftBreak}
{Kraft}, R.~P. 1967, \apj, 150, 551, \dodoi{10.1086/149359}

\bibitem[{{Kunimoto} \& {Matthews}(2020)}]{HJOccur3}
{Kunimoto}, M., \& {Matthews}, J.~M. 2020, \aj, 159, 248,
  \dodoi{10.3847/1538-3881/ab88b0}

\bibitem[{{Latham} {et~al.}(1989){Latham}, {Mazeh}, {Stefanik}, {Mayor}, \&
  {Burki}}]{LathamsWorld}
{Latham}, D.~W., {Mazeh}, T., {Stefanik}, R.~P., {Mayor}, M., \& {Burki}, G.
  1989, \nat, 339, 38, \dodoi{10.1038/339038a0}

\bibitem[{{Li} {et~al.}(2019{\natexlab{a}}){Li}, {Mustill}, \&
  {Davies}}]{scatter3}
{Li}, D., {Mustill}, A.~J., \& {Davies}, M.~B. 2019{\natexlab{a}}, \mnras, 488,
  1366, \dodoi{10.1093/mnras/stz1794}

\bibitem[{{Li} {et~al.}(2019{\natexlab{b}}){Li}, {Tenenbaum}, {Twicken},
  {Burke}, {Jenkins}, {Quintana}, {Rowe}, \& {Seader}}]{Li:DVmodelFit2019}
{Li}, J., {Tenenbaum}, P., {Twicken}, J.~D., {et~al.} 2019{\natexlab{b}},
  \pasp, 131, 024506, \dodoi{10.1088/1538-3873/aaf44d}

\bibitem[{{Lidov}(1962)}]{Kozai2}
{Lidov}, M.~L. 1962, \planss, 9, 719, \dodoi{10.1016/0032-0633(62)90129-0}

\bibitem[{{Lin} {et~al.}(1996){Lin}, {Bodenheimer}, \& {Richardson}}]{Disk1}
{Lin}, D.~N.~C., {Bodenheimer}, P., \& {Richardson}, D.~C. 1996, \nat, 380,
  606, \dodoi{10.1038/380606a0}

\bibitem[{{Lund}(2021 submitted)}]{lund2020}
{Lund}, M.~B. 2021 submitted, \mnras

\bibitem[{{Masset} \& {Papaloizou}(2003)}]{HJF:Masset2003}
{Masset}, F.~S., \& {Papaloizou}, J.~C.~B. 2003, \apj, 588, 494,
  \dodoi{10.1086/373892}

\bibitem[{{Masuda} \& {Winn}(2020)}]{2020AJ....159...81M}
{Masuda}, K., \& {Winn}, J.~N. 2020, \aj, 159, 81,
  \dodoi{10.3847/1538-3881/ab65be}

\bibitem[{{Mayor} \& {Queloz}(1995)}]{51Peg}
{Mayor}, M., \& {Queloz}, D. 1995, \nat, 378, 355, \dodoi{10.1038/378355a0}

\bibitem[{{McCormac} {et~al.}(2013){McCormac}, {Pollacco}, {Skillen}, {Faedi},
  {Todd}, \& {Watson}}]{mccormac13donuts}
{McCormac}, J., {Pollacco}, D., {Skillen}, I., {et~al.} 2013, \pasp, 125, 548,
  \dodoi{10.1086/670940}

\bibitem[{{McCully} {et~al.}(2018){McCully}, {Volgenau}, {Harbeck}, {Lister},
  {Saunders}, {Turner}, {Siiverd}, \& {Bowman}}]{McCully:2018}
{McCully}, C., {Volgenau}, N.~H., {Harbeck}, D.-R., {et~al.} 2018, in Society
  of Photo-Optical Instrumentation Engineers (SPIE) Conference Series, Vol.
  10707, Software and Cyberinfrastructure for Astronomy V, ed. J.~C. {Guzman}
  \& J.~{Ibsen}, 107070K, \dodoi{10.1117/12.2314340}

\bibitem[{{Morton}(2015)}]{isochrones}
{Morton}, T.~D. 2015, {isochrones: Stellar model grid package}.
\newblock \doeprint{1503.010}

\bibitem[{{Nagasawa} \& {Ida}(2011)}]{Kozai4}
{Nagasawa}, M., \& {Ida}, S. 2011, \apj, 742, 72,
  \dodoi{10.1088/0004-637X/742/2/72}

\bibitem[{{Nagasawa} {et~al.}(2008{\natexlab{a}}){Nagasawa}, {Ida}, \&
  {Bessho}}]{Kozai3}
{Nagasawa}, M., {Ida}, S., \& {Bessho}, T. 2008{\natexlab{a}}, \apj, 678, 498,
  \dodoi{10.1086/529369}

\bibitem[{{Nagasawa} {et~al.}(2008{\natexlab{b}}){Nagasawa}, {Ida}, \&
  {Bessho}}]{HJF:2008Nagasawa}
---. 2008{\natexlab{b}}, \apj, 678, 498, \dodoi{10.1086/529369}

\bibitem[{{Nielsen} {et~al.}(2019){Nielsen}, {Bouchy}, {Turner}, {Giles},
  {Mascare{\~n}o}, {Lovis}, {Marmier}, {Pepe}, {S{\'e}gransan}, {Udry},
  {Otegi}, {Ottoni}, {Stalport}, {Ricker}, {Vanderspek}, {Latham}, {Seager},
  {Winn}, {Jenkins}, {Kane}, {Wittenmyer}, {Bowler}, {Crossfield}, {Horner},
  {Kielkopf}, {Morton}, {Plavchan}, {Tinney}, {Zhang}, {Wright}, {Mengel},
  {Clark}, {Okumura}, {Addison}, {Caldwell}, {Cartwright}, {Collins},
  {Francis}, {Guerrero}, {Huang}, {Matthews}, {Pepper}, {Rose},
  {Villase{\~n}or}, {Wohler}, {Stassun}, {Howell}, {Ciardi}, {Gonzales},
  {Matson}, {Beichman}, \& {Schlieder}}]{TESS1}
{Nielsen}, L.~D., {Bouchy}, F., {Turner}, O., {et~al.} 2019, \aap, 623, A100,
  \dodoi{10.1051/0004-6361/201834577}

\bibitem[{{Paredes} {et~al.}(2021){Paredes}, {Henry}, {Quinn}, {Gies},
  {Hinojosa-Go{\~n}i}, {James}, {Jao}, \& {White}}]{Paredes2021}
{Paredes}, L.~A., {Henry}, T.~J., {Quinn}, S.~N., {et~al.} 2021, \aj, 162, 176,
  \dodoi{10.3847/1538-3881/ac082a}

\bibitem[{{Pecaut} \& {Mamajek}(2013)}]{2013ApJS..208....9P}
{Pecaut}, M.~J., \& {Mamajek}, E.~E. 2013, \apjs, 208, 9,
  \dodoi{10.1088/0067-0049/208/1/9}

\bibitem[{{Plavchan} {et~al.}(2020){Plavchan}, {Barclay}, {Gagn{\'e}}, {Gao},
  {Cale}, {Matzko}, {Dragomir}, {Quinn}, {Feliz}, {Stassun}, {Crossfield},
  {Berardo}, {Latham}, {Tieu}, {Anglada-Escud{\'e}}, {Ricker}, {Vanderspek},
  {Seager}, {Winn}, {Jenkins}, {Rinehart}, {Krishnamurthy}, {Dynes}, {Doty},
  {Adams}, {Afanasev}, {Beichman}, {Bottom}, {Bowler}, {Brinkworth}, {Brown},
  {Cancino}, {Ciardi}, {Clampin}, {Clark}, {Collins}, {Davison},
  {Foreman-Mackey}, {Furlan}, {Gaidos}, {Geneser}, {Giddens}, {Gilbert},
  {Hall}, {Hellier}, {Henry}, {Horner}, {Howard}, {Huang}, {Huber}, {Kane},
  {Kenworthy}, {Kielkopf}, {Kipping}, {Klenke}, {Kruse}, {Latouf}, {Lowrance},
  {Mennesson}, {Mengel}, {Mills}, {Morton}, {Narita}, {Newton}, {Nishimoto},
  {Okumura}, {Palle}, {Pepper}, {Quintana}, {Roberge}, {Roccatagliata},
  {Schlieder}, {Tanner}, {Teske}, {Tinney}, {Vanderburg}, {von Braun}, {Walp},
  {Wang}, {Wang}, {Weigand}, {White}, {Wittenmyer}, {Wright}, {Youngblood},
  {Zhang}, \& {Zilberman}}]{AUMic}
{Plavchan}, P., {Barclay}, T., {Gagn{\'e}}, J., {et~al.} 2020, \nat, 582, 497,
  \dodoi{10.1038/s41586-020-2400-z}

\bibitem[{{Pollack} {et~al.}(1996){Pollack}, {Hubickyj}, {Bodenheimer},
  {Lissauer}, {Podolak}, \& {Greenzweig}}]{Ice1}
{Pollack}, J.~B., {Hubickyj}, O., {Bodenheimer}, P., {et~al.} 1996, \icarus,
  124, 62, \dodoi{10.1006/icar.1996.0190}

\bibitem[{{Preibisch} {et~al.}(1999){Preibisch}, {Balega}, {Hofmann},
  {Weigelt}, \& {Zinnecker}}]{bin1}
{Preibisch}, T., {Balega}, Y., {Hofmann}, K.-H., {Weigelt}, G., \& {Zinnecker},
  H. 1999, \na, 4, 531, \dodoi{10.1016/S1384-1076(99)00042-1}

\bibitem[{{Queloz} {et~al.}(2001){Queloz}, {Mayor}, {Udry}, {Burnet},
  {Carrier}, {Eggenberger}, {Naef}, {Santos}, {Pepe}, {Rupprecht}, {Avila},
  {Baeza}, {Benz}, {Bertaux}, {Bouchy}, {Cavadore}, {Delabre}, {Eckert},
  {Fischer}, {Fleury}, {Gilliotte}, {Goyak}, {Guzman}, {Kohler}, {Lacroix},
  {Lizon}, {Megevand}, {Sivan}, {Sosnowska}, \& {Weilenmann}}]{CORALIE}
{Queloz}, D., {Mayor}, M., {Udry}, S., {et~al.} 2001, The Messenger, 105, 1

\bibitem[{{Ricker} {et~al.}(2015){Ricker}, {Winn}, {Vanderspek}, {Latham},
  {Bakos}, {Bean}, {Berta-Thompson}, {Brown}, {Buchhave}, {Butler}, {Butler},
  {Chaplin}, {Charbonneau}, {Christensen-Dalsgaard}, {Clampin}, {Deming},
  {Doty}, {De Lee}, {Dressing}, {Dunham}, {Endl}, {Fressin}, {Ge}, {Henning},
  {Holman}, {Howard}, {Ida}, {Jenkins}, {Jernigan}, {Johnson}, {Kaltenegger},
  {Kawai}, {Kjeldsen}, {Laughlin}, {Levine}, {Lin}, {Lissauer}, {MacQueen},
  {Marcy}, {McCullough}, {Morton}, {Narita}, {Paegert}, {Palle}, {Pepe},
  {Pepper}, {Quirrenbach}, {Rinehart}, {Sasselov}, {Sato}, {Seager},
  {Sozzetti}, {Stassun}, {Sullivan}, {Szentgyorgyi}, {Torres}, {Udry}, \&
  {Villasenor}}]{TESSRick}
{Ricker}, G.~R., {Winn}, J.~N., {Vanderspek}, R., {et~al.} 2015, Journal of
  Astronomical Telescopes, Instruments, and Systems, 1, 014003,
  \dodoi{10.1117/1.JATIS.1.1.014003}

\bibitem[{{Rodr{\'\i}guez Mart{\'\i}nez} {et~al.}(2020){Rodr{\'\i}guez
  Mart{\'\i}nez}, {Gaudi}, {Rodriguez}, {Zhou}, {Labadie-Bartz}, {Quinn},
  {Penev}, {Tan}, {Latham}, {Paredes}, {Kielkopf}, {Addison}, {Wright},
  {Teske}, {Howell}, {Ciardi}, {Ziegler}, {Stassun}, {Johnson}, {Eastman},
  {Siverd}, {Beatty}, {Bouma}, {Bedding}, {Pepper}, {Winn}, {Lund},
  {Villanueva}, {Stevens}, {Jensen}, {Kilby}, {Crane}, {Tokovinin}, {Everett},
  {Tinney}, {Fausnaugh}, {Cohen}, {Bayliss}, {Bieryla}, {Cargile}, {Collins},
  {Conti}, {Col{\'o}n}, {Curtis}, {Depoy}, {Evans}, {Feliz}, {Gregorio},
  {Rothenberg}, {James}, {Joner}, {Kuhn}, {Manner}, {Khakpash}, {Marshall},
  {McLeod}, {Penny}, {Reed}, {Relles}, {Stephens}, {Stockdale}, {Trueblood},
  {Trueblood}, {Yao}, {Zambelli}, {Vanderspek}, {Seager}, {Jenkins}, {Henry},
  {James}, {Jao}, {Wang}, {Butler}, {Thompson}, {Shectman}, {Wittenmyer},
  {Bowler}, {Horner}, {Kane}, {Mengel}, {Morton}, {Okumura}, {Plavchan},
  {Zhang}, {Scott}, {Matson}, {Mann}, {Dragomir}, {G{\"u}nther}, {Ting},
  {Glidden}, \& {Quintana}}]{HotMiss3}
{Rodr{\'\i}guez Mart{\'\i}nez}, R., {Gaudi}, B.~S., {Rodriguez}, J.~E.,
  {et~al.} 2020, \aj, 160, 111, \dodoi{10.3847/1538-3881/ab9f2d}

\bibitem[{{Sebastian} {et~al.}(2022){Sebastian}, {Guenther}, {Deleuil},
  {Dorsch}, {Heber}, {Heuser}, {Gandolfi}, {Grziwa}, {Deeg}, {Alonso},
  {Bouchy}, {Csizmadia}, {Cusano}, {Fridlund}, {Geier}, {Irrgang}, {Korth},
  {Nespral}, {Rauer}, {Tal-Or}, \& {CoRoT-team}}]{2022MNRAS.516..636S}
{Sebastian}, D., {Guenther}, E.~W., {Deleuil}, M., {et~al.} 2022, \mnras, 516,
  636, \dodoi{10.1093/mnras/stac2131}

\bibitem[{{Siverd} {et~al.}(2018){Siverd}, {Collins}, {Zhou}, {Quinn}, {Gaudi},
  {Stassun}, {Johnson}, {Bieryla}, {Latham}, {Ciardi}, {Rodriguez}, {Penev},
  {Pinsonneault}, {Pepper}, {Eastman}, {Relles}, {Kielkopf}, {Gregorio},
  {Oberst}, {Aldi}, {Esquerdo}, {Calkins}, {Berlind}, {Dressing}, {Patel},
  {Stevens}, {Beatty}, {Lund}, {Labadie-Bartz}, {Kuhn}, {Col{\'o}n}, {James},
  {Yao}, {Johnson}, {Wright}, {McCrady}, {Wittenmyer}, {Johnson}, {Sliski},
  {Jensen}, {Cohen}, {McLeod}, {Penny}, {Joner}, {Stephens}, {Villanueva},
  {Zambelli}, {Stockdale}, {Evans}, {Tan}, {Curtis}, {Reed}, {Trueblood}, \&
  {Trueblood}}]{retro1}
{Siverd}, R.~J., {Collins}, K.~A., {Zhou}, G., {et~al.} 2018, \aj, 155, 35,
  \dodoi{10.3847/1538-3881/aa9e4d}

\bibitem[{{Smith} {et~al.}(2020){Smith}, {Eigm{\"u}ller}, {Gurumoorthy},
  {Csizmadia}, {Bayliss}, {Burleigh}, {Cabrera}, {Casewell}, {Erikson}, {Goad},
  {Grange}, {Jenkins}, {Pollacco}, {Rauer}, {Raynard}, {Udry}, {West}, \&
  {Wheatley}}]{smith2020multicam}
{Smith}, A. M.~S., {Eigm{\"u}ller}, P., {Gurumoorthy}, R., {et~al.} 2020,
  Astronomische Nachrichten, 341, 273, \dodoi{10.1002/asna.202013768}

\bibitem[{{Speagle}(2020)}]{dynesty}
{Speagle}, J.~S. 2020, \mnras, 493, 3132, \dodoi{10.1093/mnras/staa278}

\bibitem[{{Stassun} {et~al.}(2019){Stassun}, {Oelkers}, {Paegert}, {Torres},
  {Pepper}, {De Lee}, {Collins}, {Latham}, {Muirhead}, {Chittidi},
  {Rojas-Ayala}, {Fleming}, {Rose}, {Tenenbaum}, {Ting}, {Kane}, {Barclay},
  {Bean}, {Brassuer}, {Charbonneau}, {Ge}, {Lissauer}, {Mann}, {McLean},
  {Mullally}, {Narita}, {Plavchan}, {Ricker}, {Sasselov}, {Seager}, {Sharma},
  {Shiao}, {Sozzetti}, {Stello}, {Vanderspek}, {Wallace}, \& {Winn}}]{TIC19}
{Stassun}, K.~G., {Oelkers}, R.~J., {Paegert}, M., {et~al.} 2019, \aj, 158,
  138, \dodoi{10.3847/1538-3881/ab3467}

\bibitem[{{Tanaka} {et~al.}(2002){Tanaka}, {Takeuchi}, \& {Ward}}]{Disk3}
{Tanaka}, H., {Takeuchi}, T., \& {Ward}, W.~R. 2002, \apj, 565, 1257,
  \dodoi{10.1086/324713}

\bibitem[{{Temple} {et~al.}(2019){Temple}, {Hellier}, {Anderson}, {Barkaoui},
  {Bouchy}, {Brown}, {Burdanov}, {Collier Cameron}, {Delrez}, {Ducrot},
  {Evans}, {Gillon}, {Jehin}, {Lendl}, {Maxted}, {McCormac}, {Murray},
  {Nielsen}, {Pepe}, {Pollacco}, {Queloz}, {S{\'e}gransan}, {Smalley},
  {Thompson}, {Triaud}, {Turner}, {Udry}, {West}, \& {Zouhair}}]{retro2}
{Temple}, L.~Y., {Hellier}, C., {Anderson}, D.~R., {et~al.} 2019, \mnras, 490,
  2467, \dodoi{10.1093/mnras/stz2632}

\bibitem[{{Tinney} {et~al.}(2001){Tinney}, {Butler}, {Marcy}, {Jones}, {Penny},
  {Vogt}, {Apps}, \& {Henry}}]{HJ3}
{Tinney}, C.~G., {Butler}, R.~P., {Marcy}, G.~W., {et~al.} 2001, \apj, 551,
  507, \dodoi{10.1086/320097}

\bibitem[{{Tokovinin} {et~al.}(2013){Tokovinin}, {Fischer}, {Bonati},
  {Giguere}, {Moore}, {Schwab}, {Spronck}, \& {Szymkowiak}}]{Tokovinin2013}
{Tokovinin}, A., {Fischer}, D.~A., {Bonati}, M., {et~al.} 2013, \pasp, 125,
  1336, \dodoi{10.1086/674012}

\bibitem[{{Trifonov} {et~al.}(2017){Trifonov}, {K{\"u}rster}, {Zechmeister},
  {Zakhozhay}, {Reffert}, {Lee}, {Rodler}, {Vogt}, \& {Brems}}]{trifonov17}
{Trifonov}, T., {K{\"u}rster}, M., {Zechmeister}, M., {et~al.} 2017, \aap, 602,
  L8, \dodoi{10.1051/0004-6361/201731044}

\bibitem[{{Twicken} {et~al.}(2018){Twicken}, {Catanzarite}, {Clarke},
  {Girouard}, {Jenkins}, {Klaus}, {Li}, {McCauliff}, {Seader}, {Tenenbaum},
  {Wohler}, {Bryson}, {Burke}, {Caldwell}, {Haas}, {Henze}, \&
  {Sanderfer}}]{Twicken:DVdiagnostics2018}
{Twicken}, J.~D., {Catanzarite}, J.~H., {Clarke}, B.~D., {et~al.} 2018, \pasp,
  130, 064502, \dodoi{10.1088/1538-3873/aab694}

\bibitem[{Virtanen {et~al.}(2020)Virtanen, Gommers, Oliphant, Haberland, Reddy,
  Cournapeau, Burovski, Peterson, Weckesser, Bright, {van der Walt}, Brett,
  Wilson, Millman, Mayorov, Nelson, Jones, Kern, Larson, Carey, Polat, Feng,
  Moore, {VanderPlas}, Laxalde, Perktold, Cimrman, Henriksen, Quintero, Harris,
  Archibald, Ribeiro, Pedregosa, {van Mulbregt}, \& {SciPy 1.0
  Contributors}}]{scipy}
Virtanen, P., Gommers, R., Oliphant, T.~E., {et~al.} 2020, Nature Methods, 17,
  261, \dodoi{10.1038/s41592-019-0686-2}

\bibitem[{{Wang} {et~al.}(2019){Wang}, {Jones}, {Shporer}, {Fulton}, {Paredes},
  {Trifonov}, {Kossakowski}, {Eastman}, {Redfield}, {G{\"u}nther}, {Kreidberg},
  {Huang}, {Millholland}, {Seligman}, {Fischer}, {Brahm}, {Wang}, {Cruz},
  {Henry}, {James}, {Addison}, {Liang}, {Davis}, {Tronsgaard}, {Worku},
  {Brewer}, {K{\"u}rster}, {Zhang}, {Beichman}, {Bieryla}, {Brown},
  {Christiansen}, {Ciardi}, {Collins}, {Esquerdo}, {Howard}, {Isaacson},
  {Latham}, {Mazeh}, {Petigura}, {Quinn}, {Shahaf}, {Siverd}, {Rodler},
  {Reffert}, {Zakhozhay}, {Ricker}, {Vanderspek}, {Seager}, {Winn}, {Jenkins},
  {Boyd}, {F{\H{u}}r{\'e}sz}, {Henze}, {Levine}, {Morris}, {Paegert},
  {Stassun}, {Ting}, {Vezie}, \& {Laughlin}}]{Wang2019}
{Wang}, S., {Jones}, M., {Shporer}, A., {et~al.} 2019, \aj, 157, 51,
  \dodoi{10.3847/1538-3881/aaf1b7}

\bibitem[{{Wheatley} {et~al.}(2018){Wheatley}, {West}, {Goad}, {Jenkins},
  {Pollacco}, {Queloz}, {Rauer}, {Udry}, {Watson}, {Chazelas}, {Eigm{\"u}ller},
  {Lambert}, {Genolet}, {McCormac}, {Walker}, {Armstrong}, {Bayliss}, {Bento},
  {Bouchy}, {Burleigh}, {Cabrera}, {Casewell}, {Chaushev}, {Chote},
  {Csizmadia}, {Erikson}, {Faedi}, {Foxell}, {G{\"a}nsicke}, {Gillen},
  {Grange}, {G{\"u}nther}, {Hodgkin}, {Jackman}, {Jord{\'a}n}, {Louden},
  {Metrailler}, {Moyano}, {Nielsen}, {Osborn}, {Poppenhaeger}, {Raddi},
  {Raynard}, {Smith}, {Soto}, \& {Titz-Weider}}]{wheatley2018ngts}
{Wheatley}, P.~J., {West}, R.~G., {Goad}, M.~R., {et~al.} 2018, \mnras, 475,
  4476, \dodoi{10.1093/mnras/stx2836}

\bibitem[{{Wittenmyer} {et~al.}(2019){Wittenmyer}, {Bergmann}, {Horner},
  {Clark}, \& {Kane}}]{witt19}
{Wittenmyer}, R.~A., {Bergmann}, C., {Horner}, J., {Clark}, J., \& {Kane},
  S.~R. 2019, \mnras, 484, 4230, \dodoi{10.1093/mnras/stz236}

\bibitem[{{Wittenmyer} {et~al.}(2018){Wittenmyer}, {Horner}, {Carter}, {Kane},
  {Plavchan}, {Ciardi}, \& {MINERVA-Australis
  consortium}}]{2018arXiv180609282W}
{Wittenmyer}, R.~A., {Horner}, J., {Carter}, B.~D., {et~al.} 2018, arXiv
  e-prints, arXiv:1806.09282.
\newblock \doarXiv{1806.09282}

\bibitem[{{Wittenmyer} {et~al.}(2011){Wittenmyer}, {Tinney}, {Butler},
  {O'Toole}, {Jones}, {Carter}, {Bailey}, \& {Horner}}]{HJOccur1}
{Wittenmyer}, R.~A., {Tinney}, C.~G., {Butler}, R.~P., {et~al.} 2011, \apj,
  738, 81, \dodoi{10.1088/0004-637X/738/1/81}

\bibitem[{{Wittenmyer} {et~al.}(2013){Wittenmyer}, {Wang}, {Horner}, {Tinney},
  {Butler}, {Jones}, {O'Toole}, {Bailey}, {Carter}, {Salter}, {Wright}, \&
  {Zhou}}]{witt13}
{Wittenmyer}, R.~A., {Wang}, S., {Horner}, J., {et~al.} 2013, \apjs, 208, 2,
  \dodoi{10.1088/0067-0049/208/1/2}

\bibitem[{{Wittenmyer} {et~al.}(2020){Wittenmyer}, {Wang}, {Horner}, {Butler},
  {Tinney}, {Carter}, {Wright}, {Jones}, {Bailey}, {O'Toole}, \&
  {Johns}}]{cooljupiters}
---. 2020, \mnras, 492, 377, \dodoi{10.1093/mnras/stz3436}

\bibitem[{{Wizinowich} {et~al.}(2000){Wizinowich}, {Acton}, {Shelton},
  {Stomski}, {Gathright}, {Ho}, {Lupton}, {Tsubota}, {Lai}, {Max}, {Brase},
  {An}, {Avicola}, {Olivier}, {Gavel}, {Macintosh}, {Ghez}, \&
  {Larkin}}]{wizinowich200}
{Wizinowich}, P., {Acton}, D.~S., {Shelton}, C., {et~al.} 2000, \pasp, 112,
  315, \dodoi{10.1086/316543}

\bibitem[{{Wright} {et~al.}(2012){Wright}, {Marcy}, {Howard}, {Johnson},
  {Morton}, \& {Fischer}}]{HJOccur2}
{Wright}, J.~T., {Marcy}, G.~W., {Howard}, A.~W., {et~al.} 2012, \apj, 753,
  160, \dodoi{10.1088/0004-637X/753/2/160}

\bibitem[{{Wyttenbach} {et~al.}(2020){Wyttenbach}, {Molli{\`e}re},
  {Ehrenreich}, {Cegla}, {Bourrier}, {Lovis}, {Pino}, {Allart}, {Seidel},
  {Hoeijmakers}, {Nielsen}, {Lavie}, {Pepe}, {Bonfils}, \&
  {Snellen}}]{2020A&A...638A..87W}
{Wyttenbach}, A., {Molli{\`e}re}, P., {Ehrenreich}, D., {et~al.} 2020, \aap,
  638, A87, \dodoi{10.1051/0004-6361/201937316}

\bibitem[{{Zhou} {et~al.}(2019{\natexlab{a}}){Zhou}, {Huang}, {Bakos},
  {Hartman}, {Latham}, {Quinn}, {Collins}, {Winn}, {Wong}, {Kov{\'a}cs},
  {Csubry}, {Bhatti}, {Penev}, {Bieryla}, {Esquerdo}, {Berlind}, {Calkins}, {de
  Val-Borro}, {Noyes}, {L{\'a}z{\'a}r}, {Papp}, {S{\'a}ri}, {Kov{\'a}cs},
  {Buchhave}, {Szklenar}, {B{\'e}ky}, {Johnson}, {Cochran}, {Kniazev},
  {Stassun}, {Fulton}, {Shporer}, {Espinoza}, {Bayliss}, {Everett}, {Howell},
  {Hellier}, {Anderson}, {Collier Cameron}, {West}, {Brown}, {Schanche},
  {Barkaoui}, {Pozuelos}, {Gillon}, {Jehin}, {Benkhaldoun}, {Daassou},
  {Ricker}, {Vanderspek}, {Seager}, {Jenkins}, {Lissauer}, {Armstrong},
  {Collins}, {Gan}, {Hart}, {Horne}, {Kielkopf}, {Nielsen}, {Nishiumi},
  {Narita}, {Palle}, {Relles}, {Sefako}, {Tan}, {Davies}, {Goeke}, {Guerrero},
  {Haworth}, \& {Villanueva}}]{spinny:HAT-P-69}
{Zhou}, G., {Huang}, C.~X., {Bakos}, G.~{\'A}., {et~al.} 2019{\natexlab{a}},
  \aj, 158, 141, \dodoi{10.3847/1538-3881/ab36b5}

\bibitem[{{Zhou} {et~al.}(2019{\natexlab{b}}){Zhou}, {Bakos}, {Bayliss},
  {Bento}, {Bhatti}, {Brahm}, {Csubry}, {Espinoza}, {Hartman}, {Henning},
  {Jord{\'a}n}, {Mancini}, {Penev}, {Rabus}, {Sarkis}, {Suc}, {de Val-Borro},
  {Rodriguez}, {Osip}, {Kedziora-Chudczer}, {Bailey}, {Tinney}, {Durkan},
  {L{\'a}z{\'a}r}, {Papp}, \& {S{\'a}ri}}]{spinny:HATS-70b}
{Zhou}, G., {Bakos}, G.~{\'A}., {Bayliss}, D., {et~al.} 2019{\natexlab{b}},
  \aj, 157, 31, \dodoi{10.3847/1538-3881/aaf1bb}

\bibitem[{{Zhou} {et~al.}(2019{\natexlab{c}}){Zhou}, {Huang}, {Bakos},
  {Hartman}, {Latham}, {Quinn}, {Collins}, {Winn}, {Wong}, {Kov{\'a}cs},
  {Csubry}, {Bhatti}, {Penev}, {Bieryla}, {Esquerdo}, {Berlind}, {Calkins}, {de
  Val-Borro}, {Noyes}, {L{\'a}z{\'a}r}, {Papp}, {S{\'a}ri}, {Kov{\'a}cs},
  {Buchhave}, {Szklenar}, {B{\'e}ky}, {Johnson}, {Cochran}, {Kniazev},
  {Stassun}, {Fulton}, {Shporer}, {Espinoza}, {Bayliss}, {Everett}, {Howell},
  {Hellier}, {Anderson}, {Collier Cameron}, {West}, {Brown}, {Schanche},
  {Barkaoui}, {Pozuelos}, {Gillon}, {Jehin}, {Benkhaldoun}, {Daassou},
  {Ricker}, {Vanderspek}, {Seager}, {Jenkins}, {Lissauer}, {Armstrong},
  {Collins}, {Gan}, {Hart}, {Horne}, {Kielkopf}, {Nielsen}, {Nishiumi},
  {Narita}, {Palle}, {Relles}, {Sefako}, {Tan}, {Davies}, {Goeke}, {Guerrero},
  {Haworth}, \& {Villanueva}}]{zhou19}
{Zhou}, G., {Huang}, C.~X., {Bakos}, G.~{\'A}., {et~al.} 2019{\natexlab{c}},
  \aj, 158, 141, \dodoi{10.3847/1538-3881/ab36b5}

\bibitem[{{Zhou} {et~al.}(2020){Zhou}, {Winn}, {Newton}, {Quinn}, {Rodriguez},
  {Mann}, {Rizzuto}, {Vanderburg}, {Huang}, {Latham}, {Teske}, {Wang},
  {Shectman}, {Butler}, {Crane}, {Thompson}, {Henry}, {Paredes}, {Jao},
  {James}, \& {Hinojosa}}]{DSTucAb}
{Zhou}, G., {Winn}, J.~N., {Newton}, E.~R., {et~al.} 2020, \apjl, 892, L21,
  \dodoi{10.3847/2041-8213/ab7d3c}

\end{thebibliography}
\bibliographystyle{aasjournal}



\end{document}